\documentclass[useAMS,usenatbib,usegraphicx]{mn2e}

\title[Core expansion in massive star clusters]{Black holes and core expansion in massive star clusters}
\author[A.~D.~Mackey et al.]{A.~D.~Mackey$^{1}$, 
M.~I.~Wilkinson$^{2}$, M.~B.~Davies$^{3}$, and G.~F.~Gilmore$^{4}$\\
$^{1}$Institute for Astronomy, University of Edinburgh, Royal Observatory, Blackford 
Hill, Edinburgh, EH9 3HJ, UK\\
$^{2}$Department of Physics \& Astronomy, University of Leicester, University Road,
Leicester, LE1 7RH, UK\\
$^{3}$Lund Observatory, Box 43, SE-221 00 Lund, Sweden\\
$^{4}$Institute of Astronomy, University of Cambridge, Madingley Road, Cambridge, 
CB3 0HA, UK}
\begin{document}

\date{Draft version \today}

\pagerange{\pageref{firstpage}--\pageref{lastpage}} \pubyear{2007}

\maketitle

\label{firstpage}

\begin{abstract}
In this study we present the results from realistic $N$-body modelling of massive star 
clusters in the Magellanic Clouds. We have computed eight simulations with $N \sim 10^5$ particles; 
six of these were evolved for at least a Hubble time. The aim of this modelling is to examine 
in detail the possibility of large-scale core expansion in massive star clusters, and search for 
a viable dynamical origin for the radius-age trend observed for such objects in the Magellanic 
Clouds. We identify two physical processes which can lead to significant and prolonged cluster 
core expansion -- mass-loss due to rapid stellar evolution in a primordially mass segregated cluster, 
and heating due to a retained population of stellar-mass black holes, formed in the supernova 
explosions of the most massive cluster stars. These two processes operate over different time-scales 
and during different periods of a cluster's life. The former occurs only at early times and cannot
drive core expansion for longer than a few hundred Myr, while the latter typically does not begin 
until several hundred Myr have passed, but can result in core expansion lasting for many Gyr.
We investigate the behaviour of each of these expansion mechanisms under different
circumstances -- in clusters with varying degrees of primordial mass segregation,
and in clusters with varying black hole retention fractions. In combination, the two processes 
can lead to a wide variety of evolutionary paths on the radius-age plane, which fully cover the
observed cluster distribution and hence define a dynamical origin for the radius-age trend
in the Magellanic Clouds. We discuss in some detail the implications of core expansion for
various aspects of globular cluster research, as well as the possibility of observationally 
inferring the presence of a significant population of stellar-mass black holes in a cluster.
\end{abstract}


\section{Introduction}
As relatively simple objects which are integral to the study of many fundamental astronomical 
processes, massive star clusters are central to a wide variety of astrophysics over all scales 
-- from star formation and stellar and binary evolution, through stellar exotica and variable stars, 
and the dynamics of self-gravitating systems, to galaxy formation and evolution, with implications 
for cosmology. In the context of this wider astrophysics however, it is clearly essential that
we understand the clusters themselves: how internal physical processes in clusters shape their 
overall characteristics (and vice versa), and how individual clusters interact with and are 
influenced by their local environments. Only when armed with this knowledge is it possible
to disentangle cluster evolutionary processes from the specific astrophysics under
investigation.

From an observational perspective, we are provided with only a limited set of massive stellar
clusters which are close enough to us that they may be fully resolved using presently available
facilities (and hence thoroughly studied on a star-by-star basis). The Galactic globular clusters, 
while constituting the closest ensemble, are not ideal for studying massive star cluster
evolution, primarily because they are almost exclusively ancient objects with ages $\sim 10-13$ 
Gyr \citep[see e.g.,][]{rosenberg:99,salaris:02,krauss:03,deangeli:05}. Therefore, while we are 
able to precisely measure the end-points of massive star cluster evolution, the long-term 
development which brought them to these observed states must be almost completely inferred. 
Fortunately, it is relatively straightforward to turn our attention to the Large and Small 
Magellanic Clouds (LMC/SMC), which both possess extensive systems of star clusters with masses
comparable to the Galactic globulars, but crucially {\it of all ages:} $10^6 \la \tau \la 10^{10}$ 
yr. These two nearby galaxies are hence of fundamental importance to studies of star cluster 
evolution, because they are the only systems in which we can directly observe snapshots of 
cluster development over the last Hubble time using a sample of fully resolved objects.

\begin{figure}
\begin{center}
\includegraphics[width=0.45\textwidth]{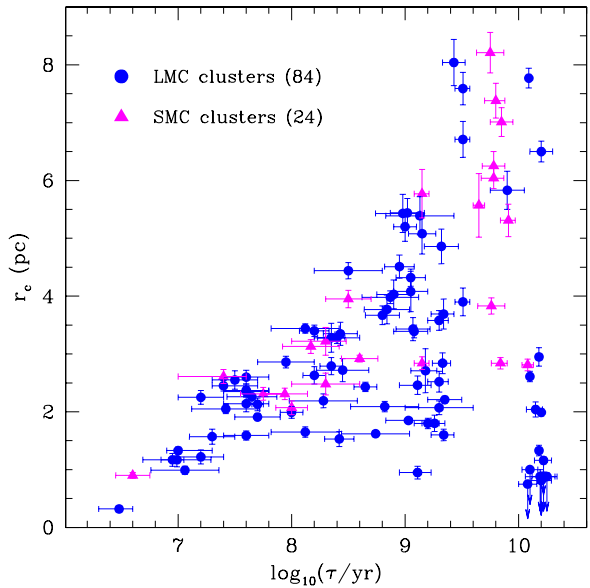}
\caption{Core radius versus age for massive stellar clusters in the Large and Small
Magellanic Clouds. This Figure includes all clusters from the HST/WFPC2 measurements 
of \citet{mackey:03a,mackey:03b} as well as the new preliminary HST/ACS measurements 
of Mackey et al. (2008, in prep.) from Program \#9891, and the recent measurements of
BS90 by \citet{rochau:07}. Core radii for several of the oldest compact clusters are 
upper limits, as indicated \citep[see][]{mackey:03a}.}
\label{f:radiusage}
\end{center}
\end{figure}

Some of the earliest studies to take advantage of this situation and investigate the structural 
evolution of massive stellar clusters were those of Elson and collaborators. In 
particular, \citet*{elson:87} constructed radial brightness and density profiles for $10$ young 
clusters in the LMC, while \citet*{elson:89} and \citet{elson:91,elson:92} extended this study 
to a larger sample of LMC clusters including much older objects.
They discovered a striking relationship between cluster core size and age -- specifically, that
the observed spread in core radius is a strongly increasing function of age. The youngest clusters 
in their sample possessed compact cores with $r_c \sim 1-2$ pc, while the oldest clusters 
exhibited a range $0 \la r_c \la 6$ pc (cf. Fig. \ref{f:radiusage}). Here, cluster core size is
parametrised by the observational core radius, $r_c$, defined as the projected radius at which the
surface density/brightness has decreased to half its central value.

The advent of the Hubble Space Telescope (HST) has allowed this discovery to be re-addressed 
observationally in significantly more detail than was possible with ground-based facilities. HST 
imaging can resolve Magellanic Cloud star clusters even in their inner cores, so that star counts 
may be conducted to very small projected radii and accurate surface density/brightness 
profiles constructed. \citet{mackey:03a} obtained structural measurements from a homogeneous 
compilation of archival Wide Field Planetary Camera 2 (WFPC2) imaging of $53$ massive LMC clusters 
spanning the full age range. We found essentially the same relationship as \citet{elson:89} -- the 
youngest massive LMC clusters possess compact cores of typical radius $\sim 1-2$ pc, but with 
increasing age the spread in core radius increases such that the oldest clusters span the range 
$0 \la r_c \la 8$ pc. \citet{mackey:03b} subsequently extended these HST measurements to $10$ SMC 
clusters, demonstrating for the first time that a radius-age trend indistinguishable from that
observed in the LMC exists for this star cluster system.

Following these two studies, we were granted HST time to conduct a snapshot survey of additional
massive LMC and SMC star clusters using the Advanced Camera for Surveys (ACS; HST program \#9891), 
with the aim of improving the sampling of the radius-age parameter space. In all, $31$ extra LMC 
and $13$ extra SMC clusters were successfully imaged, significantly enlarging the sample.
Final structural measurements for these objects are yet to be published (Mackey et al. 2008, in 
prep.); however preliminary results for the core sizes are plotted in Figure \ref{f:radiusage}, along
with the WFPC2 measurements of \citet{mackey:03a,mackey:03b}. In obtaining these new parameters,
photometric measurements were made using the pipeline described by \citet{mackey:04} and 
\citet*{mackey:06}, while radial brightness profiles were constructed following procedures
essentially identical to those described by \citet{mackey:03a} but adapted for ACS Wide Field Channel 
(WFC) imaging. Figure \ref{f:radiusage} also includes the recent HST/ACS measurements of the
SMC cluster BS90 by \citet{rochau:07}. Note that the core radii for several of the oldest, most 
compact clusters in Fig. \ref{f:radiusage} are upper limits, as indicated. This is due to severe
crowding in the HST imaging, and the possibility that several of these clusters are core-collapsed
objects \citep[see][]{mackey:03a}. All affected clusters have measured $r_c < 1$ pc.

Figure \ref{f:radiusage} represents the most complete and up-to-date information presently available
regarding the radius-age trend in the LMC and SMC star cluster systems. The upper envelope is very 
well defined for all ages up to a few Gyr. At older times than this, the full range of core radii 
observed in massive star clusters is allowed. In fact the situation is even more dramatic than was
appreciated by earlier studies. Several of the oldest clusters in the new ACS sample fall off the top 
of the diagram: the Reticulum cluster in the LMC, with age $\tau \sim 12-13$ Gyr and $r_c \sim 14.8$ 
pc; and Lindsay 1 and 113 in the SMC, with $\tau \sim 9$ Gyr and $r_c \sim 16.4$ pc, and $\tau \sim 5$ 
Gyr and $r_c \sim 11$ pc, respectively. Hence the size range observed for the oldest clusters is 
$0 \la r_c \la 17$ pc.

\begin{figure}
\begin{center}
\includegraphics[width=0.45\textwidth]{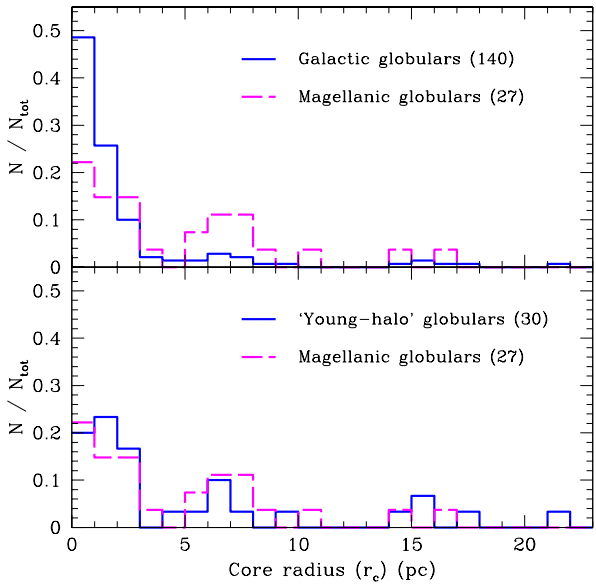}
\caption{Core radius distributions for the oldest ($\tau \ga 7$ Gyr) Magellanic Cloud
clusters from Fig. \ref{f:radiusage} (dashed lines) and for Galactic globular 
clusters (solid lines). {\bf Upper panel:} The full sample of Galactic globular clusters 
with suitable measurements of $r_c$ is plotted; objects which are members or ex-members of the 
Sagittarius dwarf galaxy are excluded. Data are from \citet{harris:96} (2003 online update) 
with new measurements as described in \citet{mackey:04} and \citet{hilker:06}.
The distributions are very similar in range and overall shape; however the Galactic globulars 
have a sharp peak at small core radii. Note that all Magellanic Cloud clusters with
core radius measurements which are upper limits already fall in the smallest $r_c$ bin.
{\bf Lower panel:} If only the Galactic ``young-halo'' 
subsystem is considered (objects which preferentially lie at Galactocentric radii 
beyond $\sim 15$ kpc) a very much closer match is observed.}
\label{f:rchist}
\end{center}
\end{figure}

These recent measurements of very extended objects are consistent with those for several old globular 
clusters in the Fornax and Sagittarius dwarf galaxies -- Fornax cluster 1, and Terzan 8 and Arp 2 
\citep{mackey:03c} -- which also have very large core radii. A number of Galactic globular clusters 
are also known to possess extended cores, as seen in Fig. \ref{f:rchist} (upper panel), which shows 
the core radius distribution of the oldest LMC and SMC clusters from Fig. \ref{f:radiusage} compared 
with that for the Galactic globular cluster system. The observed ranges in $r_c$ match well, as do 
the general shapes of the distributions. The main difference is that the distribution for Galactic 
globulars is more sharply peaked at small $r_c$. This is not surprising given that a large fraction 
of the Galactic globulars reside in the inner Galaxy, where tidal forces are expected to
rapidly destroy loosely bound clusters. Indeed, following \citet{mackey:04} 
\citep[see also][]{zinn:93,mackey:05}, if only members of the Galactic globular cluster 
``young-halo'' subsystem are considered (most of which are located at Galactocentric radii 
larger than $\sim 15$ kpc), the core radius distribution is an excellent match to that observed 
for the oldest LMC and SMC objects (Fig. \ref{f:rchist}, lower panel).
The end-points of structural evolution observed for the Galactic globulars appear quite 
consistent with the end-points of the radius-age trend observed in the Magellanic Clouds. 

The simplest interpretation of the radius-age trend is that it represents
the progression of cluster structural properties with time\footnote{An additional possibility
is that cluster formation conditions have changed significantly over the past $\sim10$ Gyr.
However, presently available constraints on this proposition are limited, and we do not 
discuss it further here.}.
In this scenario, Figures \ref{f:radiusage} and \ref{f:rchist} provide striking evidence that our 
understanding of massive star cluster evolution is incomplete, since standard models do 
not predict an order-of-magnitude expansion of the core radius over a Hubble time 
\citep[e.g.,][]{meylan:97}. Identifying the origin of the radius-age trend is therefore
of considerable importance for star cluster astrophysics, and all related fields in which
star clusters play a prominent role. 

\citet{elson:89} discussed the possibility that the increasing spread in radius with age could 
reflect inter-cluster variations in the slope of the stellar initial mass function (IMF). Clusters 
with flat IMFs possess comparatively more massive stars than those with steep IMFs. Consequently, 
they suffer more severe mass loss due to stellar evolution at early times, resulting in increased 
relative expansion. However, \citet{elson:89} found that to induce expansion along the upper 
envelope of the observed trend would require a very flat IMF slope, and the resulting early mass 
loss would be severe enough to disrupt the cluster within only a few tens of Myr. An additional 
problem with the IMF hypothesis concerns the time-scale -- the severe mass loss phase lasts for 
roughly only the first $\sim 100$ Myr of a cluster's evolution. Therefore, it cannot drive significant 
expansion over the full range of ages observed for Magellanic Cloud clusters. There is also an 
increasing body of observational evidence that the IMF in young star clusters is more-or-less 
invariant \citep[see e.g.,][]{kroupa:01,degrijs:02c}.

\citet{wilkinson:03} used $N$-body simulations of small star clusters to investigate whether
the radius-age trend could reflect core expansion induced by populations of binary stars, or
by time-varying tidal fields such as those which clusters on highly elliptical orbits
might experience. They observed similar core radius evolution for model clusters on both circular 
and elliptical orbits, and therefore concluded that the tidal fields of the Magellanic Clouds have 
not yet significantly influenced the evolution of the intermediate-age clusters in these systems.
Furthermore, while they found that the presence of large numbers of hard primordial binaries in 
their small clusters did lead to a degree of core radius expansion, the magnitude of the effect was 
insufficient to explain the observed radius-age trend.

\citet{hunter:03} suggested that rather than representing the results of dynamical evolution,
the radius-age trend might instead have its origins in a size-of-sample effect. They measured
a very large sample of Magellanic Cloud clusters with masses 
$10 \la M_{{\rm cl}} \la 10^6\,{\rm M}_\odot$ and found the signature
of such an effect in their data. On a log-abscissa plot such as Fig. \ref{f:radiusage}, older
ages correspond to larger time intervals and hence to more clusters forming in each log-time
interval. Since the star cluster mass function decreases steeply with increasing cluster mass,
this results in the maximum observed cluster mass in each log-time interval increasing with age.
The clusters in the sample of \citet{hunter:03} also showed a weak dependence of size on total mass
in that more massive clusters have larger characteristic radii. Combined with the size-of-sample
effect, this leads to a size-age distribution with an upper envelope not dissimilar to that
evident in Fig. \ref{f:radiusage}. However, it is not clear how applicable this argument is to
the cluster sample considered in Fig. \ref{f:radiusage}, because all these clusters have
masses $M \ga 10^4\,{\rm M}_\odot$, and show no coherent link between total mass and core radius
\citep[e.g.,][]{mackey:03a}. Indeed, restricting the sample of \citet{hunter:03} to clusters with 
$M \ga 10^4\,{\rm M}_\odot$, their relationship between size and mass is no longer evident.
Hence, a size-of-sample effect apparently does not explain the radius-age trend visible in 
Fig. \ref{f:radiusage}.

Finally, \citet{merritt:04} examined the formation of cores in primordially cusped clusters 
(i.e., objects which initially have $r_c \approx 0$) due to the presence of populations of massive 
stellar remnants. They used analytic calculations in combination with simplified $N$-body models
(composed of equal-mass non-evolving particles) to show that the orbits of the remnants decay 
due to dynamical friction so that they sink to the cluster centres, heating the stellar background 
in the process and turning the cusp into a core. The authors also note that further heating of the 
core may continue over a longer time-scale, due to subsequent evolution of the subsystem of massive 
remnants. The rates of core growth determined by \citet{merritt:04} are moderately successful in 
reproducing the observed radius-age trend; however their models seem to require a range of initial 
densities which is significantly larger than that found for young clusters in the Magellanic Clouds. 
It is also not clear how their results would respond to the introduction of a mass spectrum and stellar 
evolution into the simulations, or the introduction of more realistic initial conditions including 
the possibility of primordial mass segregation.

As demonstrated above, the radius-age trend is indistinguishable in the LMC and SMC, and the 
end-points of the trend are consistent with the core radius distributions of the Galactic 
globular clusters as well as of globular clusters belonging to the Fornax and Sagittarius dwarf 
galaxies \citep{mackey:03c}. These galaxies cover a very wide range of masses and morphological
types, and hence possess very different tidal fields and possible external torques. This strongly 
suggests that the radius-age trend is primarily driven by internal cluster processes, rather than 
external influences \citep[cf.][]{wilkinson:03}. To this end, we have conducted a series of 
large-scale, realistic $N$-body simulations of Magellanic Cloud clusters with the aim of 
investigating an internal dynamical origin for the radius-age trend. More specifically, we
have examined the influence of stellar-mass black holes (BHs), formed in the supernova explosions
of the most massive cluster stars, on the long-term evolution of massive stellar clusters.
We have also investigated the role played by primordial mass segregation in shaping the
early evolution of massive stellar clusters. The basic results from several of our key simulations
have been outlined in a recent {\it Letter} \citep{mackey:07}; in the present paper, we 
describe in detail the complete results of our modelling. 

\section{Numerical setup}
\subsection{$N$-body code and initial conditions}
\label{ss:nbodycode}
We use direct, realistic $N$-body modelling in order to investigate the structural and 
dynamical evolution of massive star clusters in the Magellanic Clouds. 
Simulations of this type are a powerful tool for such work because they incorporate all
the relevant physical processes with a minimum of simplifying assumptions. Recent 
technological developments mean that it is now feasible to run models with $N$ sufficiently 
large so as to be directly comparable to observed clusters. This has a number of 
advantages, discussed below.

For the present study, we have used the {\sc nbody4} code \citep{aarseth:99,aarseth:03} 
in combination with a 32-chip GRAPE-6 special purpose computer \citep{makino:03} at the 
Institute of Astronomy, Cambridge. This code uses the fourth-order Hermite scheme 
\citep{makino:91} and fast evaluation of the force and its first time derivative by the 
GRAPE-6 to integrate the equations of motion. Close encounters between stars, including
stable binary systems and hierarchies, are integrated via state-of-the-art two-body and
chain regularization schemes \citep{mikkola:93,mikkola:98}. Also included in {\sc nbody4} 
are routines for modelling the stellar evolution of both single and binary stars. For
single stars these take the form of the analytical formulae derived by \citet*{hurley:00} 
from detailed stellar evolution models, following stars from the zero-age main sequence through 
to remnant phases (such as white-dwarfs, neutron stars and black holes). Binary star evolution 
is calculated in a similar manner, following the prescription of \citet*{hurley:02} and allowing
for such phases as the tidal circularization of orbits, mass transfer, and common-envelope 
evolution. The stellar and binary evolution is calculated in time with the dynamical
integration so that interaction between the two is simulated in a consistent fashion
\citep[e.g.,][]{hurley:01,hurley:05}. The stellar evolution routines allow a spread 
in stellar masses covering the range $0.1-100\,{\rm M}_\odot$, so that one can construct 
any desired IMF for a model cluster. In addition, a uniform metallicity for 
the cluster may be selected in the range $Z = 0.0001-0.03$. A mass-loss prescription is included
such that evolving stars lose gas through winds and supernova explosions. This gas is 
instantaneously removed from the cluster. Such mass-loss can rapidly alter the gravitational 
potential of a star cluster, strongly affecting its early structural and dynamical evolution.

\begin{figure}
\begin{center}
\includegraphics[width=0.45\textwidth]{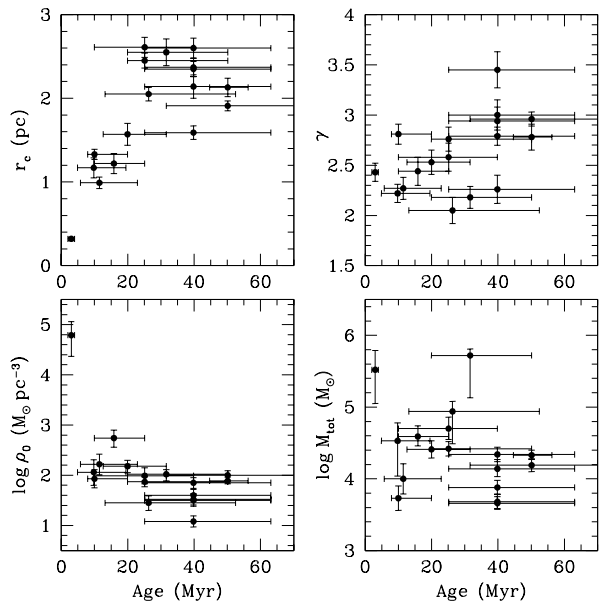}
\caption{Properties of the young massive clusters observed in the LMC and SMC. Structural 
data are taken from \citet{mackey:03a,mackey:03b}, while the central density ($\rho_0$) and 
total mass ($M_{{\rm tot}}$) estimates are taken from \citet{mclaughlin:05}, as discussed 
in the text.}
\label{f:youngclusters}
\end{center}
\end{figure}

When constructing the initial conditions for our simulated clusters, we were careful to develop
models as similar as possible to the youngest massive clusters observed in the LMC and 
SMC\footnote{Although we again note the possibility that the initial conditions for massive
clusters which formed at high redshift may be different to those for clusters forming today.}. 
In general, young massive Magellanic Cloud clusters possess radial surface brightness profiles
best described by three-parameter models of the form (\citealt*{elson:87}; EFF models hereafter):
\begin{equation}
\mu(r_p) = \mu_0 \left( 1+\frac{r_p^2}{a^2} \right) ^{-\frac{\gamma}{2}}\,\,,
\label{e:effproject}
\end{equation}
where $r_p$ is the projected radius, $\mu_0$ is the central surface brightness, $\gamma$ 
determines the power-law slope of the fall-off in surface brightness at large radii, 
and $a$ is the scale length. These models are a subset of the more general family of
models presented by \citet{zhao:96}. The scale length, $a$, is related to the observational 
core-radius, $r_c$, defined here as the projected radius at which the surface brightness has 
dropped to half $\mu_0$, by:
\begin{equation}
r_c = a(2^{\frac{2}{\gamma}} - 1)^{\frac{1}{2}} \,\,.
\label{e:rc}
\end{equation}
Some of the global properties observed for young massive Magellanic Cloud clusters are summarized 
in Fig. \ref{f:youngclusters}. In this plot, we have taken the structural parameters $r_c$ and $\gamma$ 
from \citet{mackey:03a,mackey:03b}, who constructed surface brightness profiles from HST
photometry and fit EFF models as defined above. We have also taken the central density and total mass 
estimates of \citet{mclaughlin:05}, which were computed in a more robust manner than 
those provided by \citet{mackey:03a,mackey:03b}. 

All young LMC and SMC clusters are observed to have cored (rather than cusped) profiles 
-- even the ultra-compact cluster R136 exhibits a small core (see e.g., the discussion in 
\citealt{mackey:03a}, and references therein). Their profiles are well fit by
EFF models with $\gamma \sim 2.5$: \citet{elson:87} found a median value of $\gamma = 2.6$ 
and a range $2.2 \la \gamma \la 3.2$ for their ten young LMC clusters, while the larger sample 
plotted in Fig. \ref{f:youngclusters} covers the 
range $2.05 \leq \gamma \leq 3.79$ and has a median value $\gamma = 2.67$. Excluding R136,
the young LMC and SMC clusters typically have central densities in the range 
$1.6 \la \log \rho_0 \la 2.8$ (where $\rho_0$ is in units of ${\rm M}_\odot\,{\rm pc}^{-3}$), 
and total masses in the range $4 \la \log M_{\rm{tot}} \la 5$ (where $M_{{\rm tot}}$ is in
units of ${\rm M}_\odot$). R136 is the youngest cluster in the sample, with $\tau \sim 3$ Myr, 
and also has the greatest central density: $\log \rho_0 \approx 4.8$.

Our model clusters are generated such that they initially have structural parameters in 
projection which are consistent with those observed for the youngest LMC and SMC clusters. 
This is achieved by selecting stellar positions randomly from the density distribution of
an EFF model with $\gamma = 3$. Each star is assigned a velocity drawn from a Maxwellian
distribution, where the velocity dispersion $\sigma$ is calculated using the Jeans equations
assuming an isotropic velocity distribution. In applying this generation algorithm it is 
important to know that for the EFF family of models, the deprojected density profile is 
given by:
\begin{equation}
\rho(r) = \rho_0 \left( 1+\frac{r^2}{a^2} \right) ^{-\frac{\gamma+1}{2}}\,\,,
\label{e:eff3d}
\end{equation}
where $\rho_0$ is the central volume density. From this, we can derive expressions for the
isotropic velocity dispersion as a function of radius. The $\gamma = 3$ case is the closest
value of $\gamma$ to the median $\gamma = 2.67$ observed for young LMC clusters for which
the expression for $\sigma$ is analytic (see Appendix \ref{a:profiles}).

We assign the stars in each cluster a range of masses according to the 
IMF of \citet{kroupa:01}, which is a multiple-part power-law $\xi(m) \propto m^{-\alpha_i}$,
where $\xi(m)dm$ is the number of single stars falling in the mass interval $m$ to $m+dm$.
\citet{kroupa:01} derived his IMF from a large compilation of measurements of young
stellar clusters, including many in the LMC. This is in contrast with many other widely used
IMFs -- for example, the \citet*{kroupa:93} IMF, which was derived from observations of Galactic 
field stars in the local neighbourhood and towards the Galactic poles. Therefore, we prefer 
the \citet{kroupa:01} IMF for direct modelling of Magellanic Cloud clusters.

We impose a stellar mass range of $0.1 - 100\,{\rm M}_\odot$ for our $N$-body clusters. 
The extremes of this range are set by the lowest and highest mass stars for which reliable 
stellar evolution routines are incorporated in {\sc nbody4}. Although stars more massive 
than $100\,{\rm M}_\odot$ do form in large star clusters \citep[e.g.,][]{weidner:06}, this 
upper limit is perfectly acceptable for our present models. For example, \citet{massey:98} 
found only $\sim 10-20$ stars with $M > 80\,{\rm M}_\odot$ in the extreme LMC cluster R136 
(depending on the adopted stellar evolution models), while the revised calculations in 
\citet{massey:04,massey:05}, which incorporate improved spectroscopy and modelling, suggest 
significant reductions in these estimated masses. In practice, we expect that increasing our 
upper mass limit to the inferred fundamental maximum stellar mass $\sim 150\,{\rm M}_\odot$ 
of \citet{weidner:04} would have essentially no discernible effect on the global evolution 
we observe for our models. We note that our lower mass limit means that in practice we only 
utilise the exponents $\alpha_1 - \alpha_3$ in the \citet{kroupa:01} IMF.

Our adopted IMF and stellar mass range, along with the requirement that our model clusters 
have masses commensurate with those observed for young Magellanic Cloud clusters 
(Fig. \ref{f:youngclusters}), allows the total number of stars in each given model to be 
assigned. Choosing $N \approx 10^5$ stars results in initial total cluster masses of 
$\log M_{\rm tot} \sim 4.75$.

It is only relatively recently, with the advent of special-purpose hardware, that it has been 
possible to follow models with such large $N$ over more than a Hubble time of evolution. There 
are several advantages to running simulations of this size. First, the model star clusters are 
directly comparable in terms of total mass and central density (see below) to the massive clusters 
observed in the LMC and SMC. We are therefore now moving into the regime where many of the 
scaling-with-$N$ issues which it has been necessary to account for in previous studies when applying 
the results of $N$-body simulations to the evolution of real clusters \citep[see e.g.,][]{aarseth:98} 
are circumvented. In addition, with such large $N$, fluctuations in the global evolution of the 
$N$-body model are reduced to the point where they are not significant. For small-$N$ models, 
it has been standard practice to average the results of a number of simulations to reduce such 
fluctuations, the amplitudes of which increase with decreasing $N$ 
\citep[e.g.,][]{giersz:94,wilkinson:03,heggie:06}. For large-$N$ models, this process is not 
necessary \citep[see e.g.,][]{hurley:05}\footnote{We caution, however, that small-number statistics
may still be subject to significant fluctuations between simulations -- an example in the
present work are the properties of ejected binaries, as discussed in Sections \ref{ss:pair1}
and \ref{ss:pair2}.}. Finally, with $N \sim 10^5$ we are able to 
perform detailed simulated observations of our models. This allows us to derive quantities from 
the simulations which are directly comparable to the genuine observations of LMC and SMC star
clusters. As we discuss more fully in Section \ref{s:simobs}, this step is a vital ingredient
in the analysis of realistic $N$-body models.

Star clusters in the LMC are observed at galactocentric radii between $\sim 0-14$ kpc.
We therefore evolve our model clusters in a weak external tidal field, rather than in isolation. 
This external field is incorporated by imposing the gravitational potential of a point-mass LMC 
with $M_{\rm g} = 9\times 10^9\,{\rm M}_\odot$, and placing the clusters on circular orbits of 
galactocentric radius $R_{\rm g} = 6$ kpc. \citet{wilkinson:03} give a more detailed description of
the implementation of the external field within {\sc nbody4}, which is done by integrating 
the equations of motion in an accelerating but non-rotating reference frame, centred on
the cluster's centre-of-mass. Adopting a point-mass LMC is a significant over-simplification; 
however, as noted by \citet{wilkinson:03}, the gradient of this potential is within a factor of 
two of that in the LMC mass model of \citet{vandermarel:02} at our orbital radius. 
More importantly however, our aim is not to examine the effect of tidal fields on 
the evolution of star cluster cores -- \citet{wilkinson:03} demonstrated that external fields
comparable to those experienced by Magellanic Cloud clusters do not result in strong core evolution.
Rather, we impose an external tidal field so that the gradual evaporation of stars from 
the cluster may be simulated in a self-consistent fashion, and the rates of evaporation
between different models with the same external potential and escape criterion may be
easily compared.

In an external potential, the tidal radius of a star cluster on a circular orbit may be 
estimated from the relationship \citep{king:62}:
\begin{equation}
r_t = R_{\rm g} \left( \frac{M_{\rm cl}}{3 M_{\rm g}} \right)^{\frac{1}{3}} \,\,,
\label{e:tidalradius}
\end{equation}
where $M_{{\rm cl}}$ is the cluster mass. In {\sc nbody4} stars are deemed to have escaped 
the cluster when they reach a radius $2r_t$. This is a legitimate approximation -- for 
example \citet{heggie:01} shows that although cluster stars may on occasion possess orbits 
which allow them to move far beyond $r_t$ and yet return to the cluster, in practice the 
vast majority of stars which move beyond a few $r_t$ are permanently lost. In our models 
$r_t$ is a non-static quantity (since the cluster mass is monotonically decreasing with 
time); therefore, the instantaneous value is used when assessing the above escape criterion. 
We caution that other different recipes for the implementation of tidal fields exist, which 
can lead to significantly different escape rates and thus cluster lifetimes 
\citep*[e.g.,][]{baumgardt:01,trenti:07}.

Within the $N$-body code the equations of motion are integrated in scaled units such that
$G=1$ and the virial radius and total mass of the cluster are also set to unity
\citep{heggie:86}. For a star cluster in virial equilibrium the initial energy in these
units is $-1/4$ and the crossing time is $2\sqrt{2}$. Given the total mass of the cluster
in solar masses and an appropriately chosen length-scale (which determines the conversion 
from $N$-body units to physical units) it is simple to obtain the conversion factors for 
time and velocity from $N$-body units to Myr and km$\,$s$^{-1}$, respectively.

Since the chosen length-scale sets the conversion from $N$-body units to physical
units, it controls the physical density of the cluster and hence the physical time-scale
on which internal dynamical processes occur. We assume that a model cluster initially just
fills its tidal radius. The value of this radius at time $\tau = 0$, determined via Eq. 
\ref{e:tidalradius} with $M_{{\rm cl}} = M_{{\rm tot}}$, therefore sets the ratio between 
the length scale in $N$-body units and in physical units (pc). EFF profiles formally have 
no outer bound, so when randomly generating the initial stellar positions we only accept
stars lying within the estimated tidal radius of the cluster under consideration.

The above process determined a length scale of $8.26$ pc for our model clusters.
This results in an initial central mass density of $\log \rho_0 = 2.31$ and a core radius
$r_c = 1.90$ pc for these objects, values which match well those observed for many young
Magellanic Cloud clusters (Fig. \ref{f:youngclusters}). We note that the clusters 
described here are not in any way mass segregated; however we also ran
simulations of clusters incorporating various degrees of primordial mass segregation, the 
details of which are described below in Section \ref{ss:mseg}. Those objects have the same
length scale as the clusters described here, but smaller core radii and much higher central 
densities, more in line with those of the very young LMC cluster R136. Given this correspondence 
between our models and the properties of young Magellanic Cloud clusters, we are confident 
in our selection of an appropriate length scale.

In order to allow investigation of the effects of a population of stellar-mass BHs on 
cluster evolution, we modified {\sc nbody4} to allow control of the production
of BHs in supernova explosions. For the present modelling this is implemented in a relatively
simplistic manner; however in principle the relevant code could be altered to cover more
complex formation scenarios. We define three variable parameters -- the minimum initial mass 
of BH progenitor stars, the masses of the BHs themselves, and the sizes of the natal velocity 
kicks which they receive. In each run, all stars initially more massive than $20\,{\rm M}_\odot$
produce BHs, with masses uniformly distributed in the range $8 < M_{{\rm BH}} < 12\,{\rm M}_\odot$
so that the mean BH mass is $10\,{\rm M}_\odot$. This range is consistent with dynamical
masses obtained from observations of X-ray binaries \citep[e.g.,][]{casares:06}. We generate
model clusters using the same random seed, so that they initially contain identical stellar 
populations. Our adopted IMF and total number of particles result in the formation of $198$
BHs in all clusters. 

The retention fraction of BHs in a given cluster, $f_{{\rm BH}}$, is strongly dependent on 
the natal kicks given to the BHs at formation. If a kick is too strong (i.e., $v_{{\rm kick}}$ 
larger, roughly, than the escape velocity of the cluster, $v_{{\rm esc}}$), a BH will quickly 
cross the limiting radius of the cluster and be removed from the simulation. Under our 
modifications to {\sc nbody4}, BH kicks can vary from zero ($f_{{\rm BH}} = 1$) to very large 
($f_{{\rm BH}} = 0$) and can be set as a constant, or selected randomly from a uniform distribution 
with specified limits, or a Maxwellian distribution with a specified mean. By varying these 
aspects of natal BH kicks, it is straightforward to control the BH retention fraction in any 
given model.

Although {\sc nbody4} allows the inclusion of primordial binary stars in cluster models, in
the present paper we investigate only models with no primordial binaries. The inclusion of 
such objects would introduce a very large new area of parameter space for investigation,
beyond the scope of the time available for our simulations. Even so, any complete modelling of
Magellanic-type clusters should undoubtably incorporate binary star populations as these
{\it are} observed -- for example, \citet{elson:98} observed the binary fraction in the 
young massive LMC cluster NGC 1818 to be $35 \pm 5$ per cent in the cluster core, decreasing
to $20 \pm 5$ per cent in the outer regions. We anticipate that future simulations by us 
will investigate the effects of a binary star population on the results presented in this paper. 

Finally, the youngest LMC and SMC clusters typically have metallicities not far from the solar 
value -- for example, the literature compilation in \citet{mackey:03a} suggests a range 
$-0.4 \la [$Fe$/$H$] \la 0.0$ in the LMC. Therefore, for consistency, in all simulations we set 
our clusters to have solar metallicity, $Z = 0.02$. However, we note that there is a strong
age-metallicity relationship present in both Magellanic Clouds \citep[see e.g.,][]{pagel:98},
in that older clusters are typically much more metal-poor than younger clusters. This may have 
important implications for our results. \citet{hurley:04} have demonstrated that differences in 
metallicity can result in some weak variation in the global structural and dynamical evolution 
of open clusters, mainly due to differences in stellar winds and mass loss. Furthermore, variations 
in metal abundance may have a strong effect on the number and mass of BHs produced in supernova
explosions \citep[e.g.,][]{zhang:07}. We discuss these aspects further in Section 
\ref{s:discussion}.

\subsection{Primordial mass segregation}
\label{ss:mseg}
Almost all young massive star clusters which have been observed with sufficient resolution
are seen to exhibit some degree of mass segregation. This is true for clusters in
the LMC (e.g., NGC 1805, NGC 1818, R136) and SMC (e.g., NGC 330), as well as in the Galaxy
(e.g., Orion Nebula Cluster, Arches, Quintuplet) 
\citep[e.g.,][]{degrijs:02a,degrijs:02b,sirianni:02,malumuth:94,brandl:96,hunter:96,kim:06}. 
Detailed simulations of star cluster 
formation \citep[see e.g.,][ and references therein]{bonnell:06} are consistent with these
observations, suggesting that mass segregation in young massive clusters may well be
a product of the formation process, in that more massive stars are preferentially formed at
the bottom of local potential wells where the gas density is greatest.

Irrespective of whether the observed properties of young massive clusters are
truly ``primordial'', we would like to include the possibility of very early mass 
segregation in our models in order to investigate its effects on their subsequent
evolution. To produce initially mass segregated clusters in a ``self-consistent'' 
fashion (i.e., close to virial equilibrium, with all members having appropriate
velocities) we developed the following procedure. For a given model, we first generate
a cluster as described in the previous Section. This object represents the case where 
there is no primordial mass segregation. We then implement a mass-function truncation, 
setting all stars in the cluster with masses greater than $8\,{\rm M}_\odot$ to have mass 
$8\,{\rm M}_\odot$. Next, the cluster is evolved dynamically using {\sc nbody4}, but 
with the stellar evolution routines turned off. Hence the cluster begins to dynamically 
relax and mass segregate. The degree of primordial mass segregation is then easily 
controlled using a single parameter -- the length of time, $T_{{\rm MS}}$, for which the 
cluster is ``pre-evolved''. We selected the truncation limit of $8\,{\rm M}_\odot$ empirically 
so that the pre-evolution can extend for a reasonable duration (a few hundred Myr) without 
the most massive stars sinking to the cluster centre, interacting, and ejecting each other. 
Once the desired pre-evolution time is reached, we halt the simulation, 
replace the mass-truncated stars with their original masses, and take the positions and
velocities in the pre-evolved cluster as the initial conditions for the full run
including stellar evolution. It is straightforward to read in the pre-evolved 
cluster using {\sc nbody4}, without applying any re-scaling.

Because we replace the mass-truncated stars with their original masses after the 
pre-evolution is complete, these stars (which number a few hundred in any given model)
have slightly incorrect velocities at the beginning of the simulation proper. However,
since they almost all reside in the densest part of the cluster, once the full simulation
begins these velocities change rapidly and, within the first few local dynamical times, 
become consistent with the mass distribution in the cluster. Hence this small inconsistency
has a negligible effect on the long-term evolution. We also note that during the pre-evolution
a small fraction of stars escape from the cluster. This is usually in the form of low-mass
stars drifting slowly across the limiting radius, after which they are removed from the 
simulation. This process is very gradual however, and even the clusters with the longest
pre-evolution times ($T_{{\rm MS}} = 450$ Myr) always retain more than $96$ per cent of the 
mass of the initial non-segregated object. Occasionally, despite the mass-truncation of 
stars, a massive object will interact strongly with another massive object during the 
pre-evolution, and be ejected from the cluster. Since we are very interested in how 
the most massive stars in the cluster affect its evolution, and would like to maintain
a high level of consistency between the BH populations of different model clusters,
we always replace these objects at the end of the pre-evolution period using their 
positions and velocities from a few output times before the ejection. Since this is 
necessary for at most a handful of stars per cluster, the introduced inconsistencies 
are again negligibly small.

The initial central densities and core radii of our primordially mass segregated model
clusters depend on the duration of the pre-evolution. We selected our longest pre-evolution 
times ($T_{{\rm MS}} = 450$ Myr) so that the resulting clusters possess properties
very similar to those observed for R136, which is the most compact Magellanic Cloud
cluster. These models have $r_c = 0.25$ pc and $\log \rho_0 = 4.58$ (cf. Fig. 
\ref{f:youngclusters}). In addition to these global properties, we examined in detail 
the radial variation in mass function slope for such models and compared the results 
with those observed for several young Magellanic Cloud clusters. This process is 
described in detail in Section \ref{ss:pair2}; here, we simply note the excellent 
agreement between the models and the real clusters, as verification of the validity of 
our pre-evolution algorithm. We also ran simulations using clusters with more 
intermediate pre-evolution durations ($T_{{\rm MS}} = 115$ and $225$ Myr) -- as might
be expected, these objects possess intermediate core radii and central densities: $r_c = 0.83$ 
pc and $\log \rho_0 = 2.70$, and $r_c = 0.37$ pc and $\log \rho_0 = 3.61$, respectively.

\section{``Observing'' the simulations}
\label{s:simobs}
Since the radius-age trend is defined observationally (i.e., by Fig. \ref{f:radiusage}), a 
vital ingredient in our analysis is to derive measurements from the simulations which are 
fully consistent with these observations. This requirement highlights a key advantage
in running direct, realistic $N$-body models. Because the positions, velocities,
masses and luminosities of all stars are explicitly followed, and because we do not have 
to worry about scaling our results with $N$, 
we are able to perform simulated observations of a model cluster at each output time
which lead to measured quantities that are directly comparable to those obtained for the 
real Magellanic Cloud clusters. More specifically, we calculate the {\it observational} 
core radius of each model cluster rather than using the traditional $N$-body definition 
(see below), and further, we incorporate many of the subtleties of the actual HST 
measurements which have defined Fig. \ref{f:radiusage}.

\begin{figure}
\begin{center}
\includegraphics[width=0.45\textwidth]{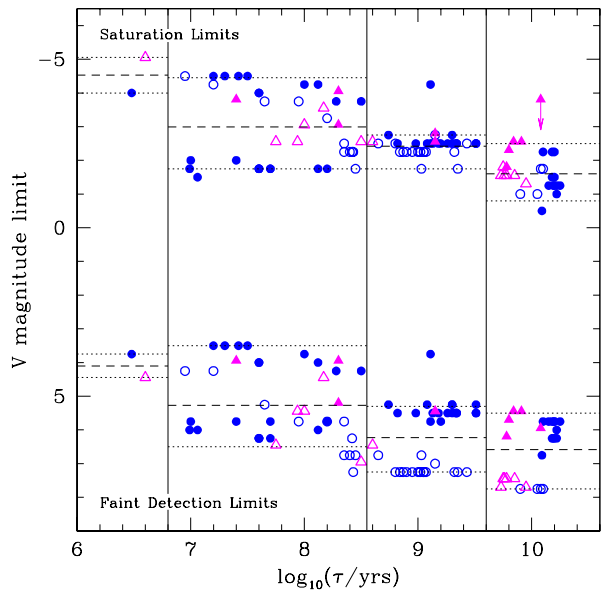}
\caption{Bright and faint stellar detection limits on the HST/WFPC2 and ACS images of LMC
and SMC clusters used for the measurements presented in Fig. \ref{f:radiusage}. Circular 
symbols mark LMC clusters, while SMC objects are triangles. Filled symbols represent the WFPC2 
imaging described in \citet{mackey:03a,mackey:03b} while open symbols are the ACS imaging 
from Mackey et al. (2008, in prep.). Clusters are split into four age bins, delineated with 
solid vertical lines. Within each age bin, the mean bright and faint detection limits are 
marked by dashed lines, while the approximate maximum scatter about each mean is marked by 
a pair of dotted lines. For ease of reference, absolute magnitudes $M_V = (-5,\,0,\,5)$ 
correspond to zero-age main sequence stellar masses of 
$M_{*} \sim (45.0,\,4.20,\,1.06)\,{\rm M}_\odot$ at solar metallicity.}
\label{f:obslimits}
\end{center}
\end{figure}

Consider Fig. \ref{f:obslimits}, where we have plotted the detection 
limits in the HST WFPC2 and ACS imaging from which Fig. \ref{f:radiusage} was 
constructed, against cluster age. The brighter limits represent saturation on the images 
(very bright stars, while recorded on the images, are generally not measured by 
photometry software), while the lower limits represent the approximate $50$ per cent detection 
completeness levels (faint stars are not always detected above background noise by photometry 
software). We have split the clusters into four age bins according to approximately constant 
detection limits -- these are delineated on the plot with solid vertical lines. Within 
each bin, we mark the mean bright and faint detection limits with dashed lines, and the 
approximate maximum scatter about these means with dotted lines.

A number of things are evident from Fig. \ref{f:obslimits}. First, for any given cluster, the 
observations sample only a portion of the range of stellar masses present in the cluster. 
Hence, the surface brightness profile, from which the structural parameters for that cluster 
are measured, is based only on the spatial distribution of stars within this range. Second, 
the sampled range varies systematically with cluster age. This is due to the fact that 
observations of star clusters in the LMC and SMC are commonly aimed at targeting stars near 
the main-sequence turn-off. Consequently, the required exposure time increases with cluster 
age, meaning that both the brighter and fainter detection limits become deeper with age. 
Looking at the two oldest bins, one can also see the increased capabilities of the ACS 
instrument compared with WFPC2. While the saturation limits are comparable for all clusters 
in these two bins, the ACS-imaged objects have faint detection limits $\sim 2$ mag fainter 
than those of the WFPC2-imaged objects -- indicative of the increased sensitivity and 
dynamic range of ACS over WFPC2.

Given the above, when deriving structural measurements from our $N$-body simulations we 
must account for the fact that in the real observations for any given cluster, only a portion 
of the stellar mass function was sampled, and the fact that this range changes with the age
of the target cluster. Since we know the details of the sampling from Fig. \ref{f:obslimits}, 
this is simple to achieve. 

To ``observe'' a model cluster, we pass the $N$-body data through a measurement pipeline 
essentially identical to that we used to obtain structural quantities for the LMC and SMC 
samples (full details of the observational pipeline may be found in \citealt{mackey:03a}). 
For the data produced at a given output time, we first convert the luminosity
and effective temperature of each star in the cluster to $UBVRI$ standard magnitudes, using 
the bolometric corrections of \citet{kurucz:92} supplemented with those of \citet*{bergeron:95}
for white dwarfs. We also convert the position and velocity of each star to physical units 
using the appropriate length and velocity scale factors (see Section \ref{ss:nbodycode}). With 
this complete, we next impose the bright and faint detection limits appropriate to the output 
time (these are the dashed mean limits in Fig. \ref{f:obslimits}). This leaves an ensemble of 
stars with which to construct a surface brightness profile, which we do following 
\citet{mackey:03a,mackey:03b}. We project the three-dimensional position of 
each star onto a plane, construct annuli of a given width about the cluster centre, and
calculate the surface brightness in each annulus. For consistency with the observational 
pipeline, we use a variety of annulus widths so that both the bright inner core and the 
fainter outer regions of the cluster are well measured. Next, we account for the fact that
both the WFPC2 and ACS cameras have fields-of-view which are considerably smaller than
the area on the sky filled by a Magellanic Cloud cluster, which might typically have 
$r_t \sim 40-50$ pc (i.e., $r_t \sim 160-200\arcsec$ at the LMC distance) 
\citep[see e.g.,][]{mateo:87,olsen:98}. This results in surface brightness profiles
generally being truncated beyond projected radii $r_p \sim 25$ pc \citep{mackey:03a,mackey:03b}.
After imposing this limit, we finally fit an EFF model of the form of Eq. \ref{e:effproject}
to the resulting surface brightness profile, and from this model derive the structural 
parameters -- in particular the core radius, $r_c$, and the power-law slope at large radii, 
$\gamma$. To reduce noise we repeat this process for three orthogonal planar projections 
at each output time and average the results.

It is worth noting the difference between the quantity $r_c$ and the `core radius' usually
defined in $N$-body simulations. This has been discussed in some detail by \citet{wilkinson:03};
however, in the interests of clarity we re-iterate a few of the most salient points.
Traditionally, observers, theorists, and numericists have employed different interpretations of
the `core radius'. That for observers is as defined above (Eq. \ref{e:rc}), as the projected 
radius at which the surface brightness (or density) has dropped to half the central value. 
Theoretically defined, the core radius is the natural scale-length of the model under 
consideration -- for example, in EFF models $a$ is the scale-length. Eq. \ref{e:rc} provides 
a general relation between $a$ and $r_c$. It should be noted however, that as a cluster evolves, 
the EFF parameters are not static, and therefore the ratio between $a$ and $r_c$ is variable with time. 

In $N$-body simulations the numerically calculated `core radius' is more correctly termed the 
density radius, $r_d$. The implementation in {\sc nbody4} is based on a quantity described by
\citet{casertano:85}, so that $r_d$ is defined as the density-weighted average of the distance
of each star from the density centre of the cluster \citep{aarseth:01}. The local density at
each star is computed from the mass within the sphere containing the six nearest neighbours.
As noted by \citet{wilkinson:03}, there is no general relationship between $r_d$ and $r_c$, 
and in fact the behaviour of $r_c$ and $r_d$ may be quite different throughout a simulation.

As a final remark, we briefly consider the appropriateness of fitting a power-law profile 
(Eq. \ref{e:effproject}), which formally has no outer limit, to a simulated cluster evolving 
in a tidal field. There are two reasons why this is acceptable. First, because of the radial 
truncation imposed to mimic the field-of-view limitations of the WFPC2 and ACS cameras,
our derived surface brightness profiles do not reach as far as the cluster tidal radius.
Following \citet{mackey:03a,mackey:03b}, it is therefore legitimate to fit EFF models to 
these observed profiles, even when a cluster is dynamically old enough to exhibit a tidal 
truncation -- in the interests of obtaining measurements of $\gamma$ which are, like those 
for $r_c$, directly comparable to the real observations, we choose to employ the same 
methodology. Even without the truncation of our radial profiles, an EFF model
would still have been the most appropriate choice. This is due to the treatment of stellar 
escapers in {\sc nbody4}, as discussed in Section \ref{ss:nbodycode}. While the tidal radius 
$r_t$ is estimated from Eq. \ref{e:tidalradius}, stars are not removed from the simulation until 
they reach $2 r_t$. Hence they are free to populate the region $r_t < r < 2r_t$, and there is 
no truncation in the density profile at (or near) $r_t$, even for dynamically old clusters.

\section{Simulations and Results}
\label{s:results}

\begin{table*}
\begin{minipage}{164mm}
\caption{Details of $N$-body runs and initial conditions. Each cluster begins with $N_0$ stars with masses summing to $M_{{\rm tot}}$, and initial central density $\rho_0$. Initial cluster structure is ``observed'' to obtain $r_c$ and $\gamma$. Each model is evolved until $\tau_{{\rm max}}$.}
\begin{tabular}{@{}lccccccccccc}
\hline \hline
Name & \hspace{0mm} & $N_0$ & $\log M_{{\rm tot}}$ & $\log \rho_0$ & $r_c$ & $\gamma$ & \hspace{0mm} & Initial MSeg & BH Retention & \hspace{0mm} & $\tau_{{\rm max}}$ \\
 & & & (${\rm M}_\odot$) & (${\rm M}_\odot\,{\rm pc}^{-3}$) & (pc) & & & ($T_{{\rm MS}}$) & ($f_{{\rm BH}}$) & & (Myr)$^1$ \\
\hline
Run 1 & & $100\,881$ & $4.746$ & $2.31$ & $1.90 \pm 0.09$ & $2.96 \pm 0.17$ & & None & $0.0$ & & $19\,987$ \\
Run 2 & & $100\,881$ & $4.746$ & $2.31$ & $1.90 \pm 0.09$ & $2.96 \pm 0.17$ & & None & $1.0$ & & $10\,668$ \\
\hline
Run 3 & & $95\,315$ & $4.728$ & $4.58$ & $0.25 \pm 0.04$ & $2.33 \pm 0.10$ & & $450$ Myr & $0.0$ & & $11\,274$ \\
Run 4 & & $95\,315$ & $4.728$ & $4.58$ & $0.25 \pm 0.04$ & $2.33 \pm 0.10$ & & $450$ Myr & $1.0$ & & $10\,000$ \\
\hline
Run 4a & & $98\,605$ & $4.738$ & $2.70$ & $0.83 \pm 0.07$ & $2.45 \pm 0.14$ & & $115$ Myr & $1.0$ & & $4\,274$ \\
Run 4b & & $97\,209$ & $4.733$ & $3.61$ & $0.37 \pm 0.05$ & $2.34 \pm 0.10$ & & $225$ Myr & $1.0$ & & $4\,457$ \\
\hline
Run 5 & & $95\,315$ & $4.728$ & $4.58$ & $0.25 \pm 0.04$ & $2.33 \pm 0.10$ & & $450$ Myr & $0.5$ & & $10\,059$ \\
\hline
Run 6 & & $100\,881$ & $4.746$ & $2.31$ & $1.90 \pm 0.09$ & $2.96 \pm 0.17$ & & None & $0.0$, NS$^2$ & & $19\,987$ \\
\hline
\label{t:runs}
\end{tabular}
\medskip
\vspace{-5mm}
\\
$^1$ As described in Section \ref{s:results}, no special significance should be attached to the listed values of $\tau_{{\rm max}}$.\\
$^2$ Run 6 is identical to Run 1, except with natal neutron star kicks set to zero so that $f_{{\rm NS}} = 1.0$. 
\end{minipage}
\end{table*}

The properties of our $N$-body runs are listed in Table \ref{t:runs}. Our main set of models 
are labelled Runs 1--4. These cover the extremes of the parameter space we are interested
in investigating, spanned by BH retention fractions $f_{{\rm BH}} = 0$ and $f_{{\rm BH}} = 1$,
and the pre-evolution durations $T_{{\rm MS}} = 0$ Myr (i.e., no primordial mass segregation)
and $T_{{\rm MS}} = 450$ Myr (strong primordial mass segregation, matching that observed in
young LMC and SMC objects). These runs are
therefore expected to represent the extremes of cluster evolution induced by variation of
the BH retention fraction and the degree of primordial mass segregation. The global 
properties of these four Runs have already been presented in a short {\it Letter} 
\citep{mackey:07}; in the present paper we examine their evolution in considerably more detail. 

In addition to our four primary runs, we performed several additional simulations in order
to sample the parameter space more completely, and in particular verify that models with
intermediate values of $f_{{\rm BH}}$ and $T_{{\rm MS}}$ exhibited evolution intermediate
between that displayed by Runs 1--4. To this end, Runs 4a and 4b explore the effects of 
primordial mass segregation in more detail, while Run 5 highlights the effects of natal kicks
on BH retention and the subsequent cluster evolution. Finally, Run 6 is used to address the 
question of whether we can reproduce the cluster evolution induced by a significant BH
population by retaining neutron stars (NSs) instead of the BHs.

For each run, we measured the initial cluster mass, central density, and observed structural
parameters $r_c$ and $\gamma$ -- these are all listed in Table \ref{t:runs}. It is important to 
re-emphasize how closely these correspond to the observed quantities for the youngest massive 
clusters in the Magellanic Clouds. This can be seen explicitly by comparing the values listed 
in Table \ref{t:runs} with the plots in Fig. \ref{f:youngclusters}. The model clusters with no 
primordial mass segregation have $r_c \sim 1.9$ pc, $\gamma \sim 3.0$, and $\log \rho_0 \sim 2.3$.
These clusters therefore appear very similar to a number of Magellanic Cloud clusters with
ages of $\sim 20$ Myr. In contrast, the heavily mass segregated model clusters have
much smaller cores and higher central densities, with $r_c \sim 0.25$ pc and $\log \rho_0 \sim 4.6$.
They also have flatter power-law fall-offs, with $\gamma \sim 2.3$. In this respect, they
strongly resemble the very compact massive young LMC cluster R136, which has an age
of $\sim 3$ Myr. The total masses of all models are very similar, in the range 
$4.728 \le \log M_{{\rm tot}} \le 4.746$. The variation is due to the pre-evolution procedure 
used to develop the mass segregated initial conditions, as described in Section \ref{ss:mseg}. 
From comparison with Fig. \ref{f:youngclusters}, it is clear that our $N$-body clusters have 
masses typical of the youngest clusters in the observed sample. We also note that the 
``observed'' integrated colours of our models at early times are consistent with measurements
for young Magellanic Cloud clusters from the literature -- for example, the integrated $(B-V)$
colours compiled by \citet{bica:96}.

Given the close correspondence between the properties of our model clusters and those observed 
for young LMC and SMC objects, we are confident that our $N$-body simulations are directly 
modelling the evolution of massive Magellanic Cloud clusters.

Output data was produced for each Run at intervals of $\Delta \tau = 1.5$ Myr for $\tau <= 100$ Myr,
and at intervals of $\Delta \tau = 15$ Myr for $\tau > 100$ Myr.
It is worth noting that no special significance should be attached to the listed values of 
$\tau_{{\rm max}}$ in Table \ref{t:runs}. The main criterion for our primary Runs (Runs 1--6, 
excluding Runs 4a and 4b) was that $\tau_{{\rm max}}$ be larger than $\sim 10$ Gyr, to
approximate the ages of the oldest Magellanic Cloud globular clusters. The listed $\tau_{{\rm max}}$
simply represent the most convenient termination points beyond this time. For interest's 
sake, Runs 1 and 6 were evolved for significantly longer periods ($\tau_{{\rm max}} = 20$ Gyr) 
than the other models, so that the clusters passed through the core-collapse phase. In contrast,
Runs 4a and 4b were evolved only as long as necessary (i.e., just long enough for the effects
of intermediate values of $T_{{\rm MS}}$ to become evident), to save on computation time.

\subsection{Runs 1 and 2: No mass segregation}
\label{ss:pair1}
We first consider the pair of simulations labelled Run 1 and Run 2. Neither of these two model 
clusters have primordial mass segregation, and both start with identical initial conditions, to 
the extent that they share the same random seed. The sole difference between them is that in 
Run 1 the natal BH kicks are set to be $v_{{\rm kick}} \approx 200$ km$\,$s$^{-1}$, 
whereas in Run 2 they are set to be zero. Thus, every BH formed in a supernova explosion in
Run 1 is provided with a sufficiently large random velocity that it very rapidly escapes from 
the cluster, so the retention fraction is $f_{{\rm BH}} = 0$. Conversely, in Run 2 all $198$ 
BHs are retained in the cluster and the retention fraction is $f_{{\rm BH}} = 1$. The purpose of 
these runs is twofold. First, Run 1 lets us consider the long-term 
evolution of our simplest cluster set-up -- no primordial mass segregation, and zero BH retention. 
This model therefore constitutes a control run against which the evolution of all our other models 
may be compared. Second, by making such a comparison, Run 2 lets us isolate the effects of a 
population of stellar-mass black holes on the structural and dynamical evolution of a massive 
star cluster.

\begin{figure*}
\begin{minipage}{175mm}
\begin{center}
\includegraphics[width=0.45\textwidth]{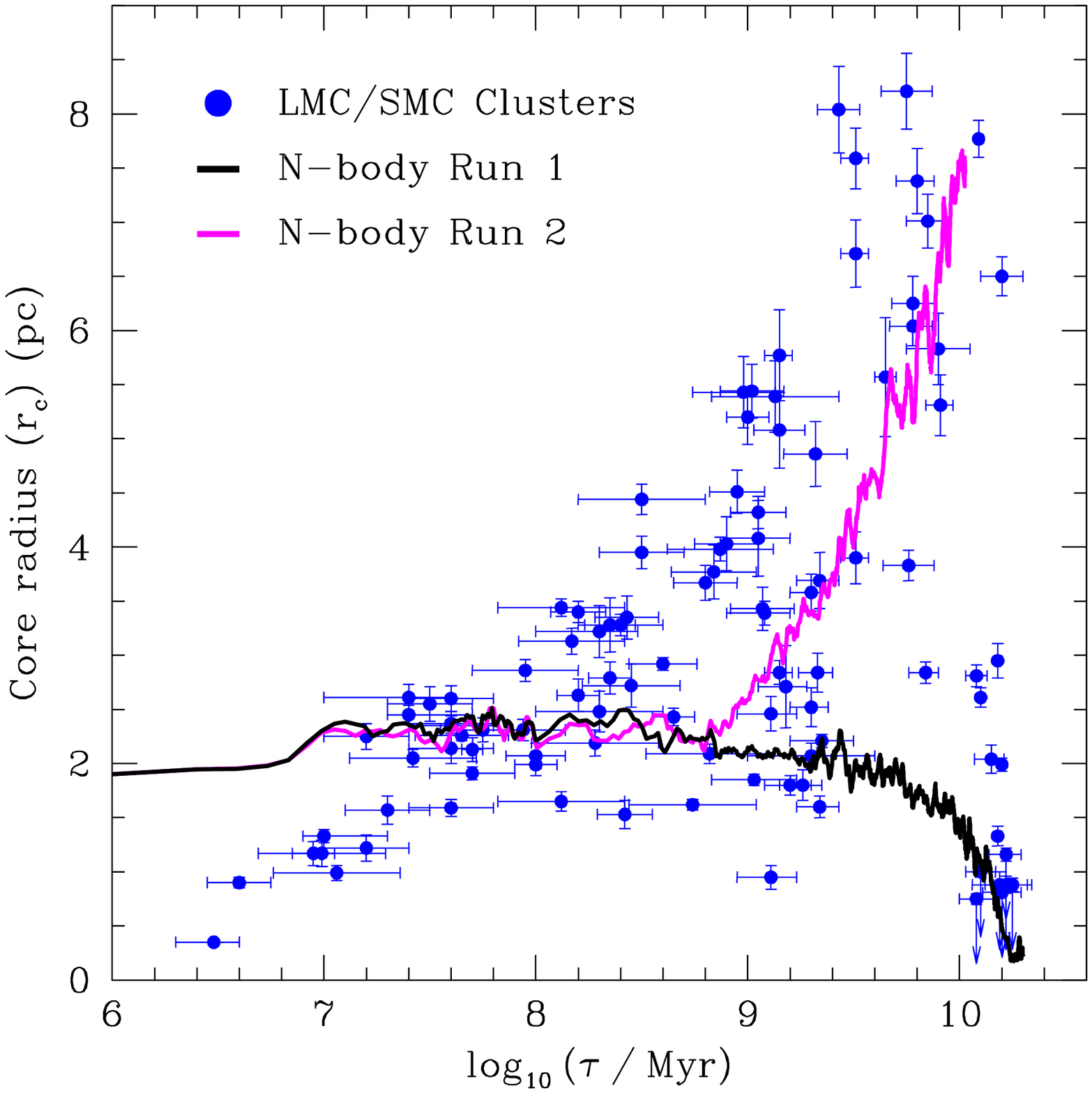}
\hspace{0mm}
\includegraphics[width=0.45\textwidth]{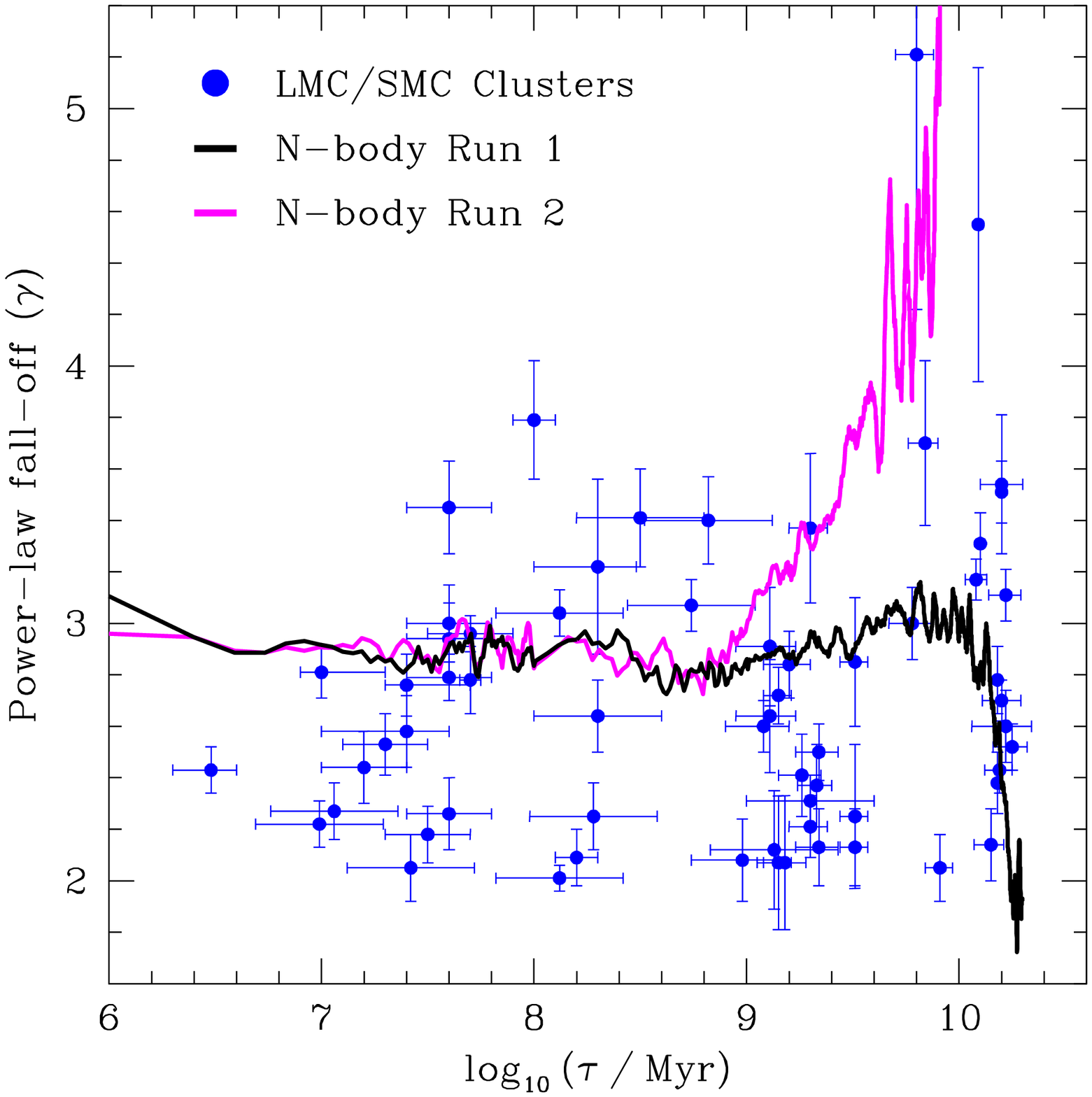}
\caption{Structural evolution of $N$-body Runs 1 and 2. Neither has primordial mass segregation;
the only difference between them is the BH retention fraction ($f_{{\rm BH}} = 0$ and $1$, 
respectively). {\bf Left panel:} Evolution of $r_c$, observed as described in Section
\ref{s:simobs}. Run 1 evolves exactly as expected for a classical massive star cluster, with the 
main trend being a slow contraction in $r_c$ as the system relaxes dynamically and moves towards 
core collapse. In stark contrast, Run 2 evolves very similarly to a point, after which strong 
expansion in the core radius is observed. The presence of $198$ stellar-mass black holes in this 
cluster thus leads to strikingly different core radius evolution. {\bf Right panel:} Evolution 
of the power-law fall-off, $\gamma$, again observed as described in Section \ref{s:simobs}.
As with $r_c$, Run 2 evolves identically to Run 1 until the BH population becomes 
dynamically active, after which the evolution strongly diverges with Run 2 developing
a steep fall-off in its outer regions. In this panel, only data points from the WFPC2 observations 
of \citet{mackey:03a,mackey:03b} are plotted. Measurements of $\gamma$ from the recent ACS 
observations of Mackey et al. (2008, in prep.) are not yet finalised.}
\label{f:evolpair1}
\end{center}
\end{minipage}
\end{figure*}

The progress of Runs 1 and 2 across the radius-age plane is displayed in Fig. \ref{f:evolpair1}
(left panel). Also shown is the evolution of these two runs in the $\gamma$-age plane (right 
panel). First consider Run 1, which behaves exactly as expected for a classical massive 
stellar cluster. At very early times, extending to roughly $\tau \sim 100$ Myr, there is a period 
of severe mass loss due to the rapid evolution of the most massive stars in the cluster. 
By $\tau \sim 100$ Myr, approximately $25$ per cent of the initial cluster mass has been lost. The 
$198$ BHs are formed in supernova explosions between $3.5$-$10$ Myr and, since they are born with 
$v_{{\rm kick}} \approx 200$ km$\,$s$^{-1}$, all are immediately ejected from the cluster. 
From Fig. \ref{f:evolpair1}, it is clear that the violent relaxation experienced by the cluster 
when $\tau \la 100$ Myr is not reflected in its core-radius evolution, presumably because 
the mass loss is distributed evenly throughout the cluster. Similarly, there is no evidence 
of the violent relaxation phase in the evolution of $\gamma$.

As the cluster grows older, the rate of mass loss decreases and the cluster settles into a 
quasi-equilibrium state, where dynamical evolution is dominated by two-body relaxation 
processes. The median relaxation time for this $N=10^5$ star cluster is given by 
$t_{rh} \approx 1.9 \times 10^5 \, M_{{\rm cl}}^{1/2} m_{*}^{-1} r_h^{3/2}$ \citep{binney:87} 
where $m_{*}$ is the typical stellar mass and $r_h$ is the 3-dimensional radius containing 
$0.5 M_{{\rm cl}}$. At $\tau = 100$ Myr, when the rapid early mass loss is mostly complete, 
$M_{{\rm cl}} \approx 43\,500 \,{\rm M}_\odot$, $m_{*} \approx 0.45 \,{\rm M}_\odot$ and 
$r_h \approx 8$ pc, so that $t_{rh} \sim 2$ Gyr. Mass segregation develops in the cluster
on roughly this time-scale: this is evident in Fig. \ref{f:evolpair1} as a gradual contraction
in $r_c$ as the most luminous stars in the magnitude range used to measure the structural 
parameters (cf. Fig. \ref{f:obslimits}) sink towards the cluster centre. As two-body
relaxation proceeds and mass segregation becomes more prominent, the core radius steadily
shrinks with time. The power-law fall-off, $\gamma$, slowly becomes steeper during
this phase; however as the core becomes increasingly more compact, so $\gamma$ becomes 
increasingly flatter after $\tau \sim 10$ Gyr.

Eventually, after many Gyr of evolution, Run 1 enters the core-collapse phase. The point 
of greatest collapse (smallest $r_c$) occurs at $\tau \approx 17.4$ Gyr, when the central 
mass density reaches $\log \rho_0 \approx 4.5$ -- a value commensurate with those inferred 
for NGC 2005 and 2019, the most likely core collapsed clusters in the LMC 
\citep[e.g.,][]{mackey:03a,mclaughlin:05}. The point of greatest collapse coincides with a 
spate of binary star formation in the core -- by $\tau = 17.5$ Gyr there are seven 
newly-formed binary stars. Subsequently, up until the end of the simulation at $\tau = 20$ 
Gyr, there is no significant change in the observed value of $r_c$. Defining the cluster age in 
terms of an integrated median relaxation time, which is necessary because $t_{rh}$ is a 
constantly evolving quantity, we find that at $\tau = 17.4$ Gyr, $8.37\,t_{rh}$ has elapsed.

\begin{figure}
\begin{center}
\includegraphics[width=0.47\textwidth]{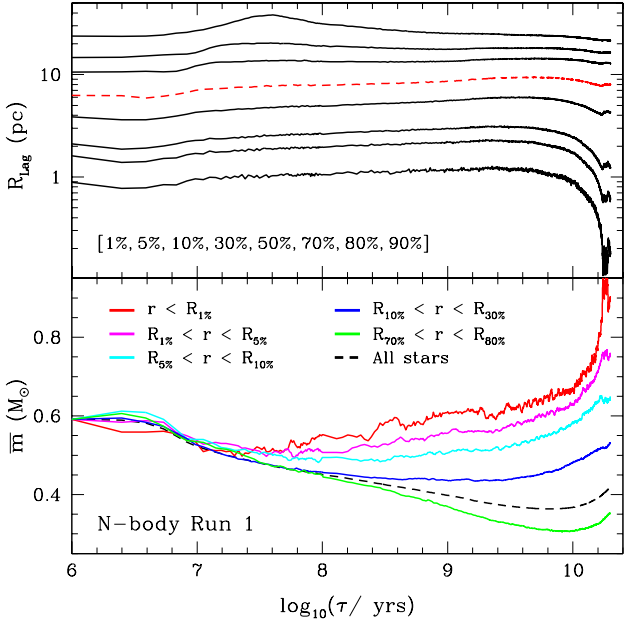}
\caption{Evolution of various Lagrangian radii (top panel) and the mean stellar mass in the
shells encompassed by selected Lagrangian radii (lower panel) for Run 1. The radii displayed 
in the top panel are, from inner to outer, the $1\%$, $5\%$, $10\%$, $30\%$, $50\% = r_h$ 
(dashed line), $70\%$, $80\%$, and $90\%$ radii. In the lower panel the shells are defined by:
$r \le R_{1\%}$, $R_{1\%} < r \le R_{5\%}$, $R_{5\%} < r \le R_{10\%}$, 
$R_{10\%} < r \le R_{30\%}$, and $R_{70\%} < r \le R_{80\%}$ (these are listed in order from 
the upper to lower solid lines at the right hand side of the panel). The dashed line is the 
mean mass for all stars in the cluster.}
\label{f:lrad1}
\end{center}
\end{figure}

Fig. \ref{f:lrad1} illustrates the evolution of Run 1 in more detail. In the top
panel a series of Lagrangian radii are plotted -- here, we define $R_{x\%}$ 
to be the 3-dimensional radius containing $x$ per cent of the {\it stellar} mass in the
cluster -- that is, {\it excluding} BHs. This exclusion is not important for Run 1, since all 
BHs are gone from the cluster by $\sim 15$ Myr; however it is crucial for examining the evolution 
of the stellar component of Runs in which $f_{{\rm BH}} > 0$. Unlike $r_c$, the innermost 
Lagrangian radii in Run 1 do show an increase in size in reaction to the early mass-loss phase; 
however, this increase is only very modest. In addition, the innermost Lagrangian radii do not 
show any sign of contraction until much later than does $r_c$ -- this is an indication of the
luminosity (and hence mass) weighting inherent in the calculation of $r_c$. The half-mass radius
of the cluster shows only a small amount of variation throughout its evolution. The outer 
radii also show only very gradual evolution. The main feature is an expansion in the $90\%$
radius during the mass-loss phase. This is due to stars in the very outer regions of the
cluster drifting beyond $r_t$ as the cluster rapidly loses mass. Eventually these objects are 
removed from the simulation (once they get beyond $2r_t$) and the $90\%$ radius slowly contracts.
This contraction continues as the cluster slowly loses mass for the rest of its lifetime,
and $r_t$ gradually shrinks accordingly. 

In the lower panel of Fig. \ref{f:lrad1}, the mean stellar mass in shells encompassed by
selected Lagrangian radii is plotted. This plot therefore shows the development of mass 
segregation in Run 1. This process is inhibited by the early violent relaxation phase,
and there is only a very small degree of segregation present in the cluster's central
regions by $\tau = 100$ Myr. Subsequently however, the stratification becomes very well
established. As expected, this occurs more rapidly in the central regions of the cluster,
where the relaxation time is shortest. By the time the core-collapse phase is reached,
there is a large degree of mass segregation present in the model cluster. One notable feature,
exhibited by both the outermost shell and the full cluster mean, is an increase in the mean 
stellar mass after $\tau \sim 10$ Gyr. This is due to the preferential removal of low-mass
cluster stars by the external tidal field. These stars typically reside in the outer
cluster regions at late times, and are hence far more susceptible to tidal effects than
are the more massive objects which inhabit the cluster core. At late times, stellar evolution
has all but slowed to a halt so that the tidal stripping of low-mass stars has a significant
effect on the mean stellar mass.

\begin{figure}
\begin{center}
\includegraphics[width=0.48\textwidth]{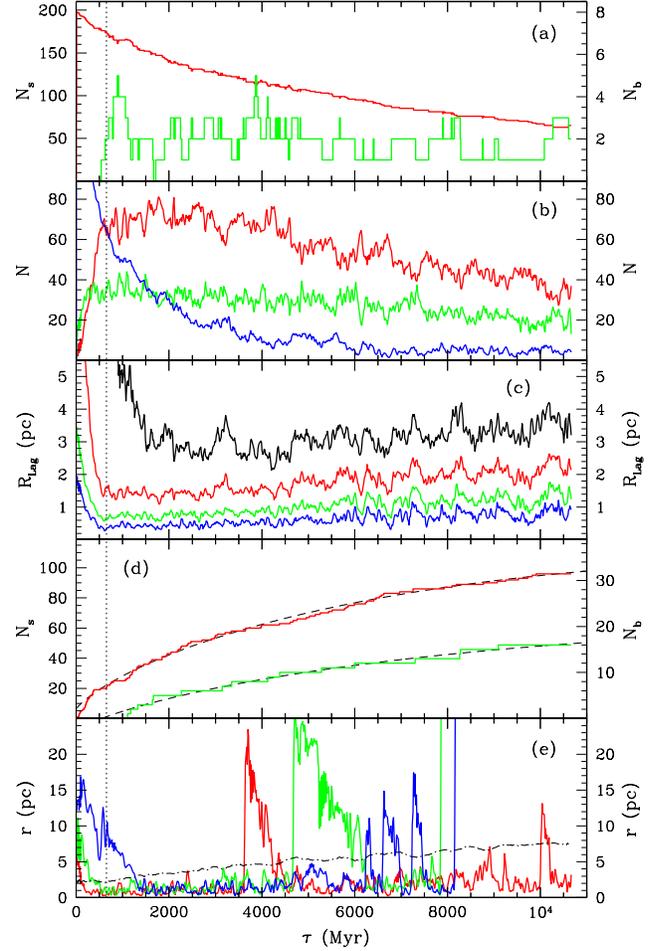}
\caption{Properties of the BH population in Run 2 as a function of time: (a) the number 
of single BHs (upper line) and binary BHs (lower line) in the cluster; (b) the number of BHs 
within the shells encompassed by the {\it stellar} Lagrangian radii (cf. Fig. \ref{f:lrad1}) 
$r \le R_{1\%}$, $R_{1\%} < r \le R_{5\%}$, and $r > R_{10\%}$ (the upper, middle and lower 
lines, respectively, at the right of the plot); (c) the {\it black hole} $10\%$, $25\%$, 
$50\%$ and $75\%$ Lagrangian radii (respectively, the innermost to outermost lines); (d) the 
cumulative numbers of escaped single BHs (upper) and binary BHs (lower), along with
fits of the form $N_e = A_0 + A_1\tau - A_2\tau\log\tau$ (dashed lines); and (e) the radial 
positions of three typical BHs. The vertical dotted line indicates $\tau = 650$ Myr, the
approximate time when core expansion begins. The evolution of $r_c$ is plotted (dot-dashed line) 
in panel (e). Note the different axis scales on either side of panels (a) and (d).}
\label{f:bhevol}
\end{center}
\end{figure}

Now consider Run 2, in which the BH kicks $v_{{\rm kick}} = 0\,$km$\,$s$^{-1}$ so that 
$f_{{\rm BH}} = 1$. From inspection of Fig. \ref{f:evolpair1}, it is clear that the evolution 
of both $r_c$ and $\gamma$ is indistinguishable from that seen in Run 1 up to 
$\log\tau \sim 8.8$, after which strong expansion of $r_c$ is observed for Run 2, in
conjunction with a significant steepening in $\gamma$. A careful comparison between
the two models reveals that this divergence begins at $\tau \approx 650$ Myr. The expansion of
$r_c$ in Run 2 continues for the remainder of the simulation, until $\tau_{{\rm max}} = 10.67$ 
Gyr. Since Runs 1 and 2 are identical apart from the kicks imparted to the BHs on their
formation, the strongly different evolution observed for these two models must be due
to the presence of the $198$ BHs in Run 2.

The properties of this BH population as a function of time are illustrated in Fig. 
\ref{f:bhevol}. As in Run 1, by $\tau \approx 100$ Myr, the most violent phase of 
stellar evolution is essentially complete. At this time, the BHs (of typical mass
$m_{{\rm BH}} = 10\,{\rm M}_\odot$) are already significantly more massive than any other 
cluster members (of typical mass $m_{*} \approx 0.45\,{\rm M}_\odot$), and are hence 
beginning to sink to the cluster centre via dynamical friction, on a time-scale of 
$\sim (m_{*} / m_{{\rm BH}})\, t_{rh} \approx 90$ Myr. 

\begin{figure}
\begin{center}
\includegraphics[width=0.47\textwidth]{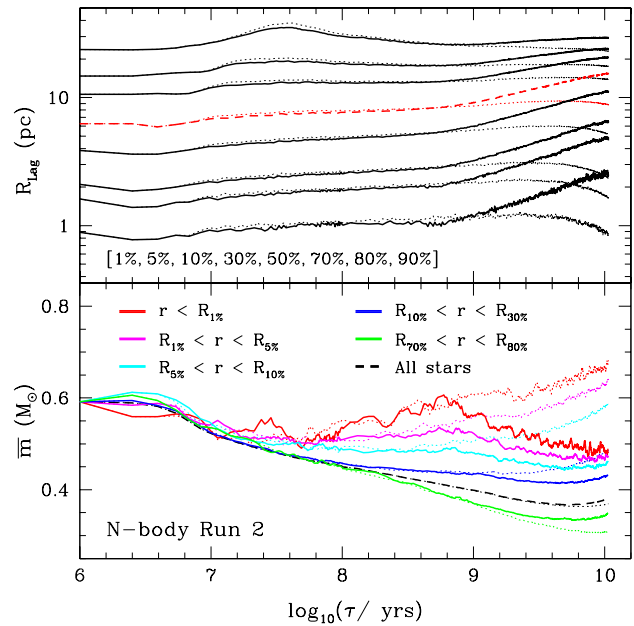}
\caption{Evolution of various Lagrangian radii (top panel) and the mean stellar mass in the
shells encompassed by selected Lagrangian radii (lower panel) for Run 2. The radii displayed 
in the top panel are, from inner to outer, the $1\%$, $5\%$, $10\%$, $30\%$, $50\% = r_h$ 
(dashed line), $70\%$, $80\%$, and $90\%$ radii. In the lower panel the shells are defined by:
$r \le R_{1\%}$, $R_{1\%} < r \le R_{5\%}$, $R_{5\%} < r \le R_{10\%}$, 
$R_{10\%} < r \le R_{30\%}$, and $R_{70\%} < r \le R_{80\%}$ (these are listed in order from 
the upper to lower solid lines at the right hand side of the panel). The dashed line is the 
mean mass for all stars in the cluster. The evolution of Run 1 is marked by dotted lines,
for comparison (note that the abscissa does not extend to such late times as are plotted
in Fig. \ref{f:lrad1}).}
\label{f:lrad2}
\end{center}
\end{figure}

This is evident from panels b and c in Fig. \ref{f:bhevol}. Panel b shows the number of BHs 
within the shells encompassed by the {\it stellar} Lagrangian radii $r \le R_{1\%}$, 
$R_{1\%} < r \le R_{5\%}$, and $r > R_{10\%}$. The evolution of these Lagrangian radii 
themselves may be seen in Fig. \ref{f:lrad2}, which is discussed in more detail below. 
Panel c shows the evolution of the {\it black hole} Lagrangian radii $B_{10\%}$, 
$B_{25\%}$, $B_{50\%}$, and $B_{75\%}$, where, by analogy with the stellar Lagrangian radii, 
$B_{x\%}$ is the 3-dimensional radius containing $x$ per cent of the BH mass in the cluster.

By $200$ Myr, the mass density of BHs at the centre of the cluster is already roughly equal 
to that of the stars, and by $400$ Myr it is about three times larger. Shortly after, this 
central BH subsystem becomes unstable to further contraction 
\citep[see][ Eq. 3-55]{spitzer:87} and decouples from the stellar core in a runaway gravothermal 
collapse, leading to a very rapidly increasing central BH density -- by $490$ Myr, the central 
density of the BH subsystem is $\sim 80$ times that of the stars. This is sufficiently dense 
for the creation of stable BH binaries in three-body interactions to be initiated -- the first 
such object is formed at $\sim 510$ Myr, and by $\approx 650$ Myr there are several (see Fig.
\ref{f:bhevol}a). At this point, the collapse of the BH subsystem is halted: the BH Lagrangian 
radii cease their inward movement and become roughly constant, while the number of BHs within 
the inner stellar Lagrangian radii also level off. It is at this time that the evolution of 
the observational structural parameters $r_c$ and $\gamma$ in Run 2 begins to strongly deviate 
from that in Run 1.

As noted above, prior to this point the evolution of Run 2 is observationally identical 
to that of Run 1. Neither the retention of the BH population at $\tau \approx 10$ Myr, nor the 
subsequent orbital decay of these objects and the resulting formation of a compact central 
BH subsystem leads to differential evolution of $r_c$. This appears at odds with the models 
presented by \citet{merritt:04}, who investigated the possibility that the radius-age trend 
results from the formation of cores in primordially cusped star clusters due to the sinking 
and central accumulation of massive stellar remnants. We attribute our differing results
to the much higher degree of central mass concentration in the cusped models of \citet{merritt:04},
which thereby respond more strongly and more rapidly to the perturbations induced by sinking 
remnants than does our initially cored, relatively low density Run 2. \citet{merritt:04}
also mentioned the possibility of additional cluster expansion due to the subsequent evolution
of the BH population, once the central subsystem had formed. As discussed below, {\it all} of
the expansion observed in our models is the result of such processes.

The number of stable BH binaries in Run 2 peaks at $5$ at $\tau \approx 890$ Myr. After this point, 
there are $0-5$ BH binaries at any given time (Fig. \ref{f:bhevol}a). Once formed, a BH binary 
undergoes superelastic collisions with other, usually single, BHs in the central core (although
BH binaries do also occasionally collide with each other). On average, as BH binaries participate 
in such interactions they become ``harder'' (more tightly gravitationally bound), with 
the released binding energy being carried off by the interacting BHs 
\citep[e.g.,][]{heggie:75, heggie:03}. In each such interaction, the binary BH also has a 
recoil velocity imparted to it, the magnitude of which is dependent on how energetic the 
interaction has been. Together, these processes result in the {\it scattering} of 
BHs outside $r_c$, often into the cluster halo. As a given binary becomes increasingly tightly
bound, so too can the collisions in which it is involved become increasingly energetic, such
that an interacting BH carries off sufficient kinetic energy that it escapes from the cluster 
altogether. Eventually the BH binary is sufficiently hard that the recoil velocity it receives
during a collision is larger than the cluster escape velocity, and the binary escapes as well. 
Hence, interactions in the central compact BH subsystem also result in the {\it ejection} of
BHs from the cluster. For clarity we will retain the italicised terminology ({\it scattering}
and {\it ejection}) henceforth.

These processes are evident in Fig. \ref{f:bhevol}e, which shows the movement of three typical 
BHs during Run 2. Each of the three is born well outside $r_c$, but all sink to the central core
via dynamical friction, as described above. Two are already present there by the time the first
BH binaries are formed. All three of the BHs are frequently scattered to $r_c$ (dot-dashed line) 
during their evolution in the cluster, and at least once each into the cluster halo. One 
is ejected from the cluster at $\tau = 7900$ Myr due to a strong interaction in the core. Another 
becomes a member of a BH binary at $\tau = 6200$ Myr, and subsequently undergoes four 
strong interactions (including one in which its partner is exchanged), with increasing recoil 
velocity each time until this is sufficient for ejection at $\tau = 8200$ Myr. 

Considering Fig. \ref{f:bhevol}c, it is clear that at any given time there are always a handful 
of BHs outside the $10$ per cent stellar Lagrangian radius. This is an indication of the ongoing 
scattering of BHs to outside $\sim r_c$, since any ejected BHs tend to escape the cluster quite 
rapidly. As is evident from Fig. \ref{f:bhevol}e, a scattered BH gradually sinks back into the 
cluster centre via dynamical friction, thus transferring its newly-gained energy to the stellar 
component of the cluster. Most is deposited within $r_c$, where the stellar density is greatest. 
The ejection of BHs also transfers energy to the cluster, since a mass $m$ escaping from a 
cluster potential well of depth $|\Phi|$ does work $m|\Phi|$ on the cluster. This mechanism is 
particularly effective in heating the stellar core, since BHs are ejected from the very centre 
of the cluster, and the energy contributed to each part of the cluster is proportional to the 
contribution which that part makes to the central potential \citep[see e.g.,][]{heggie:03}. In 
addition, here $m_{{\rm BH}}$ is significantly larger than $m_{*}$.

Together these two processes heat the stellar core of the cluster, resulting in significant 
core expansion. This becomes evident observationally at $\tau \approx 650$ Myr, and continues 
as long as the BH population is dynamically active -- in the case of Run 2, the simulation
was halted before the expansion ceased. From Fig. \ref{f:evolpair1}, the size of 
$r_c$ is roughly proportional to $\log \tau$, consistent with the shape of the upper envelope 
of the observed cluster distribution. However, in this $N$-body model the expansion begins 
too late for the evolution to trace the upper envelope exactly; rather, it runs parallel.

The evolution of the stellar Lagrangian radii in Run 2 is illustrated in Fig. \ref{f:lrad2},
along with the evolution of the mean stellar mass in the same Lagrangian shells examined earlier 
for Run 1. The progress of Run 1 is also marked on Fig. \ref{f:lrad2} for comparative purposes
(dotted lines). As noted above, the initial infall and accumulation of BHs in the cluster centre 
does not cause any differential expansion of Run 2 over Run 1 at any radii. It is only after BH 
binaries are formed and the BH population becomes dynamically active that expansion occurs in 
Run 2. This expansion is evident at all radii, although the magnitude is greatest in the central 
regions of the cluster. None of the Lagrangian radii expand by as great a factor over the 
simulation as does $r_c$. The explanation for this can be seen in the lower panel of
Fig. \ref{f:lrad2} -- the development of mass segregation amongst the stellar component in
Run 2 is severely inhibited by the activity of the BH population, compared to Run 1.
This results in a larger apparent expansion in $r_c$ than in the innermost stellar Lagrangian
radii because of the luminosity weighting inherent in the measurement of $r_c$.

The process of mass segregation in Run 2 is only suppressed after the BH population becomes 
dynamically active. Up until this point, segregation has been proceeding just as in Run 1; however, 
after $\tau \approx 650$ Myr, no further stratification occurs. Stellar evolution subsequently
reduces the mean mass in each Lagrangian shell with time. This interpretation is consistent with 
the cluster expansion processes due to BH scattering and ejection which were described above. 
In particular, the repeated BH scattering-sinking cycles constantly stir up the stellar component 
of the cluster and hence hinder the development of mass segregation, particularly in the inner 
cluster regions. The stratification which occurs before the BH population becomes dynamically
active is not reversed however -- there is still clearly a mean-mass gradient from the inner to 
the outer regions of the cluster at all times. 

A useful quantity for examining the evolution of the cluster structures in Runs 1 and 2 is
the ratio of the core radius to half-light (or half-mass) radius 
\citep[e.g.,][]{vesperini:94,trenti:07,heggie:07,hurley:07}. In regards to the latter of 
these radii, the relevant observational parameter is the projected radius containing 
half the cluster light ($r_{h,l}$). This is straightforward to calculate for the EFF family
of models. The enclosed luminosity as a function of projected radius $r_{P}$ may be obtained by
integrating Eq. \ref{ae:eff3d} within a cylinder of radius $r_{P}$ along the line of sight 
\citep[e.g., Eq. 11 in][]{mackey:03a}:
\begin{equation}
L(r_P) = \frac{2 \pi \mu_0}{\gamma - 2} \left[ a^2 - a^\gamma \left( r_P^2 + a^2 \right)^{-(\gamma - 2)/2} \right]\,\,.
\label{e:projlum}
\end{equation}
When $r_P = r_{h,l}$ the enclosed luminosity is half of the total luminosity, i.e. 
$L(r_{h,l}) / L_{\rm{tot}} = 1/2$. Substituting this into Eq. \ref{e:projlum} and rearranging
the result leads to an expression for $r_{h,l}$:
\begin{equation}
\log(r_{h,l}^2 + a^2) = \frac{2}{2-\gamma} \log \left( \frac{1}{a^\gamma} \left[ \frac{L_{\rm{tot}} (2 - \gamma)}{4 \pi \mu_0} + a^2 \right] \right)\,\,.
\label{e:halflight}
\end{equation}
Projected half-light radii, along with the ratios $r_c / r_{h,l}$ may hence be calculated for the 
LMC and SMC cluster samples of \citet{mackey:03a,mackey:03b} using their best-fitting EFF models and 
total luminosity estimates. Directly comparable quantities may also be calculated at each output time 
for our $N$-body Runs using the ``observed'' EFF models. In this
procedure, for the purposes of direct comparison we do not use $L_{\rm{tot}}$ as calculated by 
the $N$-body code, but rather the total luminosity enclosed within some limiting observational 
radius, as specified in \citet{mackey:03a}.

\begin{figure}
\begin{center}
\includegraphics[width=0.47\textwidth]{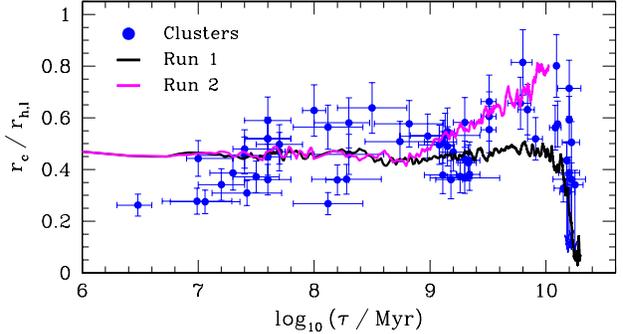}
\caption{Evolution of the ratio of core radius to projected half-light radius $r_c / r_{h,l}$
for $N$-body Runs 1 and 2, compared with measurements for LMC and SMC clusters. Note that
the measured ratios for the oldest, most compact LMC clusters are upper limits, reflecting
the upper limits to the core radius measurements for these clusters (cf. Fig. \ref{f:radiusage}).}
\label{f:evolratio1}
\end{center}
\end{figure}

The evolution of $r_c / r_{h,l}$ for Runs 1 and 2, compared with the measurements for LMC and SMC
clusters, may be seen in Fig. \ref{f:evolratio1}. For much of Run 1, this ratio is a stable
quantity at $r_c / r_{h,l} \approx 0.45$. As this model enters core collapse, however, the
ratio shrinks to become very small. This is very similar behaviour to that observed by previous
authors -- in particular \citet{hurley:07}, who measured the evolution of an identical 
(observationally defined) quantity in his large $N$-body models. Very different behaviour is
observed for Run 2, however. As soon as the BH population in this model becomes active and core 
expansion begins, $r_c / r_{h,l}$ begins to steadily increase. This presumably reflects the
increased heating efficiency of the BH population within the stellar core, as compared with the
heating efficiency at larger radii in the cluster (cf. Fig. \ref{f:lrad2}). By the end of Run 2
$r_c / r_{h,l} \sim 0.8$, matching the values observed for several of the most extended 
Magellanic Cloud clusters. These observations are consistent with the results of \citet{hurley:07},
who found that even the presence of one BH-BH binary can prevent the expected decrease in
$r_c / r_{h,l}$ -- in our models the presence of many BHs results in a significant {\it increase}
in this ratio. It has been suggested that a cluster with a large value of $r_c / r_{h,l}$ may
harbour a central IMBH \citep[see the extensive discussion presented by][]{hurley:07}; however,
our Run 2 clearly demonstrates that the presence of a population of stellar-mass BHs can also 
lead to large values of this ratio. 

Returning to Fig. \ref{f:lrad2}, one significant point of note is that although the spatial 
distribution of stellar mass 
is quite different in Run 2 compared to Run 1, the overall mean stellar mass in Run 2 (dashed line)
remains almost exactly the same as that in Run 1 throughout the simulation. Given that the initial
stellar populations in the two models were identical, this indicates that the typical mass of a 
star escaping across the tidal radius in the two runs is very similar. Calculating the mean mass
of all escaping stars in Run 1 between $\tau = 100$ Myr (when the early violent relaxation is 
essentially complete) and $\tau = 10667$ Myr (when Run 2 is terminated) reveals a value of
$0.328\,{\rm M}_\odot$, while the same calculation for Run 2 results in $0.332\,{\rm M}_\odot$.
These two values are indistinguishable, which is remarkable given the strong 
divergence in the structural evolution of the two clusters. Inspection of the distribution of 
velocities with which stars escape between $\tau = 100-10667$ Myr in each simulation reveals 
these also to be indistinguishable. Together these results imply that both models lose stars solely 
due to relaxation processes. There is only a tiny group of $\sim 20$ relatively high velocity stellar 
escapers in Run 2 (i.e., which have an escape velocity greater than that of the fastest escaper in 
Run 1) out of a total of more than $55\,000$ stellar escapers, indicating that stars interact closely 
with BH binaries only very rarely. Heating of the stellar component via close interactions between 
stars and BH binaries is therefore negligible -- the hardening of BH binaries is driven solely 
through interactions with other BHs in the central subsystem.

It is also enlightening to consider the properties of the escaping BHs in Run 2. The cumulative
number of escaped single and binary BHs is plotted in Fig. \ref{f:bhevol}d. The approximate 
time at which core expansion begins, $\tau = 650$ Myr, is marked with a vertical dotted line.
Some single BHs escape before this point -- these are BHs which are formed in the outer regions
of the cluster and drift across the tidal boundary due to the early violent fluctuations in the
cluster's gravitational potential. After $\tau = 650$ Myr, once BHs begin to be ejected solely
due to interactions in the central subsystem, it is clear that the cumulative numbers $N_{{\rm e}}$ 
of escaping single and binary BHs increase more slowly at later times -- that is, that the escape 
rates decrease with time. Hypothesising that the time derivatives of these rates vary as $-1/\tau$ 
(i.e., that $dN_{{\rm e}}/d\tau \propto -\log\tau$) suggests a fit of the form 
$N_{{\rm e}}(\tau) = A_0 + A_1\tau - A_2\tau\log\tau$ to the cumulative distributions, 
where the $A_i$ are coefficients derived in the fitting process. Best-fit curves of this form are 
also plotted in Fig. \ref{f:bhevol}d (dashed lines). Clearly these are excellent matches to  
the observed cumulative distributions, indicating that the rates of single and binary BH escape 
do indeed both have time derivatives which vary as $-1/\tau$. 

The BH escape rates decrease with time because the density of the central BH subsystem is also
decreasing with time -- this is evident from Fig. \ref{f:bhevol}c, which shows that the inner BH 
Lagrangian radii follow a generally increasing trend throughout the majority of the simulation. 
The typical number of BHs in the central subsystem falls with time because of BH ejections (Fig. 
\ref{f:bhevol}b), and these ejections also heat the BH core. Simultaneously, the stellar component 
of the cluster is becoming more extended, meaning that the gravitational potential at the centre 
of the cluster due to this component is becoming increasingly shallow. Together these processes 
lead to the density of the central BH subsystem decreasing, on average, with time. The mean BH-BH 
encounter rate also therefore decreases with time, meaning that the BH binary hardening rate 
decreases, as does the BH ejection rate, as observed.

The decreasing BH binary hardening rate also means that the BH scattering rate decreases with
time. Together with the slowing BH ejection rate, this means that the stellar core is also less 
efficiently heated with time. This is reflected in the roughly logarithmic dependence of $r_c$ 
on $\tau$. Because the BH scattering and ejection rates decrease throughout the lifetime of 
Run 2, by the end of the simulation at $\tau_{{\rm max}} = 10.67$ Gyr, there is still a
sizeable population of $65$ single BHs and $2$ binary BHs remaining in the cluster. This contrasts
strongly with the results from early, more analytic, studies of the evolution of BH subsystems
in globular clusters, which predicted depletion of any BH populations on timescales much less 
than the cluster lifetimes \citep*{kulkarni:93,sigurdsson:93}. The fact that the BH encounter 
rate decreases due to the interplay between the stellar component of the cluster and the BH 
population, as seen in our detailed numerical modelling, prolongs the life of the BH subsystem in 
a massive star cluster for much longer than previously appreciated.

\begin{figure}
\begin{center}
\includegraphics[width=0.48\textwidth]{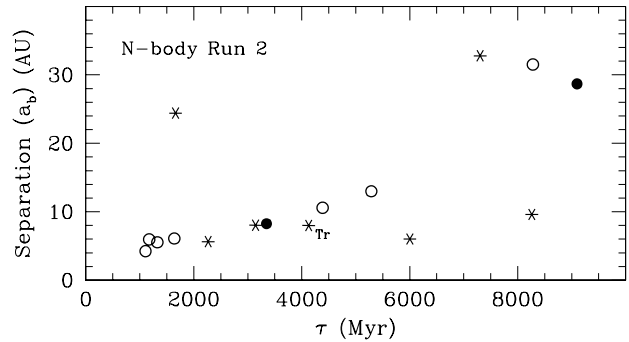}
\caption{Separations and eccentricities of the ejected BH binaries in Run 2 as a function of
cluster age. Eccentricity is represented by point style: BH binaries with $e \leq 0.8$ are
asterisks, those with $0.8 < e \leq 0.95$ are open circles, while those with $e > 0.95$
are filled circles. The asterisk marked with `Tr' is the innermost binary in the one 
ejected triple BH system (see text).}
\label{f:escbbh}
\end{center}
\end{figure}

The properties of the ejected BH binaries in Run 2 complete the picture of BH evolution in this 
model. Over the course of the simulation, $15$ 
BH binaries are ejected. Their separations ($a_b$) and eccentricities ($e$) are displayed in Fig. 
\ref{f:escbbh} as a function of cluster age. The hardest binaries are clearly ejected at the 
earliest times. This is when the cluster escape velocity ($v_{{\rm esc}}$) is largest -- that is,
when the binaries can be hardened to the greatest extent before the recoil velocity imparted during 
close interactions ejects them from the cluster. Typical separations are $a_b \approx 4-6$ AU for 
binaries ejected when $\tau < 2.5$ Gyr. As the cluster expands and loses mass, $v_{{\rm esc}}$ 
decreases and BH binaries are ejected before becoming this hard. For $\tau > 5$ Gyr, $a_b$ is 
typically $10-30$ AU. There is no strong pattern in eccentricities of ejected BH binaries -- there 
are six with $e \leq 0.8$, seven with $0.8 < e \leq 0.95$, and only two with $e > 0.95$. The maximum 
eccentricity of an ejected binary is $e = 0.972$. The BHs which are members of ejected binaries 
have a mean mass of $10.98\,{\rm M}_\odot$. This is more massive 
than the overall mean for BHs in Run 2, which have masses distributed uniformly in the
range $8 \leq m_{{\rm BH}} \leq 12\,{\rm M}_\odot$.

In addition to the $15$ BH binaries, one triple BH system is ejected from Run 2, at
$\tau \approx 4100$ Myr. This consists of a tight low-eccentricity binary ($a_b = 8$ AU, $e=0.376$)
with a single BH bound in a wider low-eccentricity orbit ($a_b = 149$ AU, $e=0.370$).

Previous studies have demonstrated that binary BHs ejected from massive star clusters can have 
orbital properties that would lead them to coalesce within a Hubble time due to the emission of
gravitational radiation \citep[see e.g.,][]{portegieszwart:00}. Such objects may therefore be 
possible candidates for detection by gravitational wave experiments. 
An approximate formula for the time-scale for a BH binary to coalesce due to the emission
of gravitational radiation is given by \citep[e.g.,][]{portegieszwart:00}:
\begin{equation}
T_{{\rm coal}} \approx 3.2\times10^8 \left(\frac{{\rm M}_\odot}{m_{{\rm BH}}}\right)^3 \left(\frac{a_b}{{\rm AU}}\right)^4 \left(1-e^2\right)^{\frac{7}{2}}\,\,\rm{Gyr}.
\label{e:grmerge}
\end{equation}
It is easy to show that none of the ejected BH binaries in Run 2 would merge
within a meaningful time-scale (here we adopt $\sim 12$ Gyr, which is the approximate age of the 
Universe minus the delay of $\sim 1.5$ Gyr before the first BH binary ejections occur in Run 2). 
The most tightly bound ejected binary has $a_b = 4.2$ AU and $e=0.839$, while the most eccentric 
ejected binary has $a_b = 8.2$ AU and $e = 0.972$. Even so, the orbital parameters of 
these objects are not vastly different from those which would lead to merging events on an interesting 
time-scale. For example, the binary with $e = 0.972$ would need $a_b = 1.06$ AU to merge in $12$ 
Gyr, while that with $a_b = 4.2$ AU would need $e = 0.994$. We consider this topic further in
Sections \ref{ss:pair2} and \ref{s:discussion}.

Recent large-scale $N$-body simulations have demonstrated comprehensively that when an 
intermediate-mass black hole (IMBH; mass $\sim$ a few$\,\times 10^3 {\rm M}_\odot$) is present 
in a massive star cluster, a central stellar density and velocity cusp develops about this object 
\citep*[e.g.,][]{baumgardt:04a,baumgardt:04b}. It is natural to ask whether a similar cusp develops 
in Run 2, where a comparable BH mass is concentrated in the cluster centre, but in the form of
many relatively small objects rather than one massive object. 

\begin{figure}
\begin{center}
\includegraphics[width=0.48\textwidth]{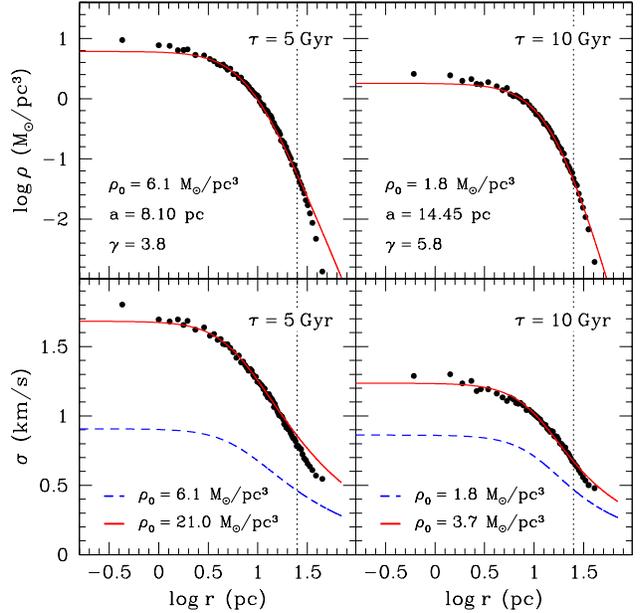}
\caption{Radial mass density profiles and 1D velocity dispersion profiles 
for Run 2 (upper and lower panels, respectively) at cluster ages of $5$ Gyr and $10$ Gyr (left and 
right panels, respectively). In all panels, data from the simulation are marked with solid black 
circles. {\it Only stars were used to derive these measurements -- BHs were excluded}. In the upper 
panels, the solid line indicates a deprojected EFF model fit using parameters set by those observed 
from the simulation at the appropriate output time. The vertical dotted lines indicate the maximum 
radius used in the construction of projected brightness profiles when obtaining these observations 
(see Section \ref{s:simobs}). In each lower panel, the dashed line represents the velocity
dispersion profile predicted by the EFF model plotted in the matching upper panel, while the solid
lines represent the same models with rescaled central densities to fit the data (see text).}
\label{f:cusp}
\end{center}
\end{figure}

Fig. \ref{f:cusp} summarises the structural and dynamical state of the stellar component of Run 2
at two output times: $\tau = 5$ Gyr and $\tau = 10$ Gyr. These are late enough that any cusp
should have had sufficient time to form 
\citep[see e.g., the time-scales in][]{baumgardt:04a,baumgardt:04b}. The top panels in 
Fig. \ref{f:cusp} show the 3-dimensional radial mass density profile of the cluster at the two
output times (solid circles). All luminous matter in the cluster was counted in each
profile (i.e., BHs were excluded). The radial bins contain $50$ stars for radii closest to the 
cluster centre, graduating to $100$ stars, then $500$ and $1000$ stars at increasingly large radii.

For comparative purposes, we have also plotted deprojected EFF models in these panels.  
These models are of the form of Eq. \ref{ae:eff3d}. In calculating them, we used the 
values of $\mu_0$, $a$, and $\gamma$ observed from the projected brightness profile at the 
appropriate time, as described in Section \ref{s:simobs}. The maximum radial extent of the 
projected brightness profiles is marked in Fig. \ref{f:cusp} by vertical dotted lines. Agreement 
between the models and data is not necessarily expected beyond these radii; in addition, 
tidal effects become important at the largest radii. For convenience (see below), 
we took $\gamma$ in the deprojected models to be the closest integer value to that observed -- 
that is, $\gamma = 4$ at $\tau = 5$ Gyr, and $\gamma = 6$ at $\tau = 10$ Gyr. In each deprojected
EFF model, the central surface luminosity density, $\mu_0$, was converted to the volume luminosity 
density $j_0$ via Eq. \ref{ae:eff3d}. For example, at $\tau = 5$ Gyr, we measured 
$\mu_0 = 0.55\,{\rm mag}\,{\rm pc}^{-2} = 51.05\,{\rm L}_\odot\,{\rm pc}^{-2}$, which
corresponds to $j_0 = 4.59\,{\rm L}_\odot\,{\rm pc}^{-3}$. To obtain a mass density from this 
value requires multiplication by a global mass-to-light ratio appropriate for the age and metal 
abundance of the cluster. We determined this empirically by fitting the deprojected EFF model 
to the measured data. The resulting mass-to-light ratios ($M/L = 1.33$ at $\tau = 5$ Gyr, and
$M/L = 2.01$ at $\tau = 10$ Gyr) are a good match for those calculated by directly summing
the mass and luminosity of all stars in the cluster, excluding BHs ($M/L = 1.35$ at $\tau = 5$ 
Gyr, and $M/L = 2.10$ at $\tau = 10$ Gyr). We note that the assumption of a globally
constant mass-to-light ratio is a reasonable approximation for Run 2 at these late times due
to the relatively low degree of mass segregation amongst the stellar component of this cluster.

It is clear from Fig. \ref{f:cusp} that no significant cusps are present in the cluster's stellar 
density profile at either time. At the most, it is possible that very marginal cusps exist, 
since the density profiles rise slightly above the EFF models (which have constant density 
cores) at the innermost few data points; however the significance of this ``density excess''
is very low. Certainly, striking cusps of the form of those observed by 
\citet{baumgardt:04a,baumgardt:04b} to develop about central IMBHs in clusters are not present.

In the lower panels of Fig. \ref{f:cusp}, we plot the 1D stellar velocity dispersion as a function
of radius at the two output times. The same radial bins as in the density 
profiles were used. Again, although the central regions of these profiles show some point-to-point 
scatter, there is no evidence at either time of a significant central velocity cusp analogous to 
the type observed by \citet{baumgardt:04a,baumgardt:04b} when an IMBH is present.

It seems likely that the absence of a stellar density and velocity cusp about the central BH 
subsystem in Run 2 is due to the fact that scattered and ejected BHs are constantly moving through 
the region where a cusp would be expected to develop. This region is hence constantly being 
disturbed so that stars cannot settle into a stable distribution about the central concentrated 
mass as they can when only a single high-mass object is present. Such a process is similar to
the destruction of cusps in galactic nuclei by supermassive black holes; except in that case
a single very massive binary BH typically does the damage \citep[e.g.,][]{merritt:01}.

Since there is no evidence from the 3D radial profiles for any large central density
or velocity cusps, the projected profiles which it is possible to observe for real clusters
will certainly show no evidence for any cusps. This is supported by the surface brightness
profiles calculated at each output time in Run 2 to measure $r_c$ and $\gamma$, which exhibit 
constant density cores as observed for the majority of LMC and SMC clusters 
\citep{mackey:03a,mackey:03b}. 

\begin{figure}
\begin{center}
\includegraphics[width=0.48\textwidth]{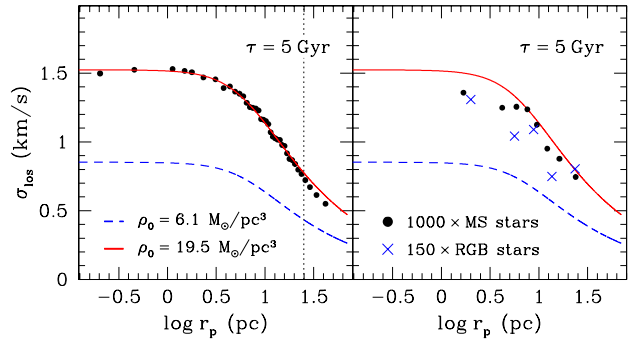}
\caption{Dispersion in the line-of-sight velocities in Run 2 at $\tau = 5$ Gyr, as a function
of projected radius. In the left panel, data from the simulation are marked with solid black 
circles. {\it Only stars were used to derive these measurements -- BHs were excluded}. The dashed
line represents the dispersion profile predicted by a Plummer sphere with central density taken
from the profile in the upper left panel of Fig. \ref{f:cusp}. The solid line represents
the same model with a rescaled central density (see text). In the right panel dispersion profiles
from two stellar sub-samples are plotted: solid dots are for a sample of $1000$ upper main
sequence stars, and crosses are for a sample of $150$ red giant branch stars. The two models 
from the left-hand panel are also marked.}
\label{f:losvel}
\end{center}
\end{figure}

Is there therefore another means by which we might infer observationally the presence of the 
significant central BH population in Run 2? The lower panels in Fig. \ref{f:cusp}, 
together with Fig. \ref{f:losvel} sketch the principles of one potentially viable method. In each 
of the lower panels in Fig. \ref{f:cusp}, we plot the stellar velocity dispersion profile predicted 
by taking the parameters $(\rho_0,a,\gamma)$ from the nicely-fitting deprojected EFF density model 
marked in the respective top panel. Since we chose integer values of 
$\gamma$, the predicted velocity dispersion profiles are analytic, and easily computed. That for 
$\gamma = 6$ is given by Eq. \ref{ae:sigma6}, while the $\gamma = 4$ case is the well known 
Plummer sphere:
\begin{equation}
\sigma^2(r) = \frac{2 \pi G \rho_0 a^2}{9 \sqrt{1+\frac{r^2}{a^2}}}\,.
\label{e:sigma4}
\end{equation}
The resulting velocity dispersion profiles are plotted with dashed lines in the lower panels
of Fig. \ref{f:cusp}. Clearly they are a very poor fit to the measured profiles. However, simply 
rescaling the central density $\rho_0$ so that the central velocity dispersion predicted by the 
deprojected EFF model is consistent with the innermost measured data points results in rather nice 
fits (solid lines), at least out to large radii where the external tidal field begins to affect the 
stellar dynamics. The required central densities at $5$ and $10$ Gyr are, respectively, 
$\approx 3.4$ and $\approx 2.1$ times those determined from the density profiles in the upper panels. 
This clearly implies that unseen matter (i.e., the BH population) is influencing the stellar dynamics 
in the cluster. By measuring the velocities of stars in an extended cluster, we might therefore be 
able to infer the presence of a retained BH population.

Such measurements are difficult. Apart from any technical intricacies, we are limited to
working in projection and, with present technology, to using line-of-sight velocities only.
The left panel of Fig. \ref{f:losvel} shows the dispersion in the line-of-sight velocities
in Run 2 at $\tau = 5$ Gyr as a function of projected radius. As previously, all luminous matter 
in the cluster was counted in the profile, but BHs were excluded. The radial bins contain $50$ 
stars for radii closest to the cluster centre, graduating to $100$ stars, then $500$ and $1000$ 
stars at increasingly large projected radii. \citet{wilkinson:02} provide an expression
for $\sigma_{{\rm los}}(r_p)$ in a Plummer sphere (their Eq. 27 and 28). In the simplest case 
of an isotropic mass-follows-light model their expression reduces to:
\begin{equation}
\sigma_{{\rm los}}^2(r_p) = \frac{\pi^2 G \rho_0 a^2}{16 \sqrt{1+\frac{r_p^2}{a^2}}}\,,
\label{e:sigmaproj}
\end{equation}
which has the same functional form as $\sigma^2(r)$, but a slightly smaller central value:
$\sigma_{{\rm los}}^2(0) \approx 0.88\sigma^2(0)$. This suggests that the dynamical signature
we observed from the lower panels of Fig. \ref{f:cusp} should still be visible in projection,
and indeed we find this to be the case. In the left panel of Fig. \ref{f:losvel} we fit a
model of the form Eq. \ref{e:sigmaproj} to the data (solid line), again leaving $\rho_0$ as a free 
parameter. In this case we find $\rho_0 = 19.5 {\rm M}_\odot$, which is very similar to the value 
required to fit the deprojected velocity data, but very different from the value implied from the 
radial density profile. We also computed a model using this latter value 
($\rho_0 = 6.1 {\rm M}_\odot$); this is the dashed line in the left panel of Fig. \ref{f:losvel}.

The difference between the two models is sufficiently large that it may be detectable in clusters
using presently available technology. In the right panel of Fig. \ref{f:losvel} we plot 
$\sigma_{{\rm los}}^2(r_p)$ determined using two samples of stars randomly selected from Run 2. 
The first is a sample of $1000$ upper main sequence stars grouped into six bins of $125$ stars, while
the second is a sample of $150$ red giant branch stars grouped into five bins of $30$ stars. 
Both samples clearly favour the model with the larger mass-to-light ratio. The red-giant sample 
is of a size which could feasibly be measured by a modern multi-object spectrograph such as 
VLT/FLAMES, although it must be borne in mind that the typical measurement errors in radial
velocities observed with such a facility will be at least comparable in magnitude to the
dispersion in a diffuse cluster ($\sim 1-2$ km$\,$s$^{-1}$) -- a sophisticated analysis would
be required to properly account for these.

While somewhat crude, our results demonstrate that in a cluster such as that modelled in Run 2, 
where there is a relatively large BH population present, {\it the stellar dynamics should imply the 
presence of significantly more mass than is evident from observations of the luminous component 
of the cluster}. This arms us with a means of searching, albeit indirectly, for BH populations 
in massive LMC and SMC star clusters. Even so, we expect such observations to be
extremely challenging due to the small velocity dispersions involved, the necessity of working
in projection, and the general sparsity (in terms of numbers of bright stars) of the extended 
clusters observed in the Magellanic Clouds.

\subsection{Runs 3 and 4: Strong mass segregation}
\label{ss:pair2}
We next consider the pair of simulations labelled Run 3 and Run 4. These are both strongly
primordially mass segregated clusters, created as described in Section \ref{ss:mseg} using a 
pre-evolution duration of $T_{{\rm MS}}=450$ Myr. This duration was selected so that Runs 3 
and 4 possess initial properties very similar to those observed for the very young, compact 
cluster R136 in the 30 Doradus complex in the LMC (see below).

Like Runs 1 and 2, Runs 3 and 4 start with identical initial conditions, to the extent that
they share the same random seed. Once again, the sole difference between them is that in
Run 3 the natal BH kicks are set to be $v_{{\rm kick}} \approx 200$  km$\,$s$^{-1}$, whereas
in Run 4 they are set to be zero -- this results in $f_{{\rm BH}} = 0$
and $f_{{\rm BH}} = 1$, respectively.

The primary aim of Runs 3 and 4 is to try and follow the evolution of models which look more
similar to the very youngest ($\tau \la 20$ Myr) massive LMC and SMC clusters than do Runs 1 
and 2. In particular, as discussed in Section \ref{ss:mseg}, a number of very young Magellanic 
Cloud clusters have been observed to be mass segregated to some degree. However, significant 
mass segregation does not develop in Runs 1 and 2 until $\tau \sim 100$ Myr or so. In addition, 
the projected brightness profiles of Runs 1 and 2 (and particularly the structural parameters 
$r_c$ and $\gamma$) do not resemble observed young LMC and SMC clusters until $\tau \approx 20$ 
Myr (e.g., Fig. \ref{f:evolpair1}). These differences mean that the observed early evolution 
of Runs 1 and 2 may not accurately reflect the processes occurring in the youngest massive 
Magellanic Cloud clusters.

Furthermore, in Run 2 we found that the BH population did not influence the structural 
evolution of the cluster until after the formation of the first BH binaries in the core at 
$\tau \approx 510$ Myr. Since Fig. \ref{f:radiusage} shows that there is clearly evolution in the 
observed radius-age trend on time-scales shorter than this, it is important to examine whether
it is possible for the BH population to become dynamically active earlier than seen in Run 2.
One might naively expect this to occur if BHs are formed preferentially at the centre
of a cluster, such as they would be in a primordially mass segregated object.

\begin{figure}
\begin{center}
\includegraphics[width=0.45\textwidth]{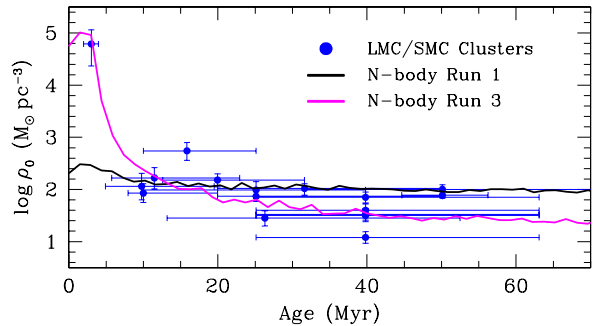}
\caption{Early evolution of the central density $\rho_0$ for Run 1 and Run 3. Run 1 has no 
primordial mass segregation while Run 3 is strongly primordially segregated. Run 3 appears 
very similar to R136 at early times; however by a few tens of Myr its central density has
decreased significantly and is a good match to LMC and SMC clusters of this age. Run 1
starts with a much lower central density, which it maintains throughout its early evolution.
Together these two models span the observed density ranges for the youngest LMC and SMC
objects.}
\label{f:densevol}
\end{center}
\end{figure}

It is important to first assess the suitability of the initial
conditions we constructed for Runs 3 and 4 before moving on to an examination of the evolution
of these Runs. One simple but useful indication is provided by the observed initial structural 
parameters $r_c$, $\gamma$, and $\rho_0$. The measured values for Runs 3 and 4 are listed in Table 
\ref{t:runs}. As described previously, these quantities are an excellent match for those determined 
for R136; see also Fig. \ref{f:youngclusters}. We note however, that R136 is nearly an order of 
magnitude more massive than our $N$-body models. Scaling issues are discussed in Section 
\ref{s:discussion}.

\begin{figure*}
\begin{minipage}{175mm}
\begin{center}
\includegraphics[width=58mm]{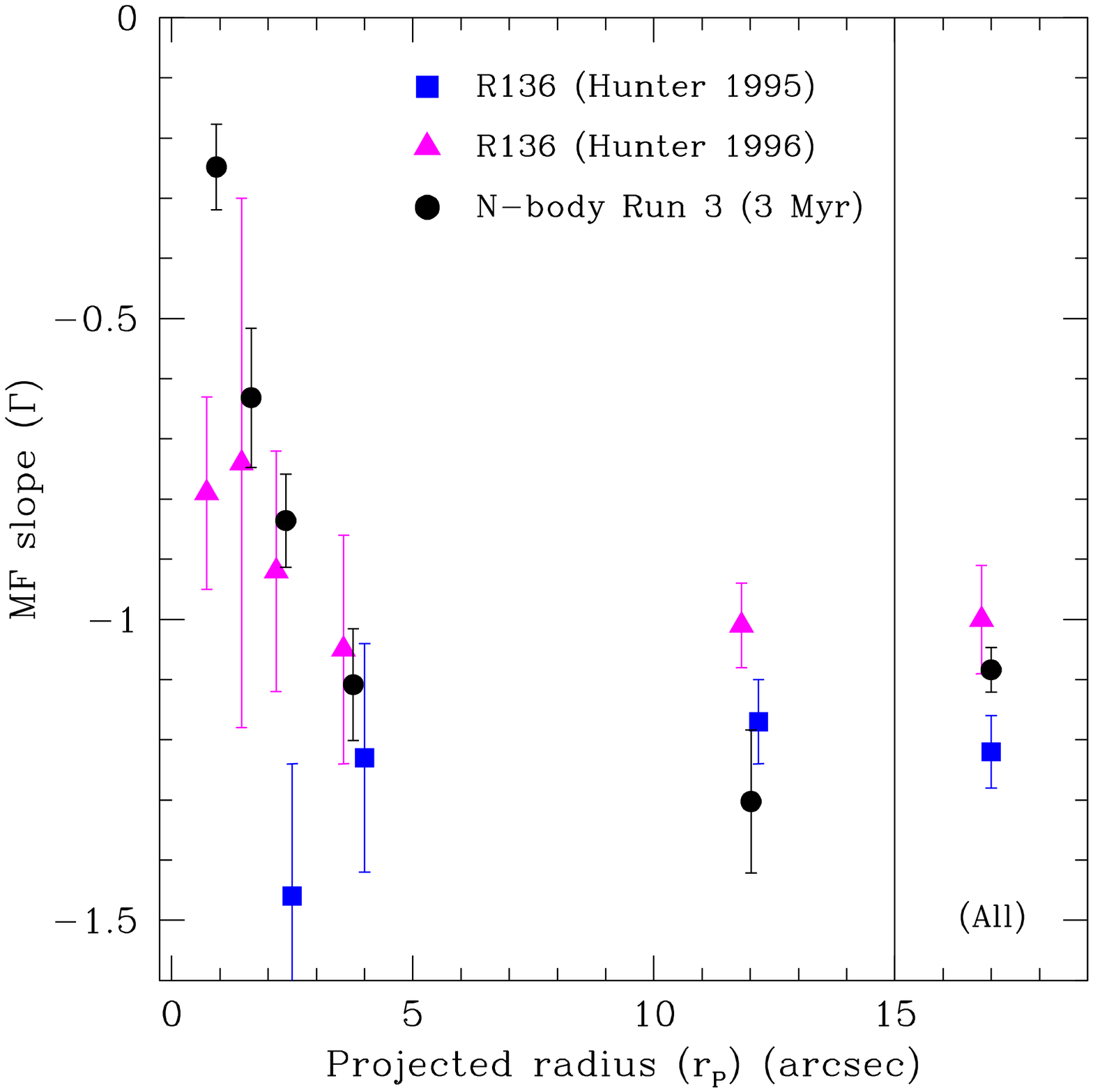}
\hspace{-2mm}
\includegraphics[width=58mm]{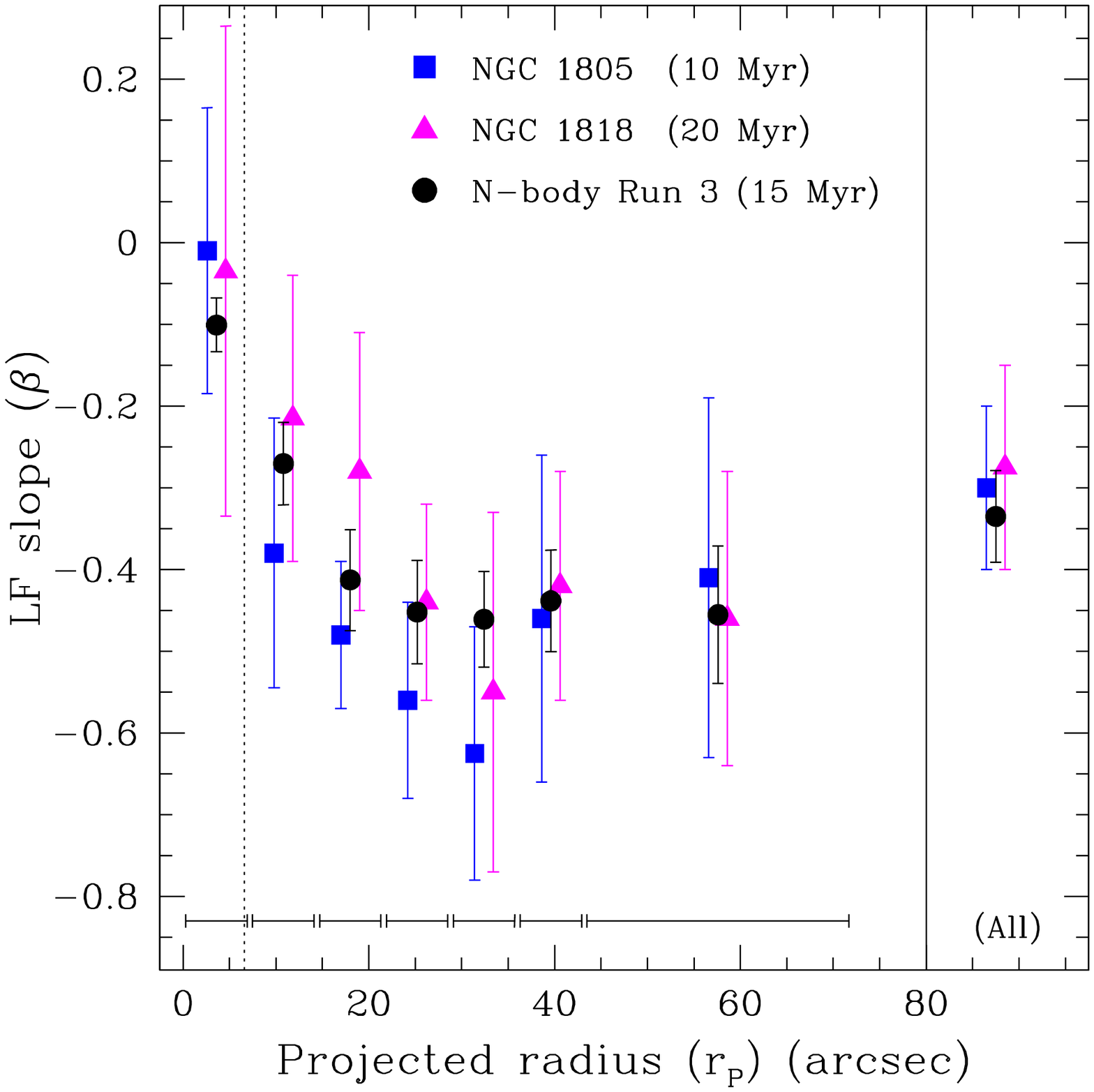}
\hspace{-2mm}
\includegraphics[width=58mm]{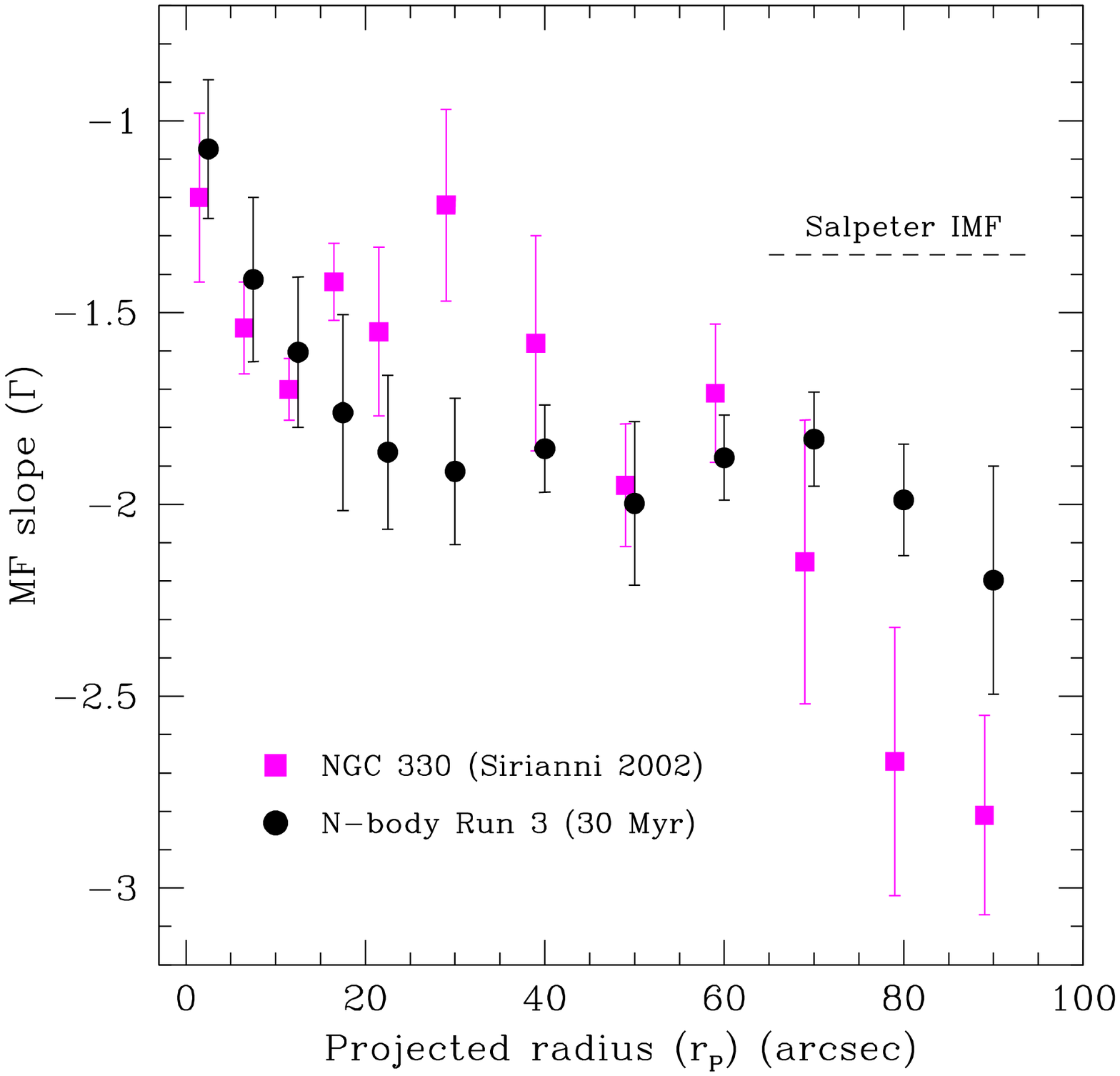}
\caption{Mass and luminosity function slopes as a function of projected radius for various
young massive LMC and SMC clusters, compared with results from simulated observations of
$N$-body Run 3. The plots have been reproduced to match those presented for each cluster by 
the original authors. {\bf Left:} Mass function slope $\Gamma$ as a function of projected 
radius in R136 in the LMC from \citet{hunter:95,hunter:96}, compared with Run 3 at age $3$ 
Myr. {\bf Centre:} Luminosity function slopes $\beta$ as a function of projected radius for 
NGC 1805 and NGC 1818 in the LMC from \citet{degrijs:02a}, compared with Run 3 at age $15$ 
Myr. {\bf Right:} Mass function slope $\Gamma$ as a function of projected radius in NGC 330 
in the SMC from \citet{sirianni:02}, compared with Run 3 at age $30$ Myr.}
\label{f:mseg}
\end{center}
\end{minipage}
\end{figure*}

One of the major differences between Runs 3 and 4 and Runs 1 and 2 is that the former have 
very much larger central densities than the latter. This is simply due to the strong central 
concentration of the most massive stars in Runs 3 and 4 as a result of the initial mass 
segregation. It is enlightening to examine the early evolution of the central densities
in these differing models -- this evolution is plotted for Runs 1 and 3 in Fig. \ref{f:densevol}.
At the start of the simulation, the density of Run 3 is directly comparable to that of R136; 
however, as the early phase of severe mass loss due to stellar evolution begins, the central 
density drops rapidly so that by $\sim 10$ Myr it matches the densities observed for other 
young LMC and SMC clusters. This rapid drop in central density implies that the central 
regions of the cluster expand during this early period of evolution, and indeed this is what 
is observed (see below). In comparison, the initial central density of Run 1 is much lower than 
that of R136, and does not change much as the rapid early stellar evolution progresses. This is 
consistent with the fact that Run 1 shows little or no central expansion during this phase. 
Together, Runs 1 and 3 span the range of central densities observed for the youngest LMC and 
SMC clusters -- we are therefore confident of the applicability of our models in this regard.

It is also possible to assess how well the primordial mass segregation 
generated in Runs 3 and 4 matches that observed in genuine young Magellanic Cloud clusters.
We do this by performing simulated observations of the radial variation of the stellar mass 
function (MF) in the models, and comparing the results to those determined from the detailed
observational studies of \citet{hunter:95,hunter:96} for R136 ($\sim 3$ Myr old); 
that of \citet{degrijs:02a} for NGC 1805 ($\sim 10$ Myr old) and NGC 1818 ($\sim 20$ 
Myr old); and that of \citet{sirianni:02} for NGC 330 ($\sim 30$ Myr old). 

\begin{figure*}
\begin{minipage}{175mm}
\begin{center}
\includegraphics[width=0.45\textwidth]{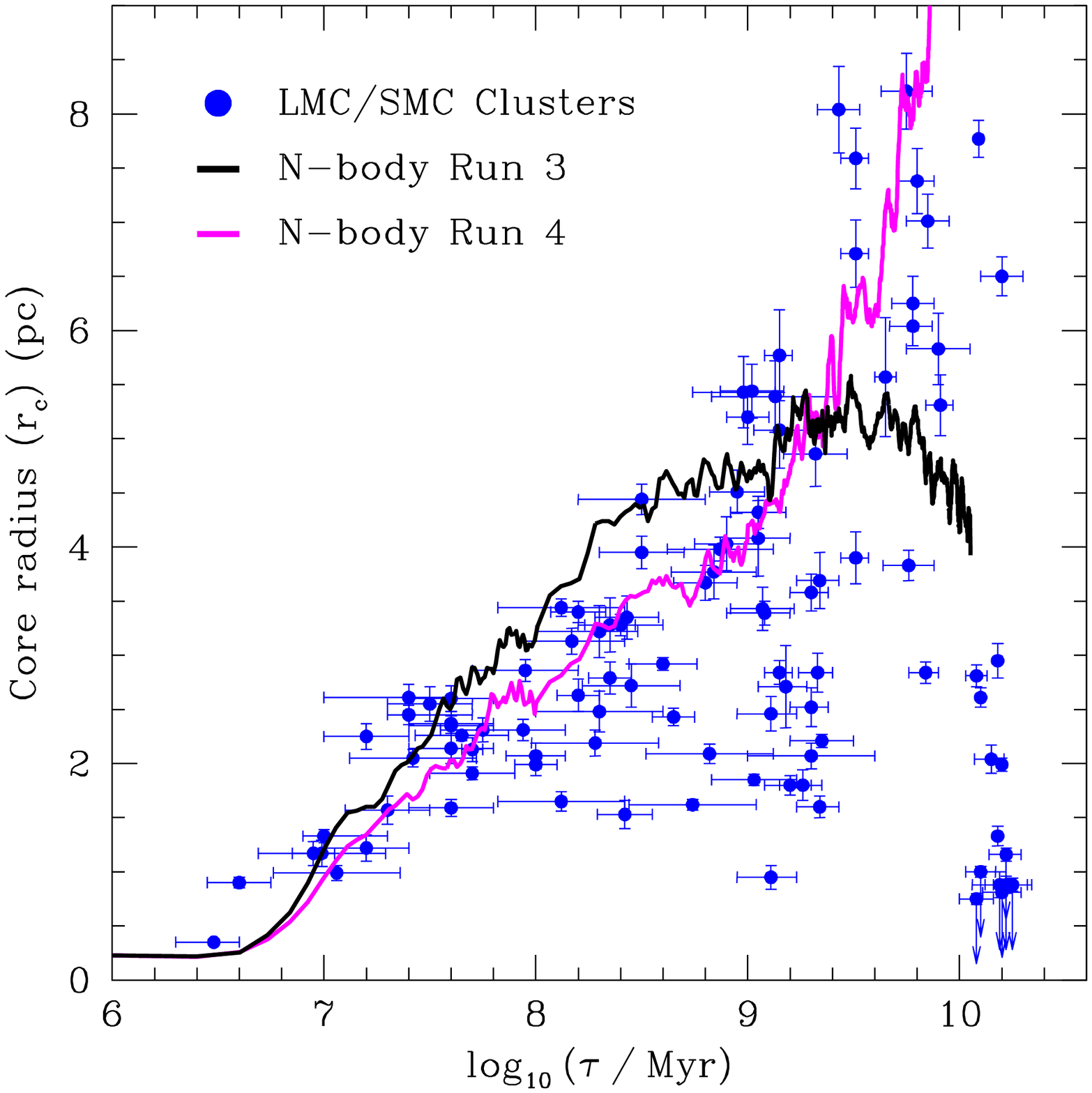}
\hspace{0mm}
\includegraphics[width=0.45\textwidth]{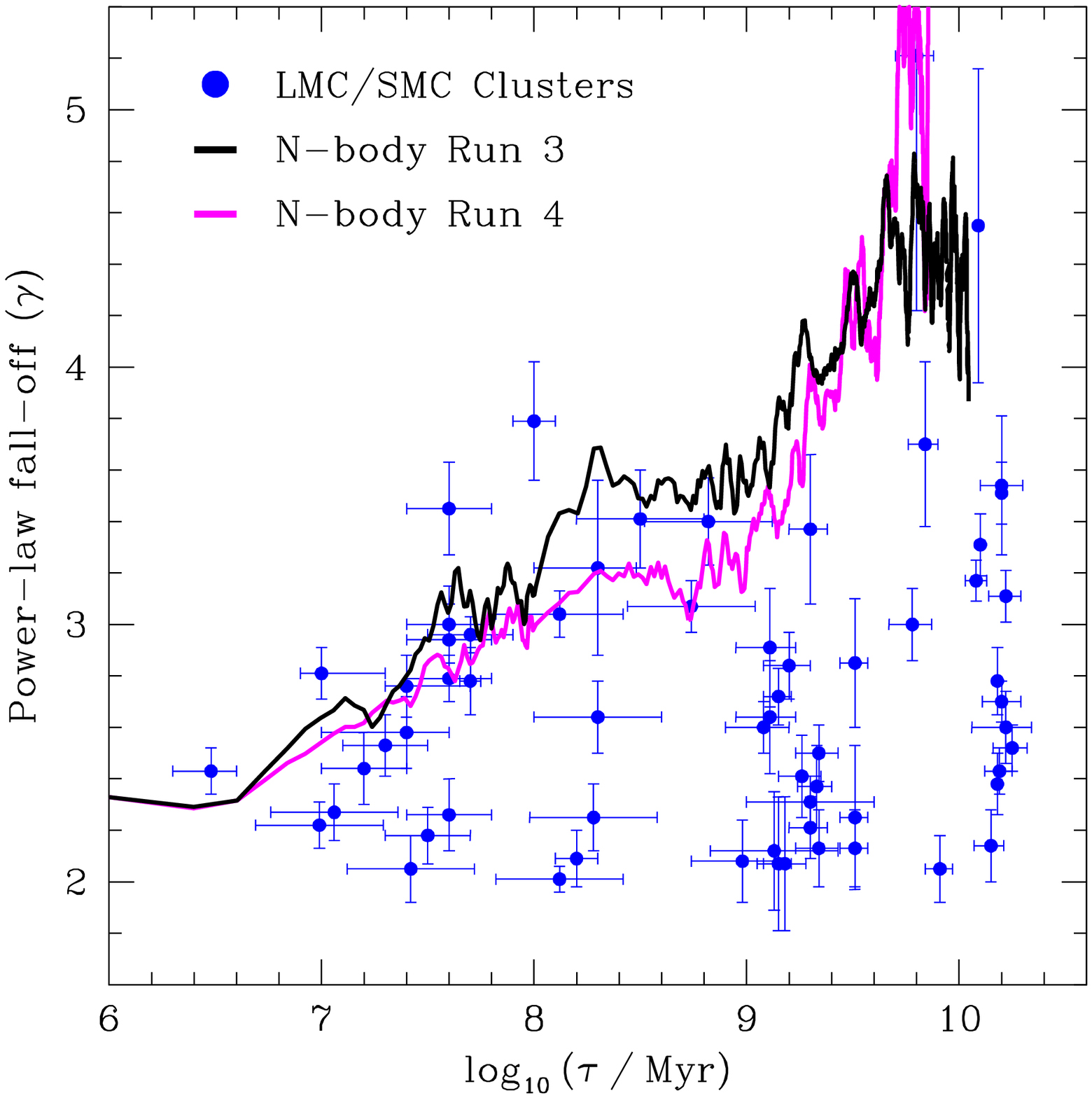}
\caption{Structural evolution of $N$-body Runs 3 and 4. Both possess significant primordial mass 
segregation; the only difference between them is the BH retention fraction ($f_{{\rm BH}} = 0$ 
and $1$, respectively). {\bf Left panel:} Evolution of $r_c$, observed as described in Section
\ref{s:simobs}. Both models experience significant expansion over the first $\sim 200$ Myr of
evolution, due to the early phase of severe mass loss due to stellar evolution. This is in
contrast to Runs 1 and 2, where the mass loss was spread throughout the cluster rather than
being centrally concentrated. Subsequently, Run 3 begins to relax dynamically and slowly contracts,
whereas the BH population retained in Run 4 becomes dynamically active, leading to further
core expansion in this model. By $\tau_{{\rm max}} = 10$ Gyr, the core radius for Run 4 has moved 
off the top of the plot, to $r_c \approx 11$ pc. {\bf Right panel:} Evolution of the power-law 
fall-off, $\gamma$, again observed as described in Section \ref{s:simobs}. Both models develop
increasingly steep $\gamma$ values as they evolve; however, that for Run 3 reaches a plateau
once the core expansion in this model ceases. In contrast, Run 4 develops a very steep fall-off
in its outer regions.}
\label{f:evolpair2}
\end{center}
\end{minipage}
\end{figure*}

In performing the simulated observations, we follow the individual cluster studies as 
closely as possible. That is, we use the same projected radial bin widths and the same stellar 
detection limits within each such bin as were used in the original observational study. This 
ensures that our measurements are closely comparable to those obtained in each. Consider, for
example, the work of \citet{sirianni:02}. These authors used five annuli of $5\arcsec$ 
width to span the range $0-25\arcsec$ in projected radius in their study of NGC 330, followed 
by ten annuli of $10\arcsec$ width to span the range $25-125\arcsec$ in projected radius. 
Ultimately, however, they decided to employ a maximum projected radius of $95\arcsec$ for their 
MF calculations, due to a significant contaminating population of field stars. Within their radial 
bins, the stellar mass limits used to calculate the MF were from the top of the main sequence
in this cluster ($\sim 7\,{\rm M}_\odot$) to the $50$ per cent completeness level -- at 
$\sim 1.3\,{\rm M}_\odot$ in the cluster centre, decreasing to $\sim 0.8\,{\rm M}_\odot$ at
a projected radius of $25\arcsec$ and beyond. When measuring our model cluster for a comparison
with the results of \citet{sirianni:02}, we took the data from the output time nearest to $30$ 
Myr, projected the positions of all stars onto a plane, converted the projected radial scale
from parsecs to arcseconds by applying the SMC distance modulus of $18.85$ assumed by 
\citet{sirianni:02}, and then applied exactly the same bin widths and mass limits per bin 
as \citet{sirianni:02} to obtain the stellar samples for MF fitting. \citet{sirianni:02}
corrected their star counts for completeness variations, so we assumed $100$ per cent 
completeness in each radial bin. At our chosen output time,
the core radius and central density of our model are within $\sim 15$ per cent of the values
measured for NGC 330 (see Figs. \ref{f:densevol} and \ref{f:evolpair2}), so the bins are sampling 
equivalent regions in the cluster.

The results of our simulated observations may be seen in Fig. \ref{f:mseg}, along with the
original measurements for R136, NGC 1805 and 1818, and NGC 330. For R136 and NGC 330, the
mass functions are represented by $\zeta(m)$, which is the number of stars per {\it logarithmic}
mass interval, as opposed to the mass function $\xi(m)$ defined in Section \ref{ss:nbodycode}. 
If the mass function $\xi(m)$ has a power-law exponent $-\alpha_i$, then the 
mass function $\zeta(m)$ will have exponent $\Gamma_i = -\alpha_i + 1$. In the case of 
NGC 1805 and 1818, we calculate and fit luminosity functions (LFs) rather than MFs, so 
as to match the work of \citet{degrijs:02a}. The assumed form of the LFs, 
defined here as the number of stars per logarithmic luminosity bin, is a power law 
of slope $\beta$.

From Fig. \ref{f:mseg} it is clear that the results obtained via our simulated observations 
of Run 3 are generally an excellent match for those measured for the real LMC and SMC clusters.
The greatest differences occur for the innermost radial bin of R136, and the outermost
radial bins of NGC 330. The former discrepancy suggests that the very centre of Run 3 (within 
$\sim 0.1$ pc) may initially be somewhat more mass segregated than R136, although we note that 
the rest of our measurements are highly consistent with the real observations of R136, and 
that the region taken by \citet{hunter:96} to obtain their measurement at the innermost radius 
is {\it extremely} crowded with bright stars. The latter discrepancy may be related to the
necessity for significant field star decontamination in the outermost regions of NGC 330 by
\citet{sirianni:02} -- again, we note that the majority our measurements of Run 3 are in
excellent agreement with those obtained by these authors for NGC 330.

Overall, these results are strongly suggestive that the initial conditions we constructed for 
Runs 3 and 4, and in particular of the algorithm we developed to generate the primordial mass 
segregation in these models, are valid. We note however, that we are not able to place any
similar observational constraints on the initial velocity distributions in these models.
As an aside, it is extremely interesting
to observe the progression of the radial mass/luminosity function profile of Run 3 from age 
$\tau = 3$ Myr to $30$ Myr, in comparison with the profiles observed for four genuine LMC and 
SMC clusters. While there has previously been nothing to link the measurements of these four
objects, the early evolution of Run 3 clearly demonstrates that a cluster which initially 
possesses a core radius, central density and radial MF profile very similar to that of R136 
can evolve to look very much like NGC 1805 and NGC 1818 after $15$ Myr and then further to look 
like NGC 330 after another $15$ Myr, simply via internal cluster dynamical processes under the 
influence of rapid mass loss due to stellar evolution.

\begin{figure}
\begin{center}
\includegraphics[width=0.47\textwidth]{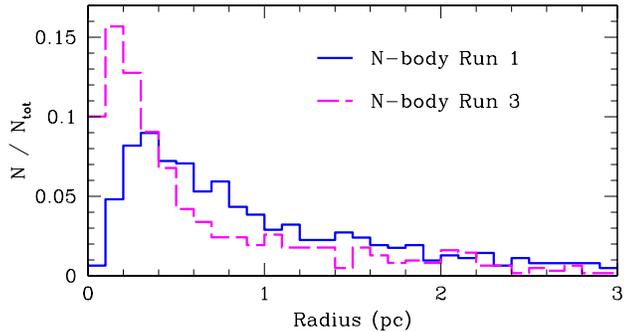}
\caption{Radial distributions of supernova explosions in Run 1 (solid line) and Run 3 
(dashed line). All supernova explosions in both models have occurred by $\tau \approx 50$ Myr.
The significantly greater central concentration of the mass loss in Run 3 compared to Run 1
is quite evident.}
\label{f:supernovae}
\end{center}
\end{figure}

In Fig. \ref{f:evolpair2}, we show the evolution of Runs 3 and 4 across the radius-age and
$\gamma$-age planes. Unlike Runs 1 and 2, both Runs 3 and 4 exhibit dramatic core expansion
right from the beginning of their evolution. This is in response to the early phase of severe
mass-loss due to stellar evolution, which in Runs 3 and 4 is highly centrally concentrated
because of the primordial mass segregation. Fig. \ref{f:supernovae} shows the radial 
distributions of all supernovae in Runs 1 and 3 -- the more centralized location of these
events in Run 3 compared with Run 1 is clearly evident. The central concentration of the 
mass-loss, together with the high central density in Runs 3 and 4, means that the amount of 
heating per unit mass lost is maximised in these models, hence leading to the observed 
dramatic core expansion. This core expansion is at least partly responsible for the rapid decrease 
in the central density of Run 3 which we noted in Fig. \ref{f:densevol} (the demise of the most 
massive cluster stars also contributes to this decrease).

It is interesting that Runs 3 and 4 do not undergo an early core collapse despite their high
central densities. Early core collapse in a massive cluster may lead to a runaway merger event,
which is one possible formation channel for a central intermediate-mass black hole (IMBH)
\citep[e.g.,][]{portegieszwart:02,portegieszwart:04}. Previous work has demonstrated that 
clusters with a very short initial median relaxation time are susceptible to early collapse -- 
\citet{portegieszwart:02} suggest $t_{rh} < 25$ Myr. It is not clear whether a similar threshold
is applicable to our primordially mass segregated models. These have very much longer initial 
median relaxation times ($t_{rh} \approx 1.2$ Gyr), but rather short central relaxation times
($t_{r0} \approx 9$ Myr). It is possible that expansion of the cluster core due to mass-loss 
from rapid stellar evolution acts against the tendency of the core to collapse more strongly
in our models than in previous models, due to the initial preferentially central location of
many massive stars.

By $\tau \approx 100$ Myr the rate of mass-loss from stellar evolution has significantly 
decreased, and by $\tau \approx 200$ Myr the core expansion in Runs 3 and 4 has largely stalled. 
Even though both models initially contain identical stellar populations, Run 3 loses 
more mass up to this point than does Run 4, because the $198$ BHs in Run 3 escape 
from the cluster immediately after formation, whereas those in Run 4 are retained. This difference 
is reflected in the larger degree of early core expansion observed in Run 3 compared to Run 4. 
Up until an age of roughly a few hundred Myr, Run 3 closely traces the observed upper 
envelope of the radius-age trend.

In both models, the early mass-loss and core expansion is accompanied by a significant
steepening in the outer power-law fall-off. This is again in contrast to the evolution
observed for Runs 1 and 2 during the early mass-loss phase, where $\gamma$ remains essentially
constant. Similarly to the core-radius evolution in Runs 3 and 4, the steepening of $\gamma$
stalls beyond $\tau \approx 100$ Myr in these models, once the rate of mass-loss has decreased. 
Furthermore, the evolution of $\gamma$ up to this point is slightly different in the two models, 
due to the retention of BHs in Run 4 and their expulsion in Run 3. 

Thereafter, Runs 3 and 4 begin to evolve quite differently. Run 3 progresses in exactly the
fashion of Run 1 -- the cluster settles into a quasi-equilibrium state where the dynamical
evolution is dominated by two-body relaxation processes, leading to the gradual development
of mass stratification. Because Run 3 is far less dense than Run 1 by this stage, its relaxation 
time-scale is much longer. By $\tau_{{\rm max}} = 10.27$ Gyr, only $3.1$ integrated median
relaxation times have elapsed in this model, compared with $4.7$ median relaxation times in Run 1 
at the same age. Hence, while Run 3 is evolving towards core collapse when the simulation is 
terminated, we would expect it to enter this phase at a much older age than observed for Run 1.

In contrast, at $\tau \approx 750$ Myr, core expansion resumes in Run 4. This continues until the
end of the simulation, which is terminated at $\tau_{{\rm max}} = 10.0$ Gyr. By this time, the
core radius of Run 4 has evolved off the top of Fig. \ref{f:evolpair2}, to reach almost
$11$ pc. This is roughly comparable in size to the core radii observed for the most extended old 
Magellanic Cloud clusters, such as Reticulum in the LMC and Lindsay 1 in the SMC (Mackey et al.
2007, in prep.).

\begin{figure}
\begin{center}
\includegraphics[width=0.48\textwidth]{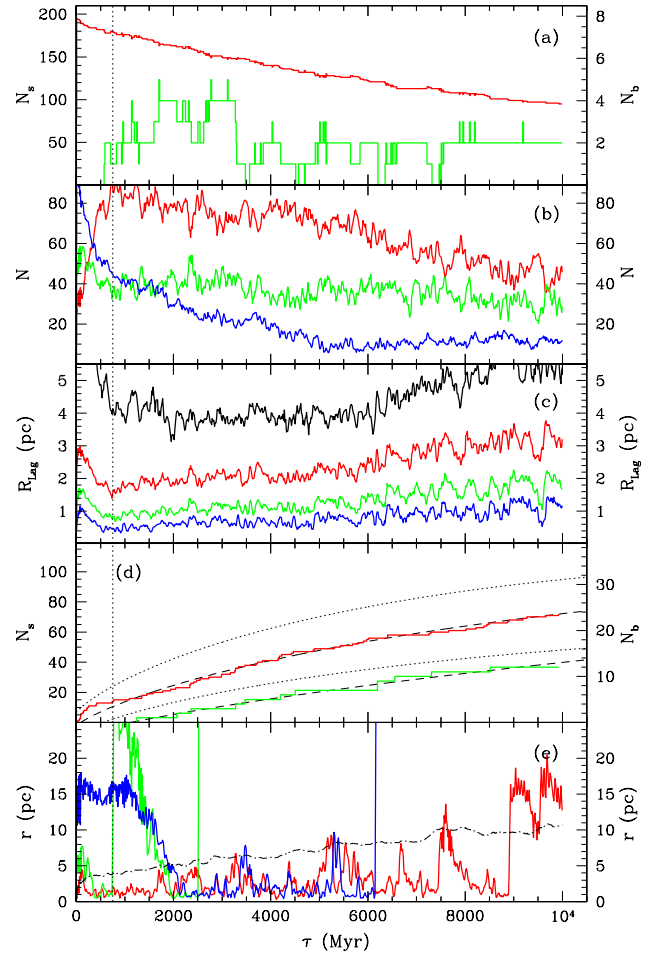}
\caption{Properties of the BH population in Run 4 as a function of time: (a) the number 
of single BHs (upper line) and binary BHs (lower line) in the cluster; (b) the number of BHs 
within the shells encompassed by the {\it stellar} Lagrangian radii $r \le R_{1\%}$, 
$R_{1\%} < r \le R_{5\%}$, and $r > R_{10\%}$ (the uppermost to lowermost lines, respectively,
at the right of the plot); (c) the {\it black hole} $10\%$, $25\%$, $50\%$ and $75\%$ Lagrangian 
radii (respectively, the innermost to outermost lines); (d) the cumulative numbers of escaped 
single BHs (upper line) and binary BHs (lower line), along with fits of the form 
$N_e = A_0 + A_1\tau - A_2\tau\log\tau$ (dashed lines) and the cumulative numbers of escaped 
single and binary BHs measured for Run 2 (dotted lines); and (e) the radial positions of three 
typical BHs. The vertical dotted line indicates $\tau = 750$ Myr, the approximate time when 
core expansion due to BH activity begins. The evolution of $r_c$ is plotted (dot-dashed line) 
in panel (e). Note the different axis scales on either side of panels (a) and (d).}
\label{f:bhevol2}
\end{center}
\end{figure}

The second, prolonged, period of core expansion in Run 4 is due to the dynamical activity of
its retained BH population. The evolution of this BH subsystem, illustrated in Fig.
\ref{f:bhevol2}, is qualitatively identical to that which we observed in Run 2. The BHs, once
formed, sink via dynamical friction and begin to accumulate at the centre of the cluster (panels
b and c). The density of this central BH subsystem increases until it becomes sufficiently high 
as to initiate the creation of stable BH binaries in three-body interactions (panel a). The first 
such object is formed in Run 4 at $\sim 570$ Myr. Subsequently, these binaries undergo superelastic
collisions with other BHs in the cluster centre, which leads to the hardening of the binaries, 
the scattering of BHs beyond $r_c$, and the ejection of BHs (panels d and e). These processes
in turn result in the observed long-term core expansion.

\begin{figure}
\begin{center}
\includegraphics[width=0.48\textwidth]{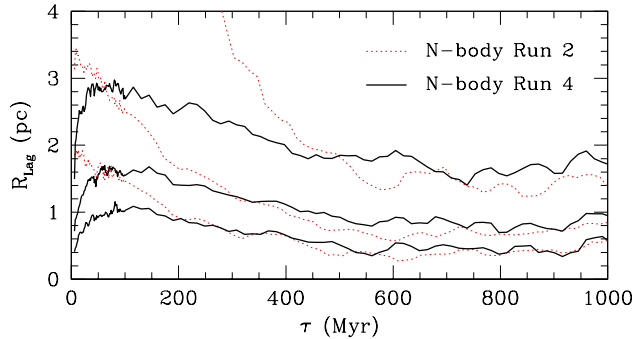}
\caption{Evolution of the $10\%$, $25\%$ and $50\%$ BH Lagrangian radii in Run 4 (solid lines).
The evolution of the same radii in Run 2 are plotted for comparison (dotted lines). This plot
clearly shows that the BH subsystem in Run 4 {\it expands} at early times along with the rest
of the core, in response to the rapid mass-loss from stellar evolution which is occurring.
Once this phase is mostly complete ($\tau \approx 100$ Myr), the evolution of the BH subsystem
is very similar to that in Run 2.}
\label{f:bhlagrad}
\end{center}
\end{figure}

One intriguing aspect of the evolution of the BH subsystem in Run 4 is that even though this
cluster was strongly primordially mass segregated, so that the majority of the BHs were
created in its very inner regions (cf. Fig. \ref{f:supernovae}), the first BH binary does not 
form until a similar time as that in the non-segregated Run 2. From Fig. \ref{f:bhevol2}, it
is also clear that the peak central BH density occurs at a similar time as it does in Run 2.
Fig. \ref{f:bhlagrad} shows an expanded view of the early evolution of the BH Lagrangian radii 
in Run 4, with those for Run 2 plotted for comparison. As expected, the majority of BHs are
formed near the centre of the cluster -- the BH Lagrangian radii are initially 
much smaller than in Run 2. However, unlike Run 2, Run 4 suffers significant early 
core expansion due to the rapid stellar evolution phase. The BH subsystem does not escape this 
-- the early centrally concentrated mass-loss is severe enough that the resulting 
expansion overcomes the natural tendency of the BHs to sink to the cluster core. This is 
reflected in the swift outward movement of the BH Lagrangian radii in Run 4, until the mass-loss 
phase is mostly complete around $\tau \approx 100$ Myr. Subsequently, the BHs do begin to sink 
to the cluster centre via dynamical friction, and the evolution of the BH Lagrangian radii in 
Run 4 closely follows that in Run 2. 

This result suggests that, contrary to expectations, the BH population in a primordially mass
segregated or centrally concentrated cluster does not become dynamically active at significantly
earlier times than does an identical population in a non-segregated cluster. In turn, this implies 
that the evolution in the radius-age trend observed on time-scales shorter than $\approx 500$ Myr 
is not due to the influence of a BH population, unless such populations are comprised of BHs with 
masses significantly in excess of $10\,{\rm M}_\odot$. 
Instead, the early evolution of the radius-age trend most probably reflects centrally concentrated 
mass-loss in dense clusters due to rapid stellar evolution. Never the less, Runs 3 and 4 clearly 
demonstrate that this process cannot alone propel clusters to the upper right corner of Fig. 
\ref{f:radiusage}, since it operates on a time-scale which is much too short. Our $N$-body models 
which possess core radii larger than $\sim 6$ pc after a Hubble time of evolution do so only 
because they have experienced prolonged core expansion due to the activity of a retained BHs, 
irrespective of whether they also experienced core expansion at ages $\tau \la 100$ Myr.

As with Run 2, by $\tau_{{\rm max}} = 10.0$ Gyr there is still a significant BH population 
remaining in Run 4: $95$ single BHs and $2$ binary BHs. In fact, this population is
rather larger than that in Run 2 at the same age. From Fig. \ref{f:bhevol2}d, it is clear 
that, while the cumulative numbers of escaped single and binary BHs in Run 4 follow the same 
functional dependence on age as in Run 2, they are, at all times, smaller than those in 
Run 2. That is, the rates of escape of BHs are always somewhat lower in Run 4 than they are 
in Run 2.

We attribute this variation to the different overall structures of Runs 2 and 4 when their 
respective BH populations become dynamically active. In Run 2 the core radius and
central density remain nearly constant from the beginning of the simulation until
this point (cf. Figs. \ref{f:evolpair1} and \ref{f:densevol}); in contrast, Run 4 undergoes 
significant early core expansion. By $\tau = 500$ Myr, Run 4 is a considerably more diffuse 
cluster than is Run 2. The shallower gravitational potential in Run 4 affects 
the distribution of the BHs within this cluster (cf. Fig. \ref{f:bhevol2}c and Fig. \ref{f:bhevol}c).
This leads to a slower interaction rate between BHs in Run 4 than in Run 2, and hence the 
observed lower BH escape rates. The same effect is primarily responsible for the BH escape
rate in a model cluster decreasing as the core radius increases (see Section \ref{ss:pair1}),
although in that case the decreasing size of the BH population contributes in addition.

\begin{figure}
\begin{center}
\includegraphics[width=0.48\textwidth]{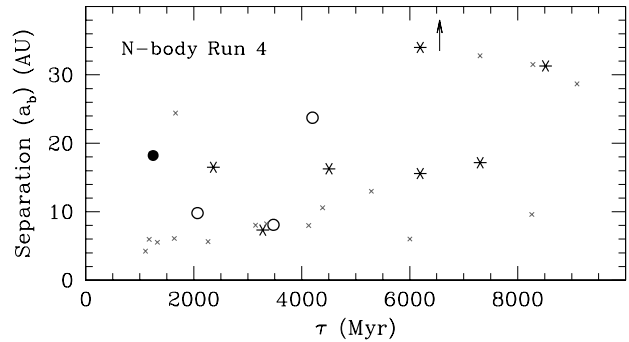}
\caption{Separations and eccentricities of the ejected BH binaries in Run 4 as a function of
cluster age. Eccentricity is represented by point style: BH binaries with $e \leq 0.8$ are
asterisks, those with $0.8 < e \leq 0.95$ are open circles, while those with $e > 0.95$
are filled circles. The ejected BH binaries from Run 2 are plotted for comparison (small
crosses). The arrow marks the ejection time of one additional Run 4 BH binary, which has
separation $a_b = 56.4$ AU and eccentricity $e = 0.609$.}
\label{f:escbbh2}
\end{center}
\end{figure}

The more diffuse nature of Run 4 also affects the properties of the ejected BH binaries in
this simulation compared with those in Run 2. In Run 4, there are $12$ BH binaries ejected
over the course of the simulation. Their separations and eccentricities are plotted in Fig. 
\ref{f:escbbh2}, along with the ejected BH binaries from Run 2 for comparison. Because Run 4
is always more diffuse than Run 2 at times when binary BHs exist, these objects are more
easily ejected in Run 4 than in Run 2. Hence, the ejected BH binaries in Run 4 are generally
not as tightly bound as those in Run 2. This can be seen in terms of the binary separations
in Fig. \ref{f:escbbh2}, which are typically larger for the ejected binaries in Run 4 than
for those in Run 2 at similar times. In addition, the ejected BH binaries in Run 4 are typically
less eccentric than those in Run 2 -- of the $12$ ejected Run 4 BH binaries, only one has
$e > 0.95$, while there are three with $0.8 < e \le 0.95$ and eight with $e \le 0.8$. The
maximum eccentricity of an ejected binary is $e = 0.981$, while the minimum is $e = 0.225$.
As in Run 2, the members of ejected binaries are typically more massive than the
average mass for the full BH population -- the mean mass of members in escaping binaries
in Run 4 is $11.20\,{\rm M}_\odot$. In Section \ref{ss:pair1}, we showed that none of the 
ejected BH binaries from Run 2 would merge due to the emission of gravitational radiation 
within a Hubble time (Eq. \ref{e:grmerge}). This is also, unsurprisingly, the case in Run 4. 

\begin{figure}
\begin{center}
\includegraphics[width=0.47\textwidth]{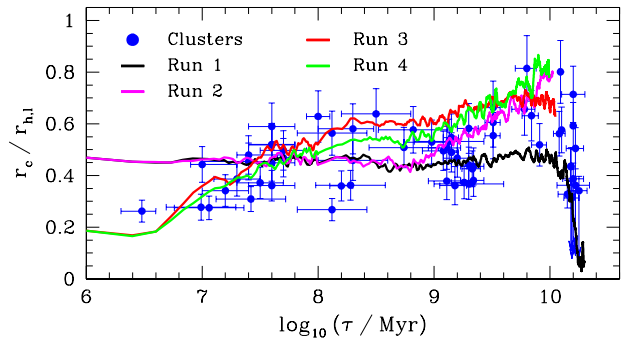}
\caption{Evolution of the ratio of core radius to projected half-light radius $r_c / r_{h,l}$
for $N$-body Runs 3 and 4, compared with measurements for LMC and SMC clusters. The evolution
of Runs 1 and 2 is also plotted, for comparative purposes.}
\label{f:evolratio2}
\end{center}
\end{figure}

Fig. \ref{f:evolratio2} shows the evolution of the ratio of core radius to projected half-light 
radius, $r_c / r_{h,l}$, for Runs 3 and 4. Runs 1 and 2 are also plotted, for comparative purposes.
The initial value of $r_c / r_{h,l}$ for Runs 3 and 4 is significantly smaller than that for Runs
1 and 2, reflecting the primordial mass segregation in these models. However, the early core 
expansion in Runs 3 and 4 results in a rapid and significant increase in $r_c / r_{h,l}$. 
Overall, Runs 3 and 4 better match the observed distribution of young and intermediate-age
Magellanic Cloud clusters than do Runs 1 and 2.
By the end of Run 3, $r_c / r_{h,l}$ is beginning to decrease as two-body relaxation begins to
dominate in this model; however, for the majority of this Run $r_c / r_{h,l} \sim 0.7$. This
demonstrates that a large observed value of $r_c / r_{h,l}$ in a physically old star cluster
need not reflect the presence of a central massive body (such as an IMBH) or a central 
accumulation of many less-massive bodies (such as stellar-mass BHs), but rather may reflect
the fact that such a cluster is not very {\it dynamically} old. Run 4, which undergoes
prolonged core expansion due to its BH population, finishes with $r_c / r_{h,l} \sim 0.8$,
matching Run 2 closely.

\subsection{Runs 4a and 4b: Variable mass segregation}
\label{ss:massseg}
As described at the beginning of Section \ref{s:results}, our four primary simulations cover 
the extremes of the parameter space we are investigating (spanned by 
$0 \le f_{{\rm BH}} \le 1$ and $0 \le T_{{\rm MS}} \le 450$ Myr), and are therefore 
expected to represent the extremes of cluster evolution induced by variation of these
particular initial conditions. However, it is important to sample intermediate values of
both $f_{{\rm BH}}$ and the degree of primordial mass segregation to check that the
parameter space is well behaved and that the models evolve as we expect (i.e., intermediate
between the extremes of Runs 1-4). 

With this in mind, we completed two additional simulations with $f_{{\rm BH}} = 1$, 
and degrees of primordial mass segregation spaced between that for Run 2 and that for Run 4.
Because the initial conditions for these new models were taken from two different output
times during the pre-evolution of Run 4, at $T_{{\rm MS}} = 115$ Myr and $225$ Myr, we denote 
them as Runs 4a and 4b, respectively. 
We only ran these models to $\tau_{{\rm max}} \sim 4.3$ Gyr, as this was more than sufficient 
to observe the progress of the clusters' evolution.

The initial properties of Runs 4a and 4b are listed in Table \ref{t:runs}.
As expected, the values of $r_c$, $\gamma$, and $\rho_0$ all lie between those
of Runs 2 and 4. The longer the duration of the pre-evolution, the smaller the initial
value of $r_c$, the higher the initial value of $\rho_0$, and the flatter the initial
value of $\gamma$. This latter, in particular, is expected due to the development of
a core-halo structure in a dynamically evolving cluster \citep[see e.g.,][]{spitzer:87}.

\begin{figure}
\includegraphics[width=0.45\textwidth]{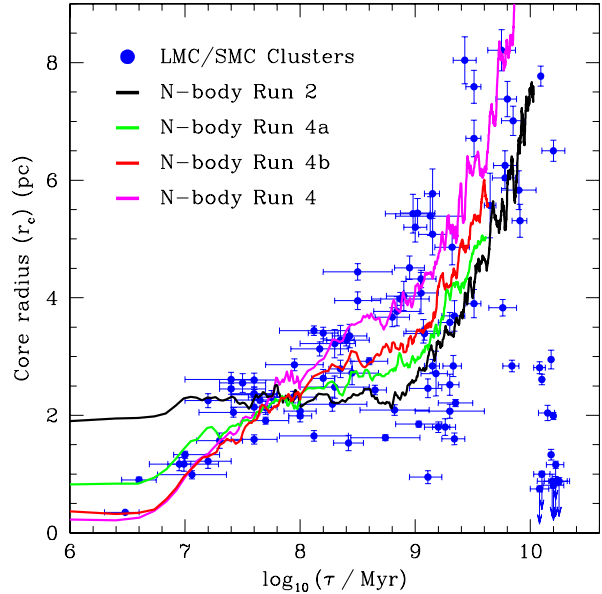}
\caption{Core radius evolution of $N$-body Runs 4a and 4b, with the evolution of Runs 2 and 4
plotted for comparison. Runs 4a and 4b, with pre-evolution durations of $T_{{\rm MS}} = 115$ Myr
and $T_{{\rm MS}} = 225$ Myr, possess primordial mass segregation of degrees intermediate
between those of Runs 2 and 4. This initial condition is the only difference between each of the 
four plotted Runs -- all four form $198$ BHs and have a BH retention fraction of $f_{{\rm BH}} = 1$.} 
\label{f:evolpair3}
\end{figure}

The core radius evolution of Runs 4a and 4b is illustrated in Fig. \ref{f:evolpair3}. Just as
with Run 4, these two models undergo two main stages of core expansion. The first, lasting
until a little after $\sim 100$ Myr, is in response to the early rapid stellar evolution. 
The second, which begins around $\tau \approx 600-800$ Myr is due to the
influence of the retained BH population. In between these two phases, there is a period during
which the core radius is roughly constant. 

As expected, the core radius evolution seen for both Runs 4a and 4b lies between the limits
defined by Runs 2 and 4. The amount of early expansion apparently varies directly with the degree 
of primordial mass segregation -- the more mass segregated a cluster, the larger the core
expansion seen at ages less than $\sim 100$ Myr. From Fig. \ref{f:evolpair2}, for Runs 3 and 4,
we saw that the amount of mass lost during the early period of rapid stellar evolution also 
influences to some extent the degree of the observed core expansion. However, all four 
models in the present case were specifically designed to possess identical populations of massive 
stars and retained BHs, and all therefore lose essentially identical amounts of mass due to stellar 
evolution at early times. The variation in the core expansion observed during this phase in Fig. 
\ref{f:evolpair3} therefore cannot be caused by differing amounts of mass-loss and must solely be 
a result of the different initial cluster structures. More centrally concentrated mass-loss, in a 
more tightly-bound cluster core, clearly leads to a greater degree of core expansion during the early 
period of a cluster's life.

After the first stage of core expansion is complete in the four model clusters, their core radii
remain roughly constant until the BH populations have accumulated at the cluster centres and started 
to form BH binaries, after which point the second phase of core expansion begins. From 
Fig. \ref{f:evolpair3}, the rates of observed expansion in this second phase are approximately 
equivalent in all four clusters, so that the tracks on the radius-age plane run close to parallel 
for the remainder of the simulations.

\begin{figure}
\begin{center}
\includegraphics[width=0.48\textwidth]{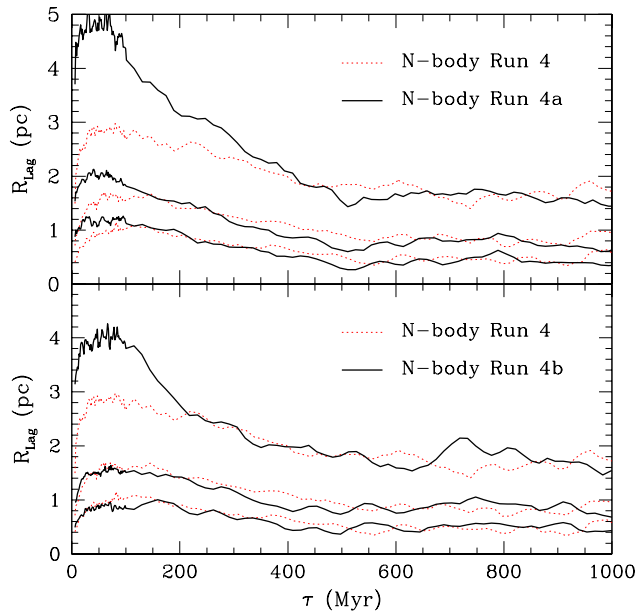}
\caption{Evolution of the $10\%$, $25\%$ and $50\%$ BH Lagrangian radii in Runs 4a (upper panel,
solid lines) and 4b (lower panel, solid lines). The evolution of the same radii in Run 4 are 
plotted for comparison in both panels (dotted lines). The BH subsystems in Run 4a and 4b both
expand at early times in response to the rapid mass-loss from stellar evolution which is occurring,
although the expansion is greater in the more heavily mass segregated Run 4b. Once the early
mass-loss phase is mostly complete ($\tau \approx 100$ Myr), the evolution of the BH subsystems
are very similar to that in Run 4 (and Run 2 -- cf. Fig. \ref{f:bhlagrad}).}
\label{f:bhlagradms}
\end{center}
\end{figure}

Fig. \ref{f:evolpair3} shows that the second stage of core expansion begins
at an approximately equivalent time in each of the four models. We already noted this fact
for Runs 2 and 4 in Section \ref{ss:pair2} and concluded that in a primordially mass segregated
cluster the BH population does not become dynamically active significantly earlier than in
a non-segregated cluster, because the strong expansion experienced by the mass segregated
cluster at early times affects the BH population sufficiently strongly to negate the
natural tendency of the BHs to sink to the cluster centre. In Fig. \ref{f:bhlagradms} we
plot the evolution of the BH Lagrangian radii in Runs 4a and 4b, with those for Run 4 plotted 
for comparison. It is clear from this plot that even though the BHs tend to form closer
to the cluster centres in more primordially mass segregated models, these models also suffer
greater degrees of early expansion, hence delaying the central accumulation of BHs.
This results in very similar evolution of the BH Lagrangian radii in Runs 2, 4, 4a, and 4b
after the early rapid stellar evolution phase is complete, and leads to the formation
of the first BH binaries at very similar ages -- $510$ Myr, $570$ Myr, $540$ Myr, and $460$ Myr,
in the four models respectively. Given that this is by nature a stochastic process, the
agreement between these times for four models with such a wide range of early structural
evolution is quite close. This observation strengthens our earlier conclusion that
primordial mass segregation in a cluster does not lead to the earlier development of a
dynamically active BH subsystem compared to a cluster which is not primordially mass segregated.

\subsection{Run 5: Intermediate BH retention}
\label{ss:kicks}
Together with Runs 4a and 4b, we computed one additional model possessing properties 
intermediate between those of our four primary Runs. In this case, instead of intermediate 
degrees of primordial mass segregation, we set up the simulation (labelled Run 5) so that 
$f_{{\rm BH}} \approx 0.5$. Its initial conditions were otherwise identical to to those in 
Runs 3 and 4 (i.e., strong primordial mass segregation set by $T_{{\rm MS}} = 450$ Myr). One 
aim of Run 5 is to check that, as should be expected, its core radius evolution lies between 
that observed for Run 3 (where $f_{{\rm BH}} = 0$) and that observed for Run 4 (where 
$f_{{\rm BH}} = 1$). More importantly however, this model explores whether the extreme case 
that $f_{{\rm BH}} \approx 1$ is {\it necessary} for significant core expansion to occur, or if 
such expansion can still develop in a system which loses a large fraction of its BHs at formation.
We set the duration of Run 5 to be roughly a Hubble time ($\tau_{{\rm max}} = 10.06$ Gyr) so that 
we could observe the full long-term core evolution of this model.

To generate a retention fraction of roughly $50$ per cent in Run 5, we examined 
the formation of BHs in Run 4 with the aim of determining a suitable distribution of natal
kicks. More specifically, we calculated the potential energy ($U_{{\rm BH}}$) of each given 
BH at the moment of its formation in Run 4, and estimated the escape velocity 
$v_{{\rm esc}} = \sqrt{-2 U_{{\rm BH}} \,/\, m_{{\rm BH}}}$. Under the assumption that
the inherited kinetic energy of the BH at formation ($K_{{\rm BH}}$) does not contribute to 
its ejection, this escape velocity corresponds to the minimum natal kick required to remove 
the BH from the cluster. However, this assumption is not always justified -- for example, the 
natal kick may be in the same direction as the inherited motion of the BH, in which case the 
minimum required velocity would be significantly smaller than the original estimate, and closer to 
$v_{{\rm esc}} = \sqrt{-2 (U_{{\rm BH}} + K_{{\rm BH}}) \,/\, m_{{\rm BH}}}$.

\begin{figure}
\begin{center}
\includegraphics[width=0.48\textwidth]{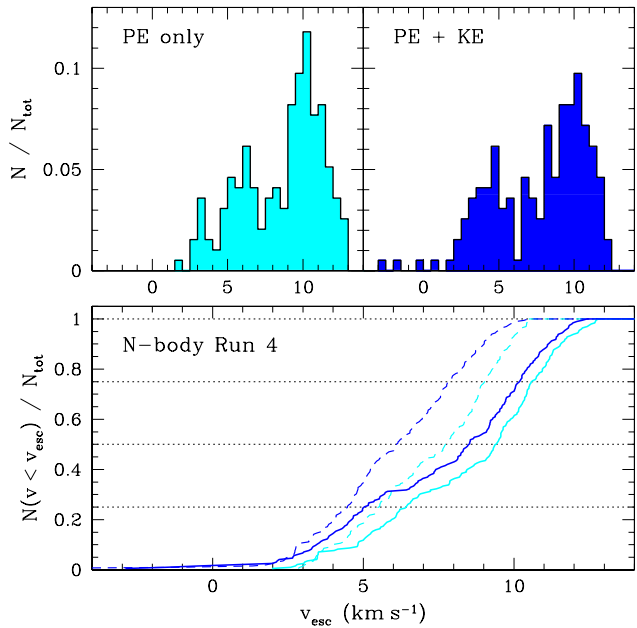}
\caption{Calculated distributions of escape velocities at formation for all $198$ BHs in Run 4.
The upper left panel is the binned distribution assuming 
$v_{{\rm esc}} = \sqrt{-2 U_{{\rm BH}} \,/\, m_{{\rm BH}}}$, while the upper right panel
is the binned distribution if the inherited kinetic energy is also included so that
$v_{{\rm esc}} = \sqrt{-2 (U_{{\rm BH}} + K_{{\rm BH}}) \,/\, m_{{\rm BH}}}$.
The distributions may be interpreted as the minimum natal kicks required to remove the 
respective BHs from Run 4 under the two different assumptions listed above. The lower panel
shows the two distributions in cumulative form (solid lines, where the distribution
including $K_{{\rm BH}}$ is to the left of that where $K_{{\rm BH}}$ is neglected). Also marked
are the equivalent distributions for Run 2, which has no primordial mass segregation (dashed 
lines).}
\label{f:bhkicks}
\end{center}
\end{figure}

The two calculated distributions of BH escape velocities in Run 4 may be seen 
in Fig. \ref{f:bhkicks}. The upper left panel is the binned distribution neglecting the
inherited kinetic energy of the BH, while the upper right panel is the binned distribution
under the assumption that the kick velocity is in the same direction as the inherited 
velocity. The lower panel shows the two distributions
in cumulative form. The most tightly bound BHs require natal kicks of order $\approx 12$ 
km$\,$s$^{-1}$ to escape the cluster, while the least tightly bound BHs require only tiny
natal kicks to escape. Note from the upper right panel that two BHs are unbound from the
cluster at their formation -- this is a result of their progenitor stars becoming unbound 
shortly before exploding as supernovae, because of the recent rapid mass-loss occurring in 
the progenitor star as well as the violent fluctuations occurring in the cluster potential
due to mass-loss from other stars.

For interest's sake, we also calculated the same distributions for Run 2, which, in contrast
to Run 4, was not primordially mass segregated. The distributions for this model are plotted
in the lower panel of Fig. \ref{f:bhkicks} as dashed lines. As might be expected, BHs
formed in the mass segregated Run 4 are significantly more tightly bound than are BHs
formed in the non-segregated Run 2. Hence, the initial structure of a cluster can have some
effect on the retention fraction of BHs. We discuss this issue in more detail in Section
\ref{s:discussion}.

The distributions observed for Run 4 in Fig. \ref{f:bhkicks} allowed us to determine a 
suitable distribution of natal kicks in Run 5 in order to set $f_{{\rm BH}} \approx 0.5$.
We did this by noting that the retention fraction $f_{{\rm BH}}$ is approximately the mean 
probability of retention calculated over the full BH subsystem -- that is, 
$f_{{\rm BH}} \approx \sum P_i(retain) \,/\, N_{{\rm BH}}$ where $N_{{\rm BH}}$ is the number of 
BHs in the subsystem and $P_i(retain)$ is the probability that the $i$-th BH will not be ejected 
by the natal kick it receives at formation. For simplicity, we set the natal kicks in Run 5 to be 
selected randomly from a uniform distribution spread between $v_{{\rm kick}} = 0$ km$\,$s$^{-1}$ 
and $v_{{\rm kick}} = v_{{\rm k,max}}$ km$\,$s$^{-1}$. In this case, for the $i$-th BH at formation
the retention probability is given by 
$P_i(retain) = P(v_{{\rm kick,i}} < v_{{\rm esc,i}}) = v_{{\rm esc,i}} \,/\, v_{{\rm k,max}}$ if
$v_{{\rm k,max}} > v_{{\rm esc,i}}$, or unity otherwise. Assuming that $v_{{\rm k,max}}$ 
is larger than $v_{{\rm esc,i}}$ for all BHs in the subsystem under consideration, in order to 
obtain a given retention fraction we require 
$v_{{\rm k,max}} = \sum v_{{\rm esc,i}} \,/\, (f_{{\rm BH}} N_{{\rm BH}})$. For Run 5 we have
that $N_{{\rm BH}} = 198$, and require that $f_{{\rm BH}} = 0.5$, and we compute the sum using 
the distributions displayed in Fig. \ref{f:bhkicks}. We found that in this scenario 
$v_{{\rm k,max}} = 17.5$ km$\,$s$^{-1}$, determined by adopting the mean of the results for the 
two measured distributions (i.e., with and without the inherited BH kinetic energy). No
physical meaning should be read into our choice of a uniform kick distribution -- we
selected it here simply for convenience. The above process could easily be generalized to
any desired probability distribution.

With the natal kick distribution described above implemented in Run 5, as expected we observed 
a significant number of BHs escaping from the cluster shortly after their formation. All $198$ BHs 
in the simulation are created by $\tau = 10$ Myr, and the first BH escapes occur at $\tau = 14$ 
Myr. By $\tau = 100$ Myr, $102$ BHs have left the cluster, and the escape rate has dropped to
approximately one BH every $100$ Myr. Subsequently, BHs appear to be leaving the cluster due to 
straightforward relaxation and evaporation processes -- by the time of the creation of the 
first stable BH binary at $\tau \approx 1350$ Myr, a further $9$ BHs have been removed. Hence, 
we estimate the BH retention fraction in this realization of Run 5 to be $f_{{\rm BH}} = 0.485$, 
but it could be as low as $f_{{\rm BH}} = 0.440$ depending on whether BHs escaping between 
$\tau \approx 100-1350$ Myr are included in the definition. We note that re-running this 
simulation with a new random seed would result in a different retention fraction, since, in 
contrast to all our previous models, the ejection of BHs due to natal kicks is a stochastic 
process in Run 5.

\begin{figure}
\begin{center}
\includegraphics[width=0.45\textwidth]{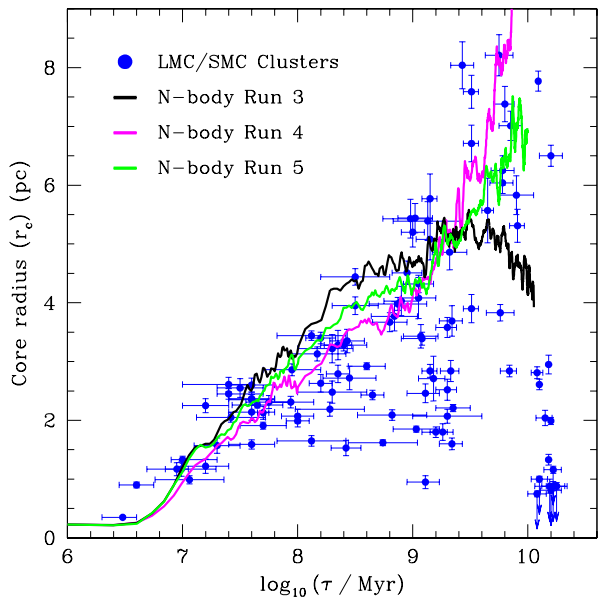}
\caption{Core radius evolution of $N$-body Run 5, with the evolution of Runs 3 and 4
plotted for comparison. All three models possess identical initial conditions, including
strong primordial mass segregation ($T_{{\rm MS}} = 450$ Myr) and the formation of $198$ BHs. 
The only difference between them is the BH retention fraction, which is zero for Run 3,
unity for Run 4, and approximately $0.485$ for Run 5 -- i.e., in Run 5 $96$ BHs are still 
present after $100$ Myr.}
\label{f:evolrun5}
\end{center}
\end{figure}

The evolution of Run 5 on the radius-age plane is plotted in Fig. \ref{f:evolrun5}, along
with the evolution of Runs 3 and 4 for comparison. All three models possess identical initial 
conditions -- the only difference between them is the BH retention fraction. At all times, 
the core radius of Run 5 lies in between those measured for Runs 3 and 4. During the first, 
early, phase of core expansion, this is a consequence of the intermediate BH retention fraction 
in Run 5 -- this model loses less mass than Run 3 but more than Run 4. The second phase of core 
expansion, due to BH activity, begins at $\tau \approx 1400$ Myr in Run 5. This is noticeably
later than in the previous models; furthermore, the rate of expansion is clearly not as great 
as is observed in Run 4 where twice as many BHs are retained.

\begin{figure}
\begin{center}
\includegraphics[width=0.48\textwidth]{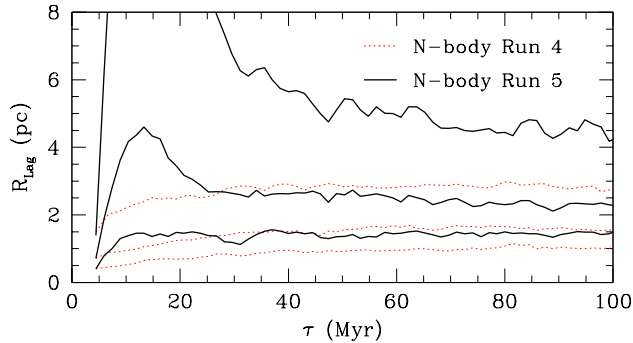}
\caption{Early evolution of the $10\%$, $25\%$ and $50\%$ BH Lagrangian radii in Run 5 (solid lines). 
The evolution of the same radii in Run 4 are plotted for comparison (dotted lines). The BH subsystem 
in Run 5 expands significantly at early times primarily due to the non-zero natal kicks, although
the BHs do also share in the general expansion of the cluster due to the rapid mass-loss from stellar 
evolution which is occurring during this period.}
\label{f:bhlagradkick}
\end{center}
\end{figure}

\begin{figure}
\begin{center}
\includegraphics[width=0.48\textwidth]{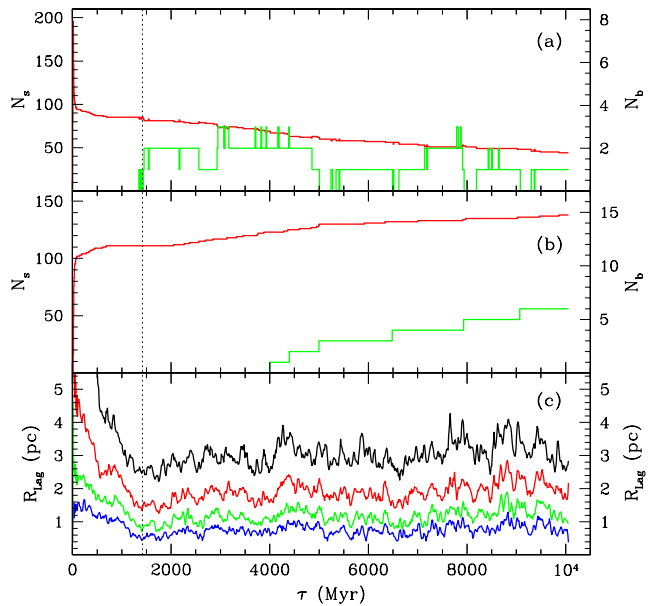}
\caption{Properties of the BH population in Run 5 as a function of time: (a) the number 
of single BHs (upper line) and binary BHs (lower line) in the cluster; (b) the cumulative numbers 
of escaped single BHs (upper line) and binary BHs (lower line); and (c) the {\it black hole} 
$10\%$, $25\%$, $50\%$ and $75\%$ Lagrangian radii (respectively, the innermost to outermost 
lines). The vertical dotted line indicates $\tau = 1400$ Myr, the approximate time when core 
expansion due to BH activity begins. Note the different axis scales on either side of panels 
(a) and (b).}
\label{f:bhevolkick}
\end{center}
\end{figure}

The evolution of the BH population in Run 5 is illustrated in Figs. \ref{f:bhlagradkick} and
\ref{f:bhevolkick}. The first of these shows the very early evolution of the $10\%$, $25\%$ 
and $50\%$ BH Lagrangian radii, compared to Run 4. Upon formation of the BHs, the BH Lagrangian 
radii are initially identical in Runs 4 and 5 because they share identical initial conditions
and random seeds. However, the Lagrangian radii in Run 5 immediately undergo a large degree
of expansion. This is primarily in response to the non-zero natal kicks given to the BHs,
although the BH subsystem does share in the general expansion of the cluster which is occurring
during this period due to the rapid mass-loss from stellar evolution. For comparison, the Run 4 
Lagrangian radii are expanding only in response to this mass-loss. The Lagrangian radii in Run 5
exhibit large bumps at early times -- these are most prominent in the $25\%$ radius at $\tau < 30$
Myr and the $50\%$ radius at $\tau < 50$ Myr. These features are due to the large number of BHs 
moving outward in the cluster on their way to escaping. The majority of these BHs have been removed 
from the cluster by $\sim 50$ Myr. After this point, the Lagrangian radii are still greatly
inflated compared to those for Run 4. This is due to the extra kinetic energy received by the 
retained BHs in Run 5 at their creation.

Fig. \ref{f:bhevolkick} shows the long-term evolution of the BH subsystem in Run 5. Because
of the extra kinetic energy the retained BHs in this model receive at birth, they take significantly
longer to sink to the cluster centre via dynamical friction than do the BHs in previous models. 
In addition, there are fewer BHs retained in Run 5, so once they have accumulated in the cluster
core, they interact more infrequently than in previous comparable models such as Run 4. The first 
BH binary does not form in Run 5 until $\tau = 1350$ Myr, much later than in our previous models. 
Binary hardening, BH scattering and BH ejection subsequently begin; however, the rates of all these 
processes are considerably reduced compared to previous simulations. The first ejection of a single 
BH after the formation of the first BH binary does not occur until $\tau = 2050$ Myr, while the 
first ejection of a BH binary does not occur until $\tau = 4000$ Myr. 

By the end of Run 5, at $\tau_{{\rm max}} = 10.06$ Gyr, only six BH binaries 
have been ejected. In common with earlier models, a significant population of BHs still remains
in Run 5 at this point, consisting of $44$ single BHs and one BH binary. The ejected BH binaries
possess properties very similar to those observed for Run 4. Two have eccentricities in the
range $0.8 < e < 0.95$ while the remaining four have $e < 0.8$. The closest ejected BH binary
has separation $a_b = 7.55$ AU, while the least tightly bound has $a_b = 61.0$ AU. The mean
mass of individual BHs in the ejected binaries is again greater than the ensemble average, at
$10.65\,{\rm M}_\odot$.

The reduced activity of the central BH subsystem in Run 5 compared with our other models explains 
the somewhat slower expansion of the core radius in this simulation. Despite this, the evolution
of $r_c / r_{h,l}$ is very similar to earlier models with retained BH populations. Once the
late, prolonged phase of core expansion begins in Run 5 (i.e., after $\tau \approx 1400$ Myr),
the locus traced by $r_c / r_{h,l}$ lies exactly on top of that for Run 4, reaching 
$r_c / r_{h,l} \sim 0.8$ by the end of the simulation. This indicates that despite the reduced
heating rate due to the BH population (and hence slower expansion of $r_c$), the distribution
of this heating within the cluster is similar to that for Runs with larger numbers of BHs.
Overall, Run 5 clearly demonstrates that complete BH retention is not necessary for significant 
and prolonged core expansion to occur -- even with half the number of retained BHs as Run 4, 
Run 5 still reaches the upper right-hand corner of the radius-age plane after $\sim 10$ Gyr of 
evolution.

\subsection{Run 6: Can neutron stars replace BHs?}
\label{ss:ns}
Finally, we computed one further model, designed to investigate whether the influence of a 
population of BHs is necessary for prolonged core expansion to develop in a cluster, or whether
such expansion can also result from similar dynamical processes involving larger numbers
of less massive stellar remnants such as neutron stars (NSs). To this end we set 
up the new simulation, named Run 6, so that it was initially identical to Run 1 -- that is, 
possessing no primordial mass segregation and retaining no BHs. However, unlike Run 1 where 
NSs were formed with large kicks so that the NS retention fraction $f_{{\rm NS}} = 0$, in Run 
6 we set the natal NS kicks to be zero in order to achieve the extreme case that 
$f_{{\rm NS}} = 1$. In all, $425$ NSs are formed in Run 6 from supernova explosions occurring
between $\tau \approx 10-43$ Myr. The masses of these NSs lie in the range 
$1.30 \leq m_{{\rm NS}} \leq 2.32\,{\rm M}_\odot$. We extended the duration
of Run 6 to be as long as that for Run 1 (i.e., $\tau_{{\rm max}} \approx 20$ Gyr), to enable 
a comparison between the full long-term development of the two models.

\begin{figure}
\begin{center}
\includegraphics[width=0.45\textwidth]{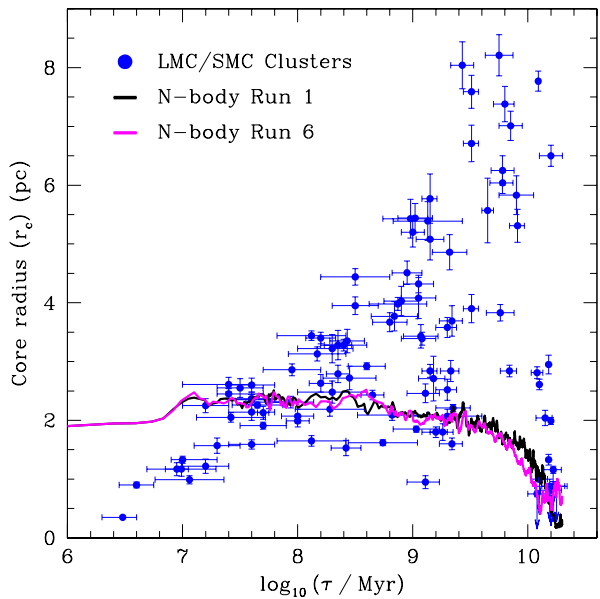}
\caption{Core radius evolution of $N$-body Run 6, with the evolution of Run 1 plotted for 
comparison. These two models possess identical initial conditions -- neither has any primordial
mass segregation, and $f_{{\rm BH}} = 0$ in both. The only difference between them is that in
Run 6 neutron stars are formed with no natal kicks so that $f_{{\rm NS}} = 1$, whereas in
Run 1 they are formed with large natal kicks so that $f_{{\rm NS}} = 0$. Hence, Run 6 retains
$425$ neutron stars, which are not present in Run 1. Unlike a BH population, the NS population
in Run 6 does not lead to core expansion, but does cause the cluster to enter the core
collapse phase at an earlier age.}
\label{f:evolrun6}
\end{center}
\end{figure}

The evolution of Run 6 on the radius-age plane may be seen in Fig. \ref{f:evolrun6}. Clearly,
the retention of a large number of NSs in this model does not result in core expansion at
any stage during the lifetime of the cluster. In fact, the evolution is remarkably similar
to that of Run 1, with the cluster undergoing many Gyr of gradual contraction before
entering the core collapse phase. By the end of the early rapid mass-loss period at roughly 
$\tau \sim 100$ Myr, the median relaxation time in Run 6 is very similar to that in Run 1 
at the same age -- i.e., $t_{rh} \sim 2$ Gyr. In the absence of any retained BHs, the NSs
are the most massive objects in the system and hence begin to sink to the cluster centre on 
a time-scale of $\sim (m_{*} / m_{{\rm NS}})\, t_{rh} \approx 500$ Myr. However, the NSs
are not very many times more massive than the otherwise most massive stars in the cluster, 
and so the central density of NSs never exceeds that of the other cluster members by a
sufficient degree that the NS subsystem is unstable to a runaway collapse 
\citep[][ Eq. 3-55]{spitzer:87}. Hence, the NS subsystem evolves quite differently to the
BH subsystems in our previous models, which did become unstable to runaway collapse.
No NS binaries are formed in the central core, and consequently, widespread scattering and 
ejection of NSs does not occur. As a result, Run 6 proceeds towards core collapse rather
than undergoing prolonged core expansion.

From Fig. \ref{f:evolrun6} it is evident that Run 6 enters the core collapse phase 
at a significantly earlier time than does Run 1 -- observationally, the point of deepest 
collapse (smallest core radius) occurs at $\tau \approx 12.8$ Gyr in Run 6, compared with 
$\tau \approx 17.4$ Gyr in Run 1. At any given age, the median relaxation time in Run 6 is 
very similar to that in Run 1, so that the point of deepest collapse in Run 6 occurs 
after significantly fewer integrated median relaxation times than in Run 1 -- $4.40\,t_{rh}$ 
as opposed to $8.37\,t_{rh}$. More enlightening is to examine the relaxation time in
the central core of each cluster, $t_{r0} \propto v_s r_c^2 m_{*0}^{-1}$, where $v_s$ is 
the velocity scale in the core and $m_{*0}$ is the mean mass of all the particles in thermal 
equilibrium in the central parts \citep[e.g.,][]{meylan:87}. Calculating for each model
the integrated number of central relaxation times which have elapsed by the time the point of 
deepest collapse occurs, the two values are within $\sim 10$ per cent of each other.
The central relaxation time in Run 6 is generally much shorter than in Run 1, due to the larger 
mean mass of the centralmost objects (predominantly NSs). 

It is also evident from Fig. \ref{f:evolrun6} that during collapse, the smallest observed 
core radius in Run 6 is larger than the smallest observed core radius in Run 1. This is
due to the luminosity cut-offs inherent in the calculation of $r_c$. In Run 1, the stars
contributing most of the light for the calculation of $r_c$ are also among the most massive 
remaining members, and hence typically reside in the innermost cluster regions during 
the core collapse phase. In Run 6 however, the centralmost members are the NSs, which do not 
contribute light for the calculation of $r_c$. The most luminous stars in Run 6 therefore 
appear to constitute a more widely distributed ``core'' during the late phases of evolution 
than do those in Run 1. The larger core radius during collapse in Run 6 is also reflected
in the evolution of $r_c / r_{h,l}$ for this model. While the behaviour of this ratio is 
very similar to that for Run 1, during collapse $r_c / r_{h,l}$ oscillates around $\sim 0.2$ 
rather than the smaller values ($r_c / r_{h,l} \la 0.05$) observed for Run 1.

Just as in Run 1, the point of deepest collapse in Run 6 coincides with a spate of binary 
formation in the core. This time, the binaries generally possess at least one NS member; 
several of them are NS-NS binaries. These are the first binary objects involving NSs to be 
created in Run 6 -- no such objects are formed at earlier times in this model.

\section{Discussion and Summary}
\label{s:discussion}

\begin{figure*}
\begin{minipage}{175mm}
\begin{center}
\includegraphics[width=120mm]{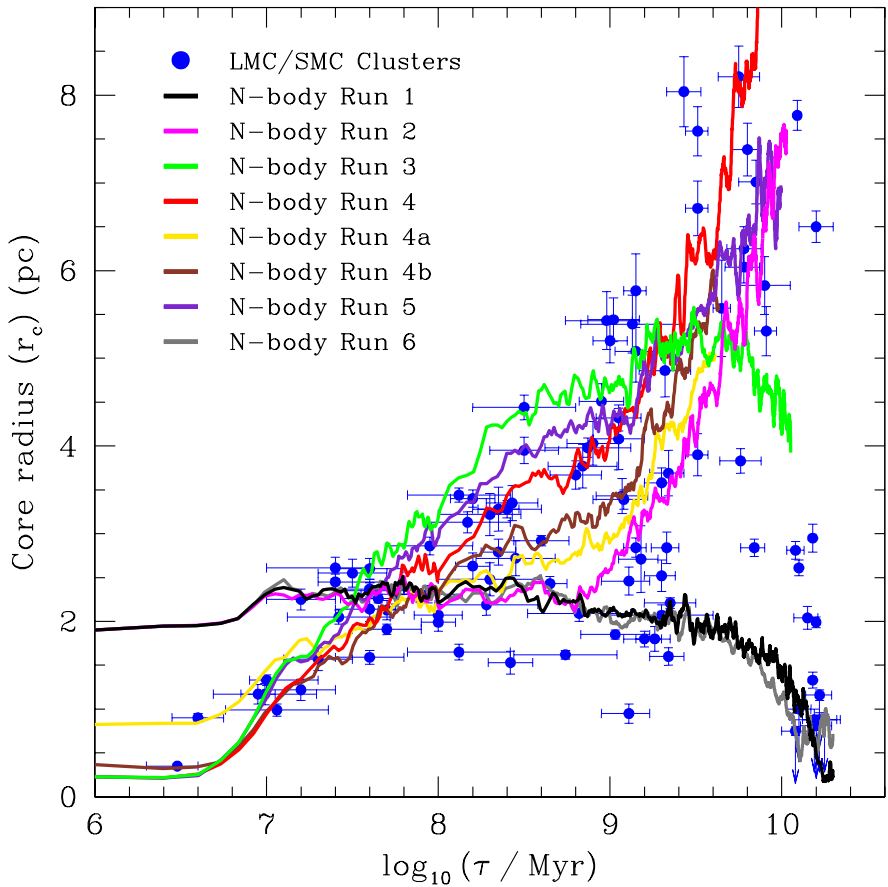}
\caption{Our full suite of eight $N$-body Runs plotted together for comparative purposes.
In combination, the two core expansion processes described in this paper lead to a wide 
variety of evolutionary paths on the $r_c$ versus age plane, which fully cover the observed 
distribution of massive Magellanic Cloud star clusters.}
\label{f:evolall}
\end{center}
\end{minipage}
\end{figure*}

In this paper we have presented an ensemble of eight large-scale $N$-body simulations aimed
at directly modelling the evolution of massive star clusters like those observed in the
Magellanic Clouds. Figure \ref{f:evolall} shows the core radius evolution of all eight
models plotted on the same set of axes, for the purposes of direct comparison.
Using these models we have identified two physical processes which 
can occur in such clusters and result in substantial core expansion -- mass-loss due to rapid 
stellar evolution in a cluster which is mass-segregated or otherwise centrally concentrated, 
and heating due to a significant population of retained stellar-mass ($\sim 10\,{\rm M}_\odot$) 
BHs formed in the supernova explosions of the most massive cluster stars. These two processes 
operate over different time-scales and at different stages in a cluster's life. The former only 
occurs during the first $\sim 100-200$ Myr after the formation of a cluster, when massive stars 
are still present. These evolve rapidly, losing a large fraction of their mass as they do so. 
The latter begins, at the earliest, several hundred Myr after the formation of the cluster 
but may remain active for at least a Hubble time beyond this starting point. In combination, 
these two processes can lead to a wide variety of evolutionary paths on the core-radius versus 
age plane, which fully cover the observed distribution of massive star clusters. They therefore 
define a physically-motivated dynamical explanation for the radius-age trend 
observed for the star cluster systems belonging to the Magellanic Clouds.

Our $N$-body modelling has allowed us to examine in more detail the behaviour of each
of these core-expansion mechanisms. As stated above, the phase of severe mass-loss due to
rapid stellar evolution is mostly complete by $\sim 100-200$ Myr into a cluster's life, by
which time all the most massive stars in the cluster have expired.  
Mass-loss due to stellar evolution still occurs after this point; however, it is from much 
less massive stars and therefore proceeds at a far more gradual rate without noticeably 
affecting the core radius evolution of the host cluster. 
Our simulations show that the amount of observed core expansion in a cluster due to the 
early mass-loss phase depends on both the degree of primordial mass segregation in the cluster, 
and the amount of mass lost in relation to the total cluster mass. In models where the former
parameter is held constant and the latter parameter is varied (e.g., Runs 3, 4, and 5), the cluster
losing the most mass expands the fastest. In models where the latter parameter is held constant
and the former parameter is varied (e.g., Runs 2, 4, 4a, and 4b), the cluster with the most
significant degree of primordial mass segregation expands the fastest. Furthermore, the early 
rapid phase of mass-loss does not cause any significant core expansion in our models unless the 
mass-loss is centrally concentrated -- models which do not possess any primordial mass
segregation exhibit essentially no early expansion. In models where early expansion occurs,
the ratio of the core radius to half-light radius $r_c / r_{h,l}$ increases significantly.
This is in contrast to models which do not undergo early expansion, where $r_c / r_{h,l}$
remains fairly constant with time. Inflated values of $r_c / r_{h,l}$ may remain for 
$\ga 10$ Gyr in some clusters (cf. Run 3), since the central and median relaxation times
in these objects become rather long.

Since the amount of mass lost in the early rapid stellar evolution phase is an important
contributor to the amount a cluster core expands during this phase, the expansion is
effectively regulated by the cluster's stellar IMF, modulated by second order effects such 
as BH retention. A steep IMF results in proportionally few high-mass members and hence a small 
amount of early expansion, whereas a flat IMF results in proportionally many high-mass members 
and hence a large amount of early expansion. In principle, therefore, significant inter-cluster 
IMF variations could lead to a variety of dramatically different paths across the radius-age 
plane at ages $\tau \la 200$ Myr, and consequently induce a large spread in the observed cluster 
distribution  \citep[e.g.,][]{elson:89}. However, there is an increasing body of evidence that 
large-scale variations in the stellar IMF between Magellanic Cloud clusters are not present 
\citep[e.g.,][]{kroupa:01,degrijs:02c}. This in turn 
suggests that inter-cluster variations in the degree of primordial mass segregation or central 
concentration may be the primary driver of the observed spread in the radius-age distribution 
at young ages. The sharp upper envelope of the distribution at ages less than a few hundred
Myr therefore most likely reflects an upper limit to the degree of primordial mass segregation
present in Magellanic Cloud clusters. Indeed, our model with an IMF matching that generally
observed for Magellanic Cloud clusters \citep{kroupa:01}, and an initial structure (including
mass segregation) matching that observed for R136 (which is the most extreme young object 
presently observed in the Magellanic Clouds) evolves along a path closely matching the
upper envelope of the radius-age distribution at early times.

One important process which can affect the early evolution of a massive star cluster, but 
which is not included in our $N$-body models, is the expulsion of gas left over from the 
star formation process. This residual gas is removed from the cluster within the first 
$\sim 10$ Myr due to the combined effect of massive stellar winds and supernova explosions. 
Just as with the early mass-loss due to stellar evolution, mass-loss due to gas expulsion can 
cause cluster core expansion, although typically on an even shorter time-scale of 
$\approx 10-20$ Myr \citep[e.g.,][]{bastian:06,goodwin:06}. The larger the mass of expelled
gas, the larger the degree of core expansion; if the mass of expelled gas is too great, the 
cluster may become unbound. In cases where the star formation efficiency is relatively high, 
and in the absence of sustained mass-loss due to stellar evolution, clusters soon settle 
into new equilibrium states with core radii generally not much larger than their initial 
values (see \citealt{goodwin:06}, Fig. 2). Therefore, gas expulsion may be affecting 
the core radius evolution of clusters younger than $\sim 50$ Myr in our radius-age plot, but
is unlikely to be of relevance to cluster evolution on longer time-scales. 
To correctly model the effects of gas loss occurring in combination with early stellar 
mass loss on the evolution of the various types of clusters studied in the present work 
will require more sophisticated codes than are presently available.

Core expansion due to the dynamical influence of a population of retained stellar-mass BHs 
in a cluster occurs over a much longer timescale than that due to early mass loss. 
Our simulations show that the BH population in a cluster only induces core expansion once the
BHs have accumulated in a sufficiently dense central subsystem that BH binaries are
created. These binaries are the catalyst for core expansion, since it is the interactions
between them and other single and binary BHs which lead to BH scattering and ejection, and 
subsequent heating of the central cluster regions. We do not observe core expansion due to
the BH population prior to the formation of BH binaries in any of our simulations. In 
particular, the sinking and accumulation of BHs in the core does not appear to affect the
observed core radius.

In our models, the time at which the first BH binaries are formed is relatively independent 
of the early evolution of the cluster. Models which are identical but for widely 
varying degrees of primordial mass segregation and hence widely varying amounts of early 
expansion (Runs 2, 4, 4a, and 4b) all form their first BH binaries at ages of roughly $\sim 500-600$ 
Myr. Even though BHs are preferentially formed very centrally in a model with a significant degree 
of primordial mass segregation compared to a model with no primordial mass segregation, the 
former object undergoes very significant early expansion compared to the latter. The BH subsystem 
does not escape this expansion, and by $\tau \sim 200$ Myr it has roughly the same 
distribution within the cluster as does the BH subsystem in the initially non-segregated model. 
The subsequent evolution of the two BH populations is very similar.
The time of formation of the first BH binaries is, however, strongly sensitive to the natal 
kicks received by the BHs at formation. In the case of non-zero kicks, retained BHs take 
longer to accumulate in the cluster centre than in the case of no kicks, due to the extra kinetic 
energy they receive at birth. In addition, non-zero natal kicks generally result 
in the expulsion of some fraction of the BH population, leading to a smaller retained BH 
subsystem and a smaller probability per unit time of BH binary formation. Although our modelling
did not test it, the time of formation of the first BH binaries is also expected to be sensitive
to the mean BH mass. More massive BHs will sink to the cluster centre much more rapidly than
less massive BHs, and hence form a dense central core at a significantly earlier time.

Once BH binaries have formed in a cluster and core expansion begins, the rate of expansion
is dependent on the number of BHs in the cluster. A cluster with fewer BHs expands more
slowly than an otherwise identical cluster with more BHs (cf. Runs 4 and 5). This is
because the interaction rate between BHs in the cluster centre is much lower for the model
with the smaller number of BHs, so that fewer BHs are scattered and ejected per unit time 
and the rate of heating of the cluster is reduced. The interaction rate between BHs in the 
cluster centre is also apparently sensitive to the density of the surrounding stellar core
-- it is significantly reduced in lower density clusters (cf. Runs 2 and 4). These observations 
have important implications for the survivability of BH subsystems within clusters. As the number 
of BHs in a cluster decreases due to the ejection of BHs after close encounters, and the central
density of the cluster decreases due to the expansion of its core, the interaction rate between 
BHs in the cluster centre also decreases. This in turn results in a lower BH ejection rate, 
allowing the BH population in a cluster to survive much longer than previously believed.
All our long-duration simulations with retained BHs still possess a sizeable BH population
after a Hubble time of evolution. As a result, some degree of core expansion is still occurring
in these models at late times.

We emphasize that even though most of our models examine the scenario 
where {\it all} BHs are retained in a cluster, such an extreme case is not {\it required} 
for core expansion to occur. We still observe significant expansion
in the more moderate case of $\sim 50$ per cent retention, although the rate of
expansion is reduced due to the factors outlined above.

It is also worth emphasising that while rapid mass-loss due to stellar evolution
is the dominant cluster core expansion process at early times ($\tau \la 200$ Myr), that
expansion ceases as the mass-loss rate slows. This process therefore {\it cannot} drive the
full observed radius-age distribution, which still exhibits a significantly increasing
spread in core size at much later times. A cluster which has expanded during its first
few hundred Myr of evolution, but which has not retained a BH population sufficient to 
induce additional late-time expansion, begins to contract again as two-body relaxation
processes take over (Run 3). Our models {\it only} achieve core sizes matching those observed 
for the most extended $\sim 10$ Gyr old Magellanic Cloud clusters if expansion due to a
retained BH population also occurs. This long-term expansion cannot be reproduced 
by other types of stellar remnants such as neutron stars (Run 6). Such remnants are not of high 
enough mass to accumulate in a central subsystem of sufficient density to allow 
frequent formation of the binary objects which drive the cluster heating.

The ratio $r_c / r_{h,l}$ evolves very similarly in all of our models where core expansion due
to a retained population of BHs occurs, once this phase has started. By $\tau \sim 10$ Gyr, the 
ratio approaches a large value of $r_c / r_{h,l} \approx 0.8$, comparable to the largest values 
observed for old Magellanic Cloud clusters, and Galactic globular clusters. This is irrespective
of the early evolution of a cluster (i.e., whether expansion due to early mass-loss occurs or
not), the time of onset of the expansion due to BHs, and the subsequent rate of this expansion.
As described in Section \ref{s:results}, these observations are compatible with those presented
recently by \citet{hurley:07}.

Several other mechanisms are known to be able to sustain large or expanding cores in massive
clusters. For example, the presence of primordial binary stars may stall core collapse, while
the presence of a central IMBH may result in cluster expansion \citep{baumgardt:04a,baumgardt:04b}. 
The heating effect of stellar-mass BHs, as considered in this paper, is far more efficient than 
the heating effect due to primordial binaries in comparable clusters. To transfer binding energy 
from primordial binaries to other cluster members requires frequent interactions and hence a
dense environment. For most of their lives, Magellanic Cloud clusters are not sufficiently
dense, as has been demonstrated from $N$-body modelling by \citet{mackey:03}\footnote{This
Ph.D. Thesis can be supplied by ADM on request.}. Furthermore, heating due to primordial binaries
is self-regulated: a dense core will expand, reducing the interaction rate and switching
the heating off until the core contracts again. Primordial binaries therefore cannot sustain 
the type of core expansion observed in our $N$-body models. It is more difficult to estimate the 
relative heating efficiency of stellar mass BHs compared to that of a central IMBH. 
\citet{baumgardt:04a,baumgardt:04b} display the evolution of the Lagrangian radii of their large 
$N$-body models, which do show significant expansion. However, it is difficult to disentangle the 
amount of heating due to mass-loss from stellar evolution from the amount due to the central IMBH. 
Based on the material presented by \citet{baumgardt:04a,baumgardt:04b}, we estimate that heating 
due to stellar-mass BHs is probably at least as efficient as that due to a central IMBH.

The scenario outlined above as a dynamical explanation for the radius-age trend observed in
the Magellanic Clouds requires significant variations in BH population size (i.e., in the BH 
retention fraction) between otherwise very similar clusters. Clusters which have 
developed very large core radii by the time they are $\ga 12$ Gyr old (e.g., the LMC clusters
NGC 1841 and Reticulum) must have managed to retain a significant BH population. Conversely, 
clusters which have entered core collapse at late times (e.g., the LMC clusters NGC 2005 and 
2019) are unlikely to have retained very many BHs -- for example, \citet{hurley:07} showed 
that even one BH binary in a cluster can prevent the collapse of its core. 

There are a number of possibilities which could lead to inter-cluster variability in the 
BH population size. First, we note that the number of BH-forming stars in a given cluster
is only a tiny fraction of the total number of stars in the cluster, so there will always be
sampling-noise variations between clusters. In addition, the formation of a BH in a supernova
explosion is sensitive to many features of the prior evolution of the progenitor star, in 
particular how much mass it loses as it evolves. Factors which introduce mass-loss
variations, such as binarity, chemical inhomogeneities or a dispersion in stellar rotation, 
are therefore likely to further accentuate the stochastic fluctuations in BH population size between 
clusters. In principle, inter-cluster variations in the stellar IMF would also strongly
affect BH population sizes; however, as we noted earlier, such variations are not observed. 
Observations do suggest that the maximum stellar mass in a cluster correlates with the total
cluster mass \citep[e.g.,][]{weidner:06}. Hence, even if the stellar IMF is universal between 
clusters, smaller clusters will have a lower maximum stellar mass and thus fewer BHs relative 
to the total cluster mass than will larger clusters. 

Natal BH kicks are also critical. At present, these are poorly constrained both by theory 
and observation. Typical estimates lie in the range $0 \la v_{{\rm kick}} \la 200$ 
km$\,$s$^{-1}$, with kicks of a few tens of km$\,$s$^{-1}$ possibly favoured 
\citep[e.g.,][ and references therein]{willems:05}. If BH natal kicks are indeed typically
a few tens of km$\,$s$^{-1}$ in magnitude, then they are roughly comparable to the escape
velocity of a massive stellar cluster. In this case, the structure of the host cluster
when the BHs are formed (i.e., before $\tau \approx 10$ Myr) can strongly affect the
retention fraction. For example, BHs formed in a dense, strongly mass segregated cluster
are more easily retained that BHs formed in a comparably massive but less dense, non-segregated
cluster (e.g., Fig. \ref{f:bhkicks}). The retention fraction will also be sensitive to
the overall initial mass of the cluster, as well as to factors which affect the very
early evolution of the cluster such as residual gas expulsion. Stellar binarity may also 
play an important role. 

It is interesting to note that theoretical models suggest BH formation to be a strongly
sensitive function of metal abundance, in that BH production is apparently more
frequent, and $m_{{\rm BH}}$ is typically greater for metal poor progenitor stars than for metal 
rich progenitor stars \citep[see e.g.,][]{zhang:07}. Hence, the BH populations formed in clusters 
of very different metallicities are likely to be quite distinct. The strong age-metallicity 
relationships observed in both the LMC and SMC mean that this factor probably cannot
cause differences between the BH population sizes in compact and extended LMC or SMC clusters 
of a given age, since such objects will have approximately equal metallicities. 
However, the LMC and SMC age-metallicity relationships do suggest that any BH populations forming 
in present-day young Magellanic Cloud clusters are likely to be quite different to those which 
may have existed in Magellanic Cloud clusters that are now $\ga 12$ Gyr old.

The possibility of large-scale and prolonged core expansion has important implications for 
the study of all massive star clusters, including the globular clusters in our Galaxy and others.
Many such objects are at least an order of magnitude more massive than the models
presented in this paper. Even so, we expect the physical processes which we have described
will still operate in larger systems. 

Neglecting any stochastic fluctuations between clusters, the number of BHs formed in a 
cluster is, to first order, dependent only on the stellar IMF and the minimum progenitor 
mass. We do not expect these to change with cluster mass, so with all other parameters held 
constant, the mass fraction of BHs should remain the same for clusters of increasing mass. 
Similarly, the mass fraction lost due to rapid stellar evolution early on in a cluster's life
should also remain the same for clusters of increasing mass. Assuming that natal BH kicks are 
also not a function of cluster mass, the BH retention fraction should increase with increasing 
cluster mass, since it is more difficult to eject BHs from the deeper gravitational well of 
a more massive cluster. Overall, we therefore expect the relative size of retained BH 
populations should be larger for more massive clusters. Given the above, more massive clusters 
have at least the same potential for core expansion as do less massive clusters.

In terms of the dynamics of the expansion, the central densities of the model clusters we have 
studied in this paper are directly comparable to the central densities of more massive objects 
such as globular clusters. The central and median relaxation times in our models are also
commensurate with those calculated for typical globular clusters. Given this, we expect 
similar dynamical processes to operate on similar time-scales in clusters larger than our
present models, so that early mass-loss and BH heating will both still be effective at
inducing core expansion. The main difference is that it becomes more difficult to eject BHs 
as the cluster mass increases. Therefore, the mean time a BH remains in a cluster will increase
with the total mass. This will increase the potential of each BH to heat the cluster through
additional scattering-sinking cycles, and will allow BH binaries to harden further than they
would do in a less massive object. BH heating in more massive clusters is hence likely
to be even more efficient than it is in less massive clusters.

We therefore predict that some degree of core expansion is possible in any massive stellar 
cluster due to the processes outlined in this paper, irrespective of the total mass of the 
cluster. In many aspects of star cluster research, this possibility is not usually considered.
However, under a wide variety of circumstances, it could have an important effect on the problem 
under consideration. As a simple example, it is well known that diffuse extended clusters are 
far more susceptible to tidal disruption than are compact clusters. Prolonged core expansion 
in clusters could result in many more such diffuse extended objects in a given population than 
would otherwise be expected. This possibility is vital to incorporate into modelling where 
destruction rates are important, such as studies designed to investigate the evolution of the 
globular cluster mass function \citep[e.g.,][]{fall:01}, the past and future dissolution of 
globular clusters in the Galactic system \citep[e.g.,][]{gnedin:97}, or whether the super star 
clusters observed in starburst galaxies will eventually evolve into objects resembling classical 
globular clusters \citep[e.g.,][]{degrijs:07}. 

Another example involves the measurement of dynamical mass estimates for young massive 
clusters in external galaxies. Such measurements are sometimes used to infer the stellar IMF 
in such clusters. \citet{bastian:06} and \citet{goodwin:06} have demonstrated that very 
young clusters ($\tau \la 50$ Myr) may be out of virial equilibrium due to the rapid expulsion 
of residual gas, so that dynamical mass measurements assuming virial equilibrium may be in
error. Our modelling has shown that significant core expansion due to stellar evolution 
occurs on a timescale close to $\sim 100$ Myr. Researchers should be aware of this additional
possibility when evaluating the properties of young clusters in external galaxies, although we 
note that it is not yet clear to what extent any signal due to such expansion would manifest 
in integrated cluster spectra. This is an avenue worthy of further investigation.

As a final example, consider the cluster half-mass radius, $r_h$. This quantity is often
assumed to be relatively stable for much of a cluster's life (cf. Fig. \ref{f:lrad1}), and 
is hence sometimes used to infer information about cluster formation \citep[e.g.,][]{vdb:04}.
However, if a cluster undergoes prolonged core expansion, $r_h$ is certainly {\it not} a 
stable quantity (Fig. \ref{f:lrad2}). Caution should therefore be exercised in the use of 
this parameter.

The possibility of core expansion may also help explain the properties of some of 
the more exotic star clusters discovered in recent years -- for example, the 
``faint fuzzies'' uncovered in several lenticular galaxies \citep{brodie:02}, 
the luminous extended clusters found in the halo of M31 \citep{huxor:05,mackey:06b},
and the diffuse star clusters located in a number of Virgo early-type galaxies \citep{peng:06}. 
All these newly-discovered clusters possess unusually extended structures compared to 
those of classical globular clusters. Core expansion, particularly the prolonged variety 
due to retained BHs, may offer a viable formation channel for these objects.

We conclude with a note on the possibility of testing observationally our prediction
of retained populations of stellar-mass BHs in some massive star clusters. While these BHs
cannot be observed directly, there are a number of means by which their presence might be
inferred in a cluster. One possibility is that a BH in a close binary with an evolved star 
is likely to be an X-ray source, as the star overflows its Roche limit and transfers gas onto 
the BH. Such BH X-ray sources are known in the field \citep[see e.g.,][]{casares:06} 
and one is known in an extra-Galactic globular cluster \citep{maccarone:07,zepf:07};
however, none have been found in Galactic globular clusters \citep{verbunt:06}. From our 
modelling, we know that clusters which do retain significant BH populations are, for most of 
their lives, objects in which the timescale for encounters between BHs and stars is very long
(due to the low stellar density in the extended core), but the timescale for encounters 
between BHs is comparatively short. Hence the creation of long-lived BH-star binaries 
is rare -- we did not observe any such objects in our simulations. It is therefore
unsurprising that no BH X-ray sources are known in the Galactic globular cluster system,
and only one is known in an external cluster.

The most promising means of inferring the presence of a BH population in a cluster is through
the dynamical effect it causes on the stellar component of the cluster. As we demonstrated
in Section \ref{ss:pair1}, unlike in the case of an IMBH, a significant stellar density
and velocity cusp does not develop about the compact central BH subsystem. None the less, the 
effect of the central concentration of unseen mass should be evident in the stellar motions
-- the velocity dispersion of the cluster should be larger than is to be expected solely from
the observed luminous mass. Observations to test this will be difficult, primarily because
the target clusters should be extended, diffuse objects with low velocity dispersions.
In addition, many will have relatively few targets suitable for spectroscopic radial velocity
measurements, such as luminous red giants. Even so, it may be possible to make sufficiently
precise observations with presently available $8-10\,$m-class facilities.

Finally, it seems likely that at least some BH binaries ejected from very massive clusters
will merge on a timescale $\la 12$ Gyr, and will therefore be sources of gravitational 
radiation detectable by interferometers such as LIGO, and in future, LISA. 
This possibility has previously been investigated in more detail by other authors 
\citep[see e.g.,][]{portegieszwart:00}. As described in Sections \ref{ss:pair1} and
\ref{ss:pair2}, the BH binaries ejected from our model clusters will not merge on a timescale
$\la 12$ Gyr; however several have orbital parameters not far from the required threshold.
A subset of the BH binaries ejected from more massive clusters than those studied here would 
almost certainly have orbital parameters well within this threshold.

\section*{Acknowledgements}
We thank Sverre Aarseth for the use of {\tt NBODY4}, and for numerous 
valuable discussions and on-going assistance. We also thank Jarrod Hurley for
providing code to calculate the absolute magnitudes of stars evolving in our
simulated clusters. We are grateful to Pete Bunclark and Mick Bridgeland for 
providing technical support with the IoA GRAPE-6 board, and to the anonymous 
referee for a thorough report which helped to improve the paper.
ADM is supported by a Marie 
Curie Excellence Grant from the European Commission under contract MCEXT-CT-2005-025869,
and is grateful for support from a PPARC Postdoctoral Fellowship during which much
of this work was completed. MIW acknowledges support from a Royal Society University 
Research Fellowship and, previously, from PPARC. MBD is a Royal Swedish Academy Research 
Fellow supported by a grant from the Knut and Alice Wallenberg Foundation. Some results 
in this paper are based 
on observations made with the NASA/ESA Hubble Space Telescope, obtained at the Space Telescope 
Science Institute, which is operated by the Association of Universities for Research in Astronomy, 
Inc., under NASA contract NAS 5-26555. These observations are associated with program \#9891.


\onecolumn
\appendix
\section{Analytic properties of the EFF family of models}
\label{a:profiles}
In this Appendix we present analytic expressions for the properties of a number of members 
of the general family of EFF models. As demonstrated in the present work, with the recent 
rapid increase in computing power and software sophistication, and hence the size and degree 
of realism feasible for $N$-body simulations, it may occur that such models represent more 
appropriate initial conditions for a given scenario than do the frequently adopted Plummer 
spheres or King models. With the formulae presented below, it is reasonably straightforward 
to develop procedures such as that described in Section \ref{ss:nbodycode} to generate the 
desired initial conditions.

\subsection{General properties}
The EFF models, after \citet*{elson:87}, are a family of models originally
defined through empirical fitting to the observed surface brightness profiles of young
massive star clusters in the LMC. These objects do not exhibit tidal cut-offs in their outer 
regions, and are therefore most appropriately
described by projected three-parameter models of the form:
\begin{equation}
\mu(r_p) = \mu_0 \left( 1+\frac{r_p^2}{a^2} \right) ^{-\frac{\gamma}{2}} ,
\label{ae:effproject}
\end{equation}
where $r_p$ is the projected radius, $\mu_0$ is the central surface brightness, 
$a$ is the scale radius, and $\gamma$ represents the power-law fall-off 
of the profile at large radii. 
These models can easily be deprojected to obtain the three-dimensional density:
\begin{equation}
\rho(r) = \rho_0 \left( 1+\frac{r^2}{a^2} \right) ^{-\frac{\gamma+1}{2}} \hspace{8mm} {\rm where} \hspace{3mm} \rho_0 = \frac{\mu_{0} \, \Gamma \left( \frac{\gamma + 1}{2} \right) }{a \sqrt{\pi} \,\, \Gamma \left( \frac{\gamma}{2} \right) } .
\label{ae:eff3d}
\end{equation}
In the above equation, $\Gamma$ is a standard gamma function. Since $\mu_0$ is the central 
surface brightness, here $\rho_0$ is the central luminosity density -- to obtain the central
mass density it is necessary to multiply by the global mass-to-light ratio $\Upsilon$. 
It can be seen that the 
three-dimensional density has exactly the same functional form as the projected density,
but with index $\gamma + 1$. By comparison with the more general spherically symmetric family of 
$(\alpha, \beta, \delta)$ models
described by \citet{zhao:96}\footnote{\citet{zhao:96} labelled these $(\alpha, \beta, \gamma)$ models;
however, we alter his nomenclature here to avoid ambiguity in the definition of $\gamma$.}, it is 
straightforward to see that the EFF models represent the subset with 
$(\alpha, \beta, \delta) = (\frac{1}{2}, \gamma+1, 0)$ and break radius $r = a$.

Assuming now that $\rho_0$ is a mass density, substituting Eq. \ref{ae:eff3d} into Eq. 2-22 from 
\citet{binney:87} and integrating yields a general expression for the gravitational potential 
of EFF models:
\begin{equation}
\Phi = -\frac{4}{3} \pi G \rho_0 \left[ \frac{3 a^2}{\gamma - 1} \left( 1+ \frac{r^2}{a^2} \right)^{-\frac{(\gamma-1)}{2}}\, + \,r^2\, _{2}F_{1}\left(\frac{3}{2}\,,\,\frac{\gamma+1}{2}\,;\,\frac{5}{2}\,;\,-\frac{r^2}{a^2} \right) \right] ,
\label{ae:effpotential}
\end{equation}
where $_{2}F_{1}(a,b;c;z)$ is Gauss's hypergeometric function. 

Similarly, the enclosed mass (or luminosity if $\rho_0$ is a luminosity density) as a function 
of radius can be derived by integrating Eq. \ref{ae:eff3d}:
\begin{equation}
M(r) = 4 \pi \int_{0}^{r} \rho(r') r'^{2} dr' \,\,= \frac{4}{3} \pi \rho_0  r^{3} \,_{2}F_{1} \left( \frac{3}{2},\frac{\gamma+1}{2};\frac{5}{2};-\frac{r^{2}}{a^{2}} \right) .
\label{ae:encmass}
\end{equation}
In the limit where $r \rightarrow \infty$, this expression converges only if $\gamma > 2$.
The asymptotic mass is given by $M_\infty = 2\pi \mu_0 \Upsilon a^2 / (\gamma -2)$.

Finally, rearranging and integrating the Jeans equations for a steady-state, spherically
symmetric, non-rotating cluster (i.e., \citet{binney:87} Eq. 4-54) yields a general
expression for the radial dependence of the isotropic velocity dispersion:
\begin{equation}
\sigma^2(r) = \frac{1}{\rho(r)} \int_r^\infty \rho(r')\frac{d\Phi(r')}{dr'} dr' ,
\label{ae:veldisp}
\end{equation}
which can, in principle, be evaluated numerically for all $\gamma$. However, for integer 
values of $\gamma$, the hypergeometric functions in Eqs. \ref{ae:effpotential} and
\ref{ae:encmass} reduce to elementary functions, resulting in straightforward analytic
expressions for $M(r)$ and $\sigma^2(r)$ which may be written into the computer code to
generate initial conditions. The best known of the analytic subset is the case when
$\gamma = 4$ -- the \citet{plummer:11} sphere. The properties of this model have been investigated
extensively by \citet{dejonghe:87} for mass-follows-light scenarios, while the more general
study of \citet{wilkinson:02} includes the possibility of a dark halo. Below, we consider
the less-well studied cases of $\gamma = 3, 5$, and\ $6$.

\subsection{The $\gamma = 3$ case}
The case where $\gamma = 3$ has been studied in passing as special cases of a general
ellipsoidal form by \citet{dezeeuw:85a,dezeeuw:85b}, who labelled the profiles 
``perfect spheres''. This particular case is of interest since it represents the EFF model 
with analytic expressions for $M(r)$ and $\sigma^2(r)$ which, in projection, is closest in 
form to the observed profiles of young massive star clusters in the LMC and SMC. \citet{elson:87}
found a median value of $\gamma = 2.6$ and a range $2.2 \la \gamma \la 3.2$ for their ten 
young LMC clusters, while the $18$ young LMC and SMC clusters plotted in Fig. 
\ref{f:youngclusters} in the present paper cover the range $2.05 \leq \gamma \leq 3.79$ 
and have a median value $\gamma = 2.67$. We started all the $N$-body simulations described 
in the present work with initial conditions generated from the following equations.

When $\gamma = 3$, the Gauss hypergeometric function in Eqs. \ref{ae:effpotential} and
\ref{ae:encmass} reduces to:
\begin{equation}
_{2}F_{1} \left( \frac{3}{2},2;\frac{5}{2};-\frac{r^{2}}{a^{2}} \right) \,\, = -\frac{3a^2}{2r^2} \left( \rule{0mm}{6mm} \left[1+\frac{r^2}{a^2}\right]^{-1} - \sqrt{-\frac{a^2}{r^2}} \, {\rm arctanh} \! \left[\sqrt{- \frac{r^2}{a^2}} \right] \right) \,\, = \frac{3a^3}{2r^3} \left( {\rm arctan} \! \left[ \frac{r}{a} \right] - \frac{r}{a} \left[1+\frac{r^2}{a^2}\right]^{-1} \right)
\label{ae:hypgeo3}
\end{equation}
Substituting into Eq. \ref{ae:encmass}, the enclosed mass $M(r)$ is then given by:
\begin{equation}
M(r) = 2 \pi \rho_0 \, a^3 \, \left(\arctan \! \left[\frac{r}{a}\right] - \frac{r}{a} \left[1+\frac{r^2}{a^2}\right]^{-1}\right) ,
\label{ae:encmass3}
\end{equation}
while carrying out the integration in Eq. \ref{ae:veldisp} provides the isotropic velocity dispersion:
\begin{equation}
\sigma^2 = -\frac{\pi G \rho_0}{8 a^2} (a^2 + r^2)^2 \left( \rule{0mm}{6mm} 3\pi^2 - \frac{1}{r (a^2 + r^2)^2} \left[ \rule{0mm}{5.5mm} 4 \left( \rule{0mm}{5mm} ar + (a^2 + r^2) \arctan \! \left[ \frac{r}{a} \right] \right) \left( \rule{0mm}{5mm} 4a^3 + 3ar^2 + 3r(a^2 + r^2) \arctan \! \left[ \frac{r}{a} \right] \right) \right] \right)
\label{ae:sigma3}
\end{equation}

\subsection{Steeper cases: $\gamma = 5$ and $\gamma = 6$}
Cluster models with steep density fall-offs do not seem to have been well studied in the literature,
if at all. For this reason, the properties of two such models are derived here. These $\gamma = 5$ and
$\gamma = 6$ cases {\it do} correspond to real objects -- old globular clusters can have observed
brightness profiles which fall off this steeply. For example, \citet{mackey:03a,mackey:03b,mackey:03c} 
found the LMC globular cluster NGC 1841 to have $\gamma = 4.55 \pm 0.61$, the SMC cluster NGC 339 to
have $\gamma = 5.21 \pm 0.99$, and the diffuse cluster 1 in the Fornax dSph to have 
$\gamma = 7.52 \pm 0.64$. It may well be desirable in future to model the evolution of clusters such 
as these. In this case, the equations below will allow suitable initial conditions to be simply 
constructed.

If $\gamma = 5$, the hypergeometric function in Eqs. \ref{ae:effpotential} and \ref{ae:encmass} 
reduces to:
\begin{eqnarray}
\lefteqn{ _{2}F_{1} \left( \frac{3}{2},3;\frac{5}{2};-\frac{r^{2}}{a^{2}} \right) \,\, = \frac{3a^2}{8r^2} \left(\sqrt{-\frac{r^2}{a^2}} (a^2+r^2)^2 \right)^{-1} \left( \rule{0mm}{6mm} a^2 \sqrt{-\frac{r^2}{a^2}}(r^2-a^2) + (a^2+r^2) \,{\rm arctanh} \! \left[\sqrt{-\frac{r^2}{a^2}}\right] \right) } \nonumber \\
& & \hspace{23.3mm} = \frac{3a^3}{8r^3} \left( \frac{ar^3 - a^3r}{(a^2+r^2)^2} + \arctan \! \left[\frac{r}{a} \right] \right) 
\label{ae:hypgeo5}
\end{eqnarray}
Substituting into Eq. \ref{ae:encmass}, as before, yields the enclosed mass $M(r)$:
\begin{equation}
M(r) = \frac{1}{2} \pi \rho_0 \, a^3 \left(\frac{a r^3 - a^3 r}{(a^2 + r^2)^2 } + \arctan \left[\frac{r}{a}\right] \right) ,
\label{ae:encmass5}
\end{equation}
while evaluating Eq. \ref{ae:veldisp} provides the isotropic velocity dispersion as a function of radius:
\begin{eqnarray}
\lefteqn{ \sigma^2(r) = \frac{\pi G \rho_0}{384 a^4r(a^2+r^2)} \left[ \rule{0mm}{6mm} 5a^8r(64-9\pi^2) + 4a^6r^3(173-45\pi^2) + 30a^4r^5(20-9\pi^2) - 180a^2r^7(\pi^2-1) - 45\pi^2r^9 \right. } \nonumber \\
& & \hspace{35mm} \left. + \,\,12(a^2+r^2)^2 \arctan\left[\frac{r}{a}\right] \left(16a^5 + 50a^3r^2 + 30ar^4 + 15r(a^2+r^2)^2 \arctan\left[\frac{r}{a}\right] \rule{0mm}{5mm} \right) \rule{0mm}{6mm} \right] .
\label{ae:sigma5}
\end{eqnarray}

The $\gamma = 6$ case is of particular interest as its properties are comparable in simplicity to
those of the widely used Plummer sphere ($\gamma = 4$). With $\gamma = 6$, the hypergeometric function 
in Eqs. \ref{ae:effpotential} and \ref{ae:encmass} can be written:
\begin{equation}
_{2}F_{1} \left( \frac{3}{2},\frac{7}{2};\frac{5}{2};-\frac{r^{2}}{a^{2}} \right) \,\, = \frac{a^3}{5} \left[\frac{5a^2+2r^2}{(a^2+r^2)^{\frac{5}{2}}} \right]
\label{ae:hypgeo6}
\end{equation}
which leads to the following straightforward expressions for the enclosed mass and isotropic 
velocity dispersion:
\begin{eqnarray}
M(r) = \frac{4}{15} \pi \rho_0 \, a^3r^3 \left[\frac{5a^2+2r^2}{(a^2+r^2)^{\frac{5}{2}}} \right] \\
\label{ae:encmass6}
\sigma^2(r) = \frac{1}{75} \pi G \rho_0 \, a^3 \left[\frac{11a^2 + 5r^2}{(a^2+r^2)^{\frac{3}{2}}} \right] .
\label{ae:sigma6}
\end{eqnarray}

In principle, it is possible to continue deriving similar analytic expressions for increasing integer 
values of $\gamma$; however we note that the expressions become correspondingly more complicated 
as $\gamma$ increases.

\label{lastpage}

\end{document}